\documentclass[aps,physrev,superscriptaddress,nofootinbib,preprintnumbers]{revtex4-2}

\usepackage{xcolor}
\usepackage{amsmath}
\usepackage{esvect} 
\usepackage{amsthm} 
\usepackage{amsfonts} 
\usepackage{hyperref} 
\usepackage{graphicx} 
\usepackage{soul} 
\usepackage{threeparttable} 
\usepackage{slashed}
\usepackage[bottom]{footmisc} 
\usepackage{placeins}

\newcommand{\inBraket}[1]{\left<#1\right>} 
\newcommand{\inCurve}[1]{\left(#1\right)}
\newcommand{\inSquare}[1]{\left[#1\right]}
\newcommand{\inCurly}[1]{ \left\{#1\right\} }
\newcommand{\inAbs}[1]{\left|#1\right|}
\newcommand{\lrarrow}[1]{\overset\leftrightarrow{#1}}

\newcommand{\melmom}[0]{\inBraket{x^2}}
\DeclareMathOperator{\Tr}{Tr} 
\newcommand{\gammapol}[0]{ \Gamma_{\text{pol}} }
\newcommand{\shortOP}[0]{\mathcal{O}^X}
\newcommand{\tskip}[0]{\tau_{\text{skip}}}
\newcommand{\corrTWO}[0]{C_{2\text{pt}}(t)}
\newcommand{\corrTHREE}[0]{ C_{3\text{pt}}^{\shortOP}(t,\tau) }

\newcommand*\Diff[1]{\mathop{}\!\mathrm{d^#1}\!}

\hypersetup{
    colorlinks,
    linkcolor={blue!60!blue},
    citecolor={blue!80!blue},
    urlcolor={blue!60!black}
}

\begin{document}

\preprint{DESY-26-061}

\title{Third moments of nucleon unpolarized, polarized, and transversity parton distribution functions from physical-point lattice QCD}



\author{Emilio Taggi}
\email[]{e.taggi@fz-juelich.de}
\affiliation{Jülich Supercomputing Center \& Center for Advanced Simulation and Analytics (CASA), Forschungszentrum Jülich, 52428 Jülich, Germany}
\affiliation{Helmholtz-Institut für Strahlen- und Kernphysik, Rheinische Friedrich-Wilhelms-Universität Bonn, 53115 Bonn, Germany}

\author{Michael Engelhardt}
\affiliation{Department of Physics, New Mexico State University, Las Cruces, NM 88003, USA}

\author{Jeremy R. Green}
\affiliation{John von Neumann-Institut für Computing NIC, Deutsches Elektronen-Synchrotron DESY, Platanenallee 6, 15738 Zeuthen, Germany}

\author{Stefan Krieg}
\affiliation{Jülich Supercomputing Center \& Center for Advanced Simulation and Analytics (CASA), Forschungszentrum Jülich, 52428 Jülich, Germany}
\affiliation{Helmholtz-Institut für Strahlen- und Kernphysik, Rheinische Friedrich-Wilhelms-Universität Bonn, 53115 Bonn, Germany}

\author{Stefan~Meinel}
\affiliation{Department of Physics, University of Arizona, Tucson, AZ 85721, USA}

\author{John W. Negele}
\affiliation{Center for Theoretical Physics---a Leinweber Institute, MIT, Cambridge, MA 02139, USA}

\author{Andrew Pochinsky}
\affiliation{Center for Theoretical Physics---a Leinweber Institute, MIT, Cambridge, MA 02139, USA}

\author{Marcel Rodekamp}
\affiliation{Institut für Theoretische Physik, Universität Regensburg, 93053 Regensburg, Germany}

\author{Sergey Syritsyn}
\affiliation{Department of Physics and Astronomy, Stony Brook University, Stony Brook, NY 11794, USA}
\affiliation{RIKEN/BNL Research Center, Brookhaven National Laboratory, Upton, NY 11973, USA}


\date{August 21, 2026}

\begin{abstract}
Using forward matrix elements of local leading-twist operators, we present a determination of the isovector third Mellin moments $\melmom$ of nucleon unpolarized, polarized, and transversity parton distribution functions.
Two lattice QCD ensembles at the physical pion mass are used, which were generated using a tree-level Symanzik-improved gauge action and 2+1 flavor tree-level improved Wilson Clover fermions coupling via 2-level HEX-smearing.
Leveraging a wide set of operators, two extraction methods for the matrix elements, and the automatic inclusion of model uncertainties via bootstrapped model averages, we extract values of the third Mellin moments.
This is the first direct calculation of these observables performed at the physical pion mass.
\end{abstract}


\maketitle


\section{Introduction\label{sec:introduction}}

Knowledge of parton distribution functions (PDFs) is crucial for making theoretical predictions and interpreting the results of high-energy-physics experiments at hadron colliders.
For this reason, several experimental groups have focused on
extracting PDFs from global fits of collider data \cite{DeRoeck:2011na, Forte:2013wc, Blumlein:2012bf, Perez:2012um, Ball:2012wy, Gao:2017yyd, Ethier:2020way}.
Numerous efforts also come from the lattice QCD community \cite{Lin:2017snn, Constantinou:2022yye, Lin:2025hka}.
Although a complete determination of PDFs is not possible using the standard tools of Euclidean lattice QCD, a first-principles computation can determine the values of the four first Mellin moments of PDFs.
Such a calculation provides useful constraints on the global fits of experimental data.
The first \cite{Alexandrou:2023qbg, Jang:2023zts, Walker-Loud:2019cif, Alexandrou:2019brg, Gupta:2018qil, Chang:2018uxx, Berkowitz:2017gql, Bhattacharya:2016zcn, Djukanovic:2024krw, Tsuji:2023llh, Bali:2023sdi, QCDSFUKQCDCSSM:2023qlx, Tsuji:2022ric, Djukanovic:2022wru, Park:2021ypf, Blossier:2013ioa, Hasan:2019noy, Harris:2019bih, Shintani:2018ozy, Ishikawa:2018rew, Liang:2018pis, Yamanaka:2018uud, Liu:2021irg, Abramczyk:2019fnf, Alexandrou:2022dtc, Bhattacharya:2015wna, Bhattacharya:2015esa} and second \cite{Alexandrou:2022dtc,Mondal:2020cmt,Alexandrou:2020sml,Alexandrou:2019ali,Djukanovic:2024krw,Rodekamp:2023wpe,Mondal:2021oot,Mondal:2020ela,Harris:2019bih,Yang:2018nqn,Green:2012ud,LHPC:2010jcs,Aoki:2010xg,Bali:2018zgl,Alexandrou:2017oeh,Abdel-Rehim:2015owa,Bali:2014gha} isovector Mellin moments have been calculated with high precision in the past, and both quantities appear in the most recent FLAG review \cite{FlavourLatticeAveragingGroupFLAG:2024oxs}.
However, the same level of precision has not yet been achieved for the isovector third Mellin moments.

Direct computations of the isovector third Mellin moments using dynamical fermions have been performed in earlier studies. 
In recent years, Ref.~\cite{Burger:2021knd} computed $\inBraket{x^2}_{\Delta u^+ - \Delta d^+}$ using a subset of the CLS ensembles \cite{Bruno:2014jqa}.
The next most recent direct calculation was from 2010 \cite{LHPC:2010jcs}, which computed $\inBraket{x^2}_{u^- - d^-}$ and $\inBraket{x^2}_{\Delta u^+ - \Delta d^+}$ using $2+1$ flavors of staggered sea quarks and domain wall valence quarks.
Refs. \cite{Gockeler:2004vx} and \cite{LHPC:2002xzk} computed $\inBraket{x^2}_{u^- - d^-}$ using $2$ flavors of clover quarks and 2 flavor of Wilson quarks, respectively.
Ref. \cite{Gockeler:2005vw} computed $\inBraket{x^2}_{\Delta u^+ - \Delta d^+}$ using $2$ flavors of clover quarks, and using a non-perturbative renormalization scheme.
No previous direct computations of $\inBraket{x^2}_{\delta u^- - \delta d^-}$ are available, making ours the first direct calculation of the isovector third transversity Mellin moment.
For a broader overview of prior results, we refer the reader to Table B.10 of Ref. \cite{Lin:2017snn}.

This work builds upon the techniques employed in \cite{Rodekamp:2023wpe} for the calculation of the isovector second Mellin moments $\inBraket{x}$.
We provide a determination of the isovector third Mellin moments $\melmom$ of unpolarized, polarized, and transversity PDFs.
To achieve this, we extract matrix elements at finite momentum for a large number of local twist-two operators.
Notably, using non-zero values of momentum is essential in this case, because the operators required to extract the third Mellin moments have vanishing matrix elements at zero momentum.
We observed that the amount of excited-state contamination varies significantly among the different operators and momenta employed.
Given the large number of operators involved, we developed a procedure to perform a model average automatically in order to accurately extract the matrix elements from all the available data.

Our approach, based on local twist-two operators, is limited to the computation of the first four Mellin moments.
For higher moments, power-divergent mixing arises and the same methodology cannot be applied anymore.
To overcome this issue, novel methods have been developed.
In recent years, these methods have successfully been used to compute Mellin moments.
In Refs. \cite{Francis:2025rya, Francis:2025pgf}, leveraging the technique of gradient flow \cite{Shindler:2023xpd}, the first five Mellin moments for the pion have been computed.
Ref. \cite{Detmold:2025lyb} computed the fourth Mellin moment for the pion and demonstrated that the heavy-quark OPE (HOPE) method is a viable approach to the study of higher Mellin moments.
Using Quasi-PDF or Pseudo-PDF approaches \cite{Cichy:2018mum}, the first few Mellin moments of the proton have been computed for unpolarized \cite{Alexandrou:2021oih, Gao:2022uhg, Fan:2020nzz}, polarized \cite{Fan:2020nzz, Alexandrou:2021oih, Gao:2026wlz}, and transversity \cite{Alexandrou:2021oih, HadStruc:2021qdf, Gao:2023ktu, Pang:2024kza} PDFs.
Except for the gradient flow method, these new methodologies rely on kinematic suppression of higher-twist effects, which is not an issue in the approach we use. 
More importantly, these new methodologies are not affected by power divergent mixing.
Thus, they are expected to play a central role in future first-principle investigations of PDFs.
We then hope that the determination of the isovector third Mellin moments reported in this work can prove to be a useful quantity in the future to assess the compatibility of different lattice QCD calculations.

The rest of the paper is divided into the following sections.
In Sec.~\ref{sec:methodology}, we explain the methodology utilized for the determination of the Mellin moments from lattice data.
In Sec.~\ref{sec:extraction}, we present the available data, and we discuss the extraction of the moment for a subset of all the available operators.
In Sec.~\ref{sec:results}, we summarize the results we obtained and we provide a final estimate for the values of the isovector third Mellin moments.
A summary of the current work is provided in Sec.~\ref{sec:summary}.
Technical details are located in the appendices: Appendix \ref{app:op_gen} explains in details how the operators utilized in the analysis were chosen;
Appendix \ref{app:op_res} contains the intermediate results for each single operator;
Appendix \ref{app:model_avg} focuses on the technical details of the model-averaging procedure we used;
Appendix \ref{app:renormalization} presents the calculation of the renormalization factors;
and Appendix~\ref{app:cont_est} contains a functional-form study of the continuum estimate.

\section{Methodology\label{sec:methodology}}

\subsection{Choice of operators}\label{subsec:operators}

In order to compute third Mellin moments of PDFs, we consider forward matrix elements of local operators of the following kind:
\begin{equation}\label{eq:op_definition1}
    \shortOP_{\alpha, \mu , \nu}(q) = \bar q \Gamma^X_\alpha \lrarrow{D}_\mu \lrarrow{D}_\nu  q, \quad X = V, A, T .
\end{equation}
The moments corresponding to unpolarized, polarized, and transversity PDFs are associated, respectively, with the vector, axial and tensorial channels, which we distinguish by the use of the symbol $X=V,A,T$.
More specifically, this means that the Dirac structure $\Gamma^X_\alpha$, is given by $\gamma_\alpha$, $\gamma_\alpha \gamma_5$ and $\sigma_{\alpha_1 \alpha_2} = i/2 \inSquare{\gamma_{\alpha_1},\gamma_{\alpha_2}}$, respectively, for the three possible channels (in the tensorial case, $\alpha$ is a compound index).
To make sure that our operators have definite transformation properties under charge conjugation, we make use of the forward-backward acting covariant derivative $\lrarrow{D} = \frac{1}{2} \inCurve{ \overset{\rightarrow}{D} - \overset{\leftarrow}{D}}$, implemented on the lattice by means of a symmetric finite difference scheme.
Furthermore, in order to deal only with connected diagrams, we consider the flavor combination corresponding to the isovector channel.
Thus, we look at the difference between operators with up and down quarks, $ \shortOP_{\alpha, \mu , \nu }\inCurve{q=u} -  \shortOP_{\alpha, \mu , \nu }\inCurve{q=d}\equiv \shortOP_{\alpha, \mu , \nu}$.

The particular definition given in Eq.~\eqref{eq:op_definition1} comes from continuum physics.
On the lattice, the operators defined in \eqref{eq:op_definition1} form a closed set under the group of lattice symmetries, the hypercubic group $H(4)$. 
In order to obtain results that renormalize multiplicatively, we have to select operators that belong to irreducible representations (irreps) of $H(4)$ where mixing with lower or equal dimensional operators does not occur.
After identifying these irreps, we were able to obtain an explicit expression for the operators, forming them by means of the group-theoretical recipe given in \cite{Sakata:1974hd} and the matrix representations for the generators of $H(4)$ provided in \cite{Baake:1981qe}.
The operators we obtained are specific combinations of operators of the form \eqref{eq:op_definition1}.
They have been classified according to their trace condition, index symmetry, and charge-conjugation parity, and they match exactly those provided in \cite{Gockeler:1996mu}; more details on this matter can be found in Appendix \ref{app:op_gen}.
Adopting the common notation, and denoting the $a$-th $b$-dimensional irrep of $H(4)$ with the symbol $\tau^{(b)}_a$, we can now list the operators we have selected for our analysis together with the irrep they belong to.
For the unpolarized ($V$) and polarized ($A$) case, there is only one irrep for each case that is not affected by mixing, so the available operators for these two channels are
\begin{equation}\label{eq:vector_axial_op}
\begin{alignedat}{1}
V1. \qquad \tau^{(4)}_2 \qquad O^{V}_{\{2, 3, 4\}}, \qquad \qquad \qquad \qquad A1. \qquad \tau^{(4)}_3 \qquad O^{A}_{\{2, 3, 4\}}, \\%
V2. \qquad \tau^{(4)}_2 \qquad O^{V}_{\{1, 3, 4\}}, \qquad \qquad \qquad \qquad A2. \qquad \tau^{(4)}_3 \qquad O^{A}_{\{1, 3, 4\}}, \\%
V3. \qquad \tau^{(4)}_2 \qquad O^{V}_{\{1, 2, 4\}}, \qquad \qquad \qquad \qquad A3. \qquad \tau^{(4)}_3 \qquad O^{A}_{\{1, 2, 4\}}, \\%
V4. \qquad \tau^{(4)}_2 \qquad O^{V}_{\{1, 2, 3\}}, \qquad \qquad \qquad \qquad A4. \qquad \tau^{(4)}_3 \qquad O^{A}_{\{1, 2, 3\}}, \\%
\end{alignedat}
\end{equation}
where with the curly brackets $\inCurly{\dots}$ we have denoted the symmetrization of the indices.
For the transversity ($T$) case, three different irreps are free of mixing effects, and the operators we can use are
\begin{equation}\label{eq:tensor_op}
\begin{alignedat}{1}
 T1. \qquad \tau^{(3)}_2 \qquad &2O^{T}_{1, 2, \{3, 4\}}+O^{T}_{1, 3, \{2, 4\}}+O^{T}_{1, 4, \{2, 3\}}-O^{T}_{2, 3, \{1, 4\}}-O^{T}_{2, 4, \{1, 3\}},\\
 T2. \qquad \tau^{(3)}_3 \qquad &O^{T}_{1, 2, \{1, 2\}}-O^{T}_{1, 3, \{1, 3\}}+O^{T}_{2, 3, \{2, 3\}},\\
 T3. \qquad \tau^{(3)}_3 \qquad &2O^{T}_{1, 2, \{1, 2\}}+O^{T}_{1, 3, \{1, 3\}}-3O^{T}_{1, 4, \{1, 4\}}-O^{T}_{2, 3, \{2, 3\}}+3O^{T}_{2, 4, \{2, 4\}},\\
 T4. \qquad \tau^{(6)}_2 \qquad  &O^{T}_{1, 3, \{1, 4\}}+O^{T}_{1, 4, \{1, 3\}}-O^{T}_{2, 3, \{2, 4\}}-O^{T}_{2, 4, \{2, 3\}},\\
 T5. \qquad \tau^{(6)}_2 \qquad &O^{T}_{1, 2, \{1, 4\}}+O^{T}_{1, 4, \{1, 2\}}+O^{T}_{2, 3, \{3, 4\}}-O^{T}_{3, 4, \{2, 3\}},\\
 T6. \qquad \tau^{(6)}_2 \qquad &O^{T}_{1, 2, \{2, 4\}}-O^{T}_{1, 3, \{3, 4\}}-O^{T}_{2, 4, \{1, 2\}}+O^{T}_{3, 4, \{1, 3\}},\\
 T7. \qquad \tau^{(6)}_2 \qquad &O^{T}_{1, 2, \{1, 3\}}+O^{T}_{1, 3, \{1, 2\}}+O^{T}_{2, 4, \{3, 4\}}+O^{T}_{3, 4, \{2, 4\}},\\
 T8. \qquad \tau^{(6)}_2 \qquad &O^{T}_{1, 2, \{2, 3\}}-O^{T}_{1, 4, \{3, 4\}}-O^{T}_{2, 3, \{1, 2\}}-O^{T}_{3, 4, \{1, 4\}},\\
 T9. \qquad \tau^{(6)}_2 \qquad &O^{T}_{1, 3, \{2, 3\}}-O^{T}_{1, 4, \{2, 4\}}+O^{T}_{2, 3, \{1, 3\}}-O^{T}_{2, 4, \{1, 4\}}.\\
\end{alignedat}
\end{equation}
As we will explain in more details in a moment, the operators we have selected all have a non-zero kinematic factor, and hence also matrix element, for at least one of the two momenta available.

Regarding the renormalization, given what we have said before, it is multiplicative and each irrep has its own renormalization factor $Z^{\tau^{(b)}_a}$.
Details regarding the computation of the renormalization factors can be found in Appendix \ref{app:renormalization}.
From now on, we will use the shorthand notation $\shortOP$ to denote one of the operators in Eqs.~\eqref{eq:vector_axial_op} and \eqref{eq:tensor_op}, and  $Z_{DDX}^\rho$ to denote the related renormalization factor, with $\rho$ being the relevant irrep of $H(4)$.
%
%

\subsection{Matrix elements}\label{subsec:matele}

Our focus is the forward matrix elements of the selected operators between nucleon states $N(p,\lambda)$, where $p$ is the 4-momentum of the nucleon and $\lambda$ its spin.
As is known \cite{Wandzura:1977qf, Jaffe:1989xx, Jaffe:1991ra, Capitani:1994qn, Gockeler:1995wg, Gockeler:1996hg, LHPC:2002xzk}, these matrix elements are related to the third Mellin moments $\melmom$ by the following expression:
\begin{equation}\label{eq:matele}
    \inBraket{ N(p,\lambda) \inAbs{ \mathcal{S} \shortOP_{ \alpha,\mu, \nu} } N(p,\lambda') } = \melmom \bar u_{N(p,\lambda)} \mathcal{S} \, \Gamma^X_\alpha ip_\mu ip_\nu u_{N(p,\lambda')} ,
\end{equation}
where $\mathcal{S}$, formally defined in Appendix \ref{app:renormalization}, selects the twist-two part of the operators by taking the symmetric traceless combination;
symmetrization and trace subtraction leave the operators in Eqs.~\eqref{eq:vector_axial_op} and \eqref{eq:tensor_op} unchanged.
The established method to extract these matrix elements from lattice data relies on the computation of the ratios of three-point and two-point correlation functions \cite{Gockeler:2005cj,Alexandrou:2016tjj}.
In our case, the relevant correlation functions are given by
\begin{equation}\label{eq:2pcorr}
    \corrTWO = \int \! \Diff{3} \vec{y} \, e^{-i \vec{p} \cdot \vec{y} } \Tr \inCurly{ \gammapol \inBraket{ \chi \inCurve{ \vec{y}, t} \bar \chi \inCurve{ \vec{0}, 0} } }, 
\end{equation}
\begin{equation}\label{eq:3pcorr}
    \corrTHREE = \int \! \Diff{3} \vec{y} \Diff{3} \vec{z} \inSquare{  e^{-i \vec{p}\,' \cdot \vec{y} } e^{ i \inCurve{ \vec{p}\, ' - \vec{p} \, } \cdot \vec{z} } \Tr \inCurly{ \gammapol \inBraket{ \chi \inCurve{ \vec{y}, t} \shortOP \inCurve{\vec{z}, \tau} \bar \chi \inCurve{ \vec{0}, 0} } } } . 
\end{equation}
In the above expressions, $\vec{p}$ and $\vec{p}\, '$ are, respectively, the momentum at the source and at the sink, $\gammapol = P_+ \inSquare{1 - i\gamma_1\gamma_2}$, with $P_+ = \inCurve{1+\gamma_4}/2$, is the polarization matrix we use, and $\chi$ is a nucleon interpolating operator, whose explicit expression is
\begin{equation}\label{eq:nucleon_op}
    \chi_\alpha = \epsilon_{abc} \inCurve{\tilde{u}^T_a C \gamma_5 P_+ \tilde{d}_b} \tilde{u}_{c, \alpha},
\end{equation}
where $\tilde{u}$ and $\tilde{d}$ are smeared quark fields.
As is clear from Eq.~\eqref{eq:matele}, we are interested in forward matrix elements, so in the following we will consider the case $\vec{p} = \vec{p}\,'$ only.

The ratio of three-point and two-point correlation functions is defined as
\begin{equation}\label{eq:ratio_corr_definition}
    R^{\shortOP}(t,\tau) \equiv \frac{ \corrTHREE }{ \corrTWO } .
\end{equation}
The matrix element is obtained as the following limit of the ratio
\begin{equation}\label{eq:ratio_limit}
    \mathcal{M}^{\shortOP} \equiv \lim_{t-\tau,\tau \rightarrow\infty} R^{\shortOP}(t,\tau) .
\end{equation}
The associated kinematic factor is derived from the ground-state contribution to the ratio.
Its expression is given by
\begin{equation}\label{eq:kin_factor}
    K^{\shortOP}(p) = - \frac{1}{2 E_N(p)} \frac{ \Tr \inCurly{ \gammapol \inCurve{-i \slashed{p} + m_N} \inSquare{ C_{\alpha\mu\nu}^{\shortOP} \Gamma^X_\alpha p_\mu p_\nu }  \inCurve{-i \slashed{p} + m_N} } }{ \Tr \inCurly{ \gammapol \inCurve{-i \slashed{p} + m_N} } } .
\end{equation}
In this expression, $E_N(p)$ is the energy of the nucleon state, $m_N$ is its mass, and
$C_{\alpha\mu\nu}^{\shortOP}$ are the Clebsch-Gordan coefficients needed to take the appropriate combination of operators of the form \eqref{eq:op_definition1} such that the final result renormalizes multiplicatively.
These coefficients can easily be read from Eqs.~\eqref{eq:vector_axial_op} and \eqref{eq:tensor_op}, and in the context of this work they have been generated starting from an explicit matrix representations of the generators of $H(4)$; more details on this matter are available in Appendix \ref{app:op_gen}.
As previously noted, the operators selected for this study are such that the corresponding kinematic factors, calculated using Eq.~\eqref{eq:kin_factor}, are non zero for at least one of the two available momenta.

The matrix element and the kinematic factor are the two ingredients needed to compute the third Mellin moments.
The relation between them is given by
\begin{equation}
\mathcal{M} = \melmom K.
\end{equation}
The kinematic factor can easily be computed using Eq.~\eqref{eq:kin_factor}.
The matrix element instead has to be extracted from data.
This extraction can be performed in two different ways.

The first way (referred to below as method 2 for consistency with Ref. \cite{Rodekamp:2023wpe}) relies on the spectral decomposition of the ratio.
If we perform the decomposition up to the first excited state, and neglect terms that decay more rapidly than $e^{-t \Delta E}$, we can then rewrite the ratio in the following way:
\begin{equation}\label{eq:ratio_decomposition}
    R^{\shortOP}(t,\tau) = \mathcal{M}^{\shortOP} 
  + R_{1} e^{-\tfrac{t}{2}\,\Delta E} \cosh \! \inSquare{ \left(\tfrac{t}{2} - \tau\right) \Delta E }
  + R_{2} e^{-t \Delta E}, 
\end{equation}
where $\Delta E = E_1 - E_0$ is the energy difference between the first excited state and the ground state, and the parameters $R_1$ and $R_2$ describe the amount of excited-state contamination. 
These three parameters depend on the specific operator $\shortOP$ that has been chosen, and the superscript has been omitted for the sake of readability.
By calculating the ratio of the correlators, and then fitting it to Eq.~\eqref{eq:ratio_decomposition}, it is possible to extract the matrix element related to the operator under consideration.

From this fit, one could also extract the energy gap $\Delta E$.
However, the right-hand side of Eq.~\eqref{eq:ratio_decomposition} is a truncated sum.
The exponentials in its expression effectively approximate contributions from multiple excited states around the first finite-volume energy level.
Hence, the fitted energy gap will not equal in general the true energy difference between the ground state and first excited state. 

A second approach to the matrix-element extraction (referred to below as method 1) revolves around the computation of the summed ratio
\begin{equation}\label{eq:sumratio_def}
    S^{\shortOP}\!\inCurve{t,\tskip} \equiv a \sum_{\tau = \tskip}^{t - \tskip} R^{\shortOP}(t, \tau) = \mathcal{M}^{\shortOP} \times \inCurve{t + a - 2 \tskip} + \text{excited states}.
\end{equation}
The summed ratio has an excited-state contamination that is exponentially suppressed with $t \Delta E$, while instead for the ratio the suppression goes with $t \Delta E/2$ \cite{Knippschild:20118n, Donnellan:2011/a}. 
The parameter $\tskip/a$ represents the number of time values we skip around the source and the sink.
Neglecting the excited-state contamination, and fixing the value $\tskip$, we can parametrize the summed ratio as a linear function in the values of the source-sink separation $t$,
\begin{equation}\label{eq:sumratio_linear}
    S^{\shortOP}\!\inCurve{t,\tskip}  = \mathcal{M}^{\shortOP} \times t + \text{const.}
\end{equation}
The matrix element can then be extracted as the slope from a linear fit.
%
%

\subsection{Data analysis}\label{subsec:data_analysis}

A complete description of the data analysis procedure is given here.
Furthermore, the steps of the extraction of the third Mellin moments for the unpolarized, polarized, and transversity channels are explained.

\textbf{(1)} The first step is the choice of the appropriate operators.
As previously discussed, we have chosen the operators given in Eqs.~\eqref{eq:vector_axial_op} and \eqref{eq:tensor_op}.
It is worth reiterating that these operators all possess a non-zero kinematic factor for at least one of the available momenta, and they all undergo multiplicative renormalization. Furthermore, they are linearly independent of each other.

\textbf{(2)} The second step consists of the computation of the ratios $R^{\shortOP}(t,\tau)$, the summed ratios $S^{\shortOP}\!\inCurve{t,\tskip}$, and the kinematic factors $K^{\shortOP}$, for each of the selected operators.
To compute the kinematic factors, we use the value of the ground-state energy that we have obtained by performing a fit of the two-point correlator $\corrTWO$.

\textbf{(3)} A crucial step is then the extraction of the matrix elements using one of the two methods previously outlined.
\textit{Method 1} consists of the determination of the matrix element from the slope of a linear regression of the summed ratio, as expressed in Eq.~\eqref{eq:sumratio_linear}.
The linear regression can be repeated with varying initial and final values of $t$, thereby yielding an additional estimate of the matrix element each time.
The results from the various linear regressions have been averaged into one value using the Akaike information criterion \cite{Jay:2020jkz, Neil:2022joj, Neil:2023pgt}.
\textit{Method 2} consists of the extraction of the matrix element through a direct fit of the ratio based on Eq.~\eqref{eq:ratio_decomposition}.
 Given the significant variability of excited-state contamination among the different operators and momenta, three kinds of fully correlated fits have been performed, and the results have been averaged using again the Akaike information criterion; more details on this matter can be found in Appendix \ref{app:model_avg}.
At this point, various determinations of the matrix elements are available.
We label them by the compound index $j=(\shortOP, \vec{p}, \mathfrak{m})$, containing the selected operator, the momentum, and the method ($\mathfrak{m}=1,2$) employed for the matrix element extraction.
For each $j$ associated with a non zero value of the kinematic factor, we can subsequently compute one value of the bare third Mellin moment
 \begin{equation}
     \mathfrak{X}^2_j = \frac{\mathcal{M}_j}{K^{\shortOP}(p)} .
 \end{equation}

\textbf{(4)} Using the renormalization factors, we compute the renormalized moments as $x^2_j = Z_{DDX}^{\rho} \mathfrak{X}^2_j$.
In order to account for the uncertainty on the renormalization factors, we carry out the renormalization separately for each bootstrap resample.
The resamples of the renormalization factors have been generated by sampling from a Gaussian distribution with mean and standard deviation given by the mean and standard deviation of $ Z_{DDX}^{\rho}$.
We computed the renormalization factors in the RI-(S)MOM scheme, and  then matched them to the $\overline{\mathrm{MS}}$ scheme at scale 2~GeV; all the details are provided in Appendix \ref{app:renormalization}.

\textbf{(5)} Separately for the three different channels, we combine the results we have obtained into one central value, defined as the following weighted average
\begin{equation}\label{eq:final_average}
    \inBraket{x^2}_X = \sum_j w_j x^2_j ,
\end{equation}
where the sums run over all the available values of $j$ for the given channel.
The weights in the average are proportional to the inverse of the variance of the renormalized results, $w_j \propto 1/\sigma_j^2$, and they are normalized in such a way that the weights associated to each of the two methods sum separately to $1/2$.
The statistical uncertainties $\sigma_j$ are calculated using a bootstrap resampling over $x_j^2$.
Bootstrap resampling is further employed to estimate the statistical uncertainty of the final average, with the weights $w_j$ held fixed across all resamples.
The choice of weights $w_j$ is the same as the one adopted in our previous work \cite{Rodekamp:2023wpe}.
The rationale behind this choice is to place the two different extraction methods for the matrix element on an equal footing, and to avoid excessive dependence on the single method that may yield smaller uncertainties.
Different choices for the weights have been explored, and the final results were found not to change significantly.

\textbf{(6)} As a last point, we estimate the size of the remaining uncertainty due to the excited-state contamination.
We provide a proxy of the systematic error by computing the weighted variance of the different results
\begin{equation}\label{eq:final_systematic}
    \sigma^2_{\text{syst}} = \sum_j w_j \inCurve{x_j^2 - \inBraket{x^2}_X}^2 .
\end{equation}
This quantity is not computed separately for each bootstrap resample, but is calculated directly on the central values.
%

\subsection{Simulation parameters}

We use a tree-level Symanzik-improved gauge action with $2+1$ flavor tree-level improved Wilson--Clover fermions coupling via 2-level HEX smearing.
Detailed information about the simulation setup can be found in Refs. \cite{BMW:2010skj, BMW:2010ucx, Hasan:2019noy}.
Key simulation parameters are summarized in Table \ref{table:sim_parms}.

Two ensembles, coarse and fine, at the physical pion mass are used.
These ensembles correspond to lattice spacings of $0.1163(4)\,\mathrm{fm}$ and $0.0926(6)\,\mathrm{fm}$, respectively.
Furthermore, we consider two different momenta, $\vec{p}_{\text{low}} = (-2,0,0)\frac{2\pi}{L}, \, \vec{p}_{\text{high}}  =(2,2,2)\frac{2\pi}{L}$ for the coarse ensemble, and $\vec{p}_{\text{low}}  = (-1,0,0)\frac{2\pi}{L}, \, \vec{p}_{\text{high}} = (2,2,2)\frac{2\pi}{L}$ for the fine ensemble.
For the low momentum, as described in \cite{Hasan:2019noy}, the smearing is done using Wuppertal smearing \cite{Gusken:1989qx}, $\tilde{q} \propto (1 + \alpha H)^N q$ with $H$ being the nearest-neighbor gauge-covariant hopping matrix, at $\alpha = 3$ and $N = 60, 100$ for the coarse and fine ensemble, respectively.
For the high momentum, the smearing is altered by introducing additional phase factors, according to the methodology of Ref. \cite{Bali:2016lva} (Momentum Wuppertal smearing).
In the notation of the reference, we used $\vec{k} = (0.8,0.8,0.8)\frac{2\pi}{L}$ and $\vec{k} = (1,1,1)\frac{2\pi}{L}$ for the coarse and fine ensemble, respectively.
This modification improves statistical precision in the high‑momentum data.

For each ensemble, two-point and three-point correlation functions are calculated.
These calculations involve source-sink separations ranging from approximately $0.3\,\mathrm{fm}$ to $1.4\,\mathrm{fm}$ for the coarse ensemble and approximately $0.7\,\mathrm{fm}$ to $1.5\,\mathrm{fm}$ for the fine ensemble.
Furthermore, we make use of all-mode-averaging (AMA) \cite{Blum:2012uh,Shintani:2014vja}.
For the low-momentum dataset, the details have already been described in \cite{Hasan:2019noy}.
For the high-momentum dataset, we apply AMA with 96 and 64 sources, respectively on the coarse and on the fine ensemble.
In both cases, there is one high-precision source.

In our previous work~\cite{Hasan:2019noy}, the autocorrelation of the gauge configurations was tested and no sizable effect was observed.
In addition, we performed a bin-size analysis using delayed binning~\cite{Spiegel:2024dec} and found no evidence of significant autocorrelation.

%
\begin{table*}[ht!]
\caption{\label{table:sim_parms}%
Details of the used ensembles.
For each ensemble, the main simulation parameters are listed.
Furthermore, the source-sink separations $t$ and momenta $\vec{p}$ used in the calculation of the ratios are displayed, together with the number of gauge configurations $N_{\mathrm{cfg}}$ available for each combination.}
\begin{ruledtabular}
\begin{tabular}{lcccccccc}
Ensemble & Size & $\beta$ & $a$ [fm] & $m_\pi$ [MeV] & $m_\pi L$ & $t/a$ & $\vec{p}\,\,[2\pi/L]$ & $N_{\mathrm{cfg}}$ \\
\hline
\hline
Coarse & $48^4$ & 3.31 & 0.1163(4) & 136(2) & 3.9 &       6,8,10      & $(2,2,2)$  & 212 \\
       &        &      &           &        &     & 3,4,5,6,7,8,10,12 & $(-2,0,0)$ & 212 \\
\hline
Fine   & $64^4$ & 3.50 & 0.0926(6) & 133(1) & 4.0 & 8,10,12  & $(2,2,2)$  & 221 \\
       &        &      &           &        &     & 10,13,16 & $(-1,0,0)$ & 427 \\
\end{tabular}
\end{ruledtabular}
\end{table*}
%

\section{Extraction of moments\label{sec:extraction}}

We now present the available data and provide a detailed explanation of the process by which the matrix elements are extracted.
This is accomplished by focusing on a single operator for each channel, while the data and results for the complete set of operators are relegated to Appendix \ref{app:op_res}.
The selected operators are $V2$, $A2$, and $T5$, the expressions for which can be found in Eqs.~\eqref{eq:vector_axial_op} and \eqref{eq:tensor_op}.
For kinematical reasons, $V2$ can only be accessed with the high momentum, while $A2$ and $T5$ can be accessed with both the high and the low momenta.
A summary of which operator can be used with which configuration is given in Appendix \ref{app:op_res}.

In Figure \ref{fig:ratio_highlight}, we show the ratios of these three operators, normalized to the kinematic factor, $ \bar R(t, \tau) =  R(t, \tau) / K$.
This normalization is such that from the values of the plateau in these plots one can directly read the values for the bare moments.
Note that the largest source-sink separation values (for each lattice and momentum) have been excluded from this plot due to their high noise levels.

As evident from the various panels in Figure \ref{fig:ratio_highlight}, the degree of excited-state contamination exhibits considerable variation across different operators, momenta, and lattices.
Consequently, to extract a value for the matrix element reliably from all the ratio data, multiple fully correlated (across all source-sink separations) fits have been done.
For each combination of lattice, operator, and momentum, the fits have been done varying both the functional form and the number of included data points.
Specifically, three types of functional forms have been employed to perform three distinct types of fits:
\emph{(i)} a fit with 0 excited states, i.e., a fit to a constant, obtained by setting $R_1$ and $R_2$ to zero in Eq.~\eqref{eq:ratio_decomposition};
\emph{(ii)} a fit with 1 excited state, where only $R_2$ has been set to zero;
and \emph{(iii)} a fit with 1 excited state, where $R_1$ and $R_2$ are both non-zero.
These three types of fits have 1, 3, and 4 free parameters, respectively, for which Bayesian priors were used.
The various fits performed have subsequently been averaged using the Akaike information criterion \cite{Jay:2020jkz, Neil:2022joj, Neil:2023pgt}, as described in greater detail in Appendix \ref{app:model_avg}.
A visualization of the fit curve obtained from the model average procedure for all operators can be found in Appendix \ref{app:op_res}.

By fitting the ratio data (normalized with respect to the kinematic factor), we derive a single value for the bare moment for each operator and momentum.
Specifically, for the operators depicted in Figure \ref{fig:ratio_highlight}, this implies that we obtain one value per lattice for the operator $V2$, whereas we obtain two values per lattice for both  $A2$ and $T5$.

\begin{figure*}[ht!]
\includegraphics[width=\textwidth]{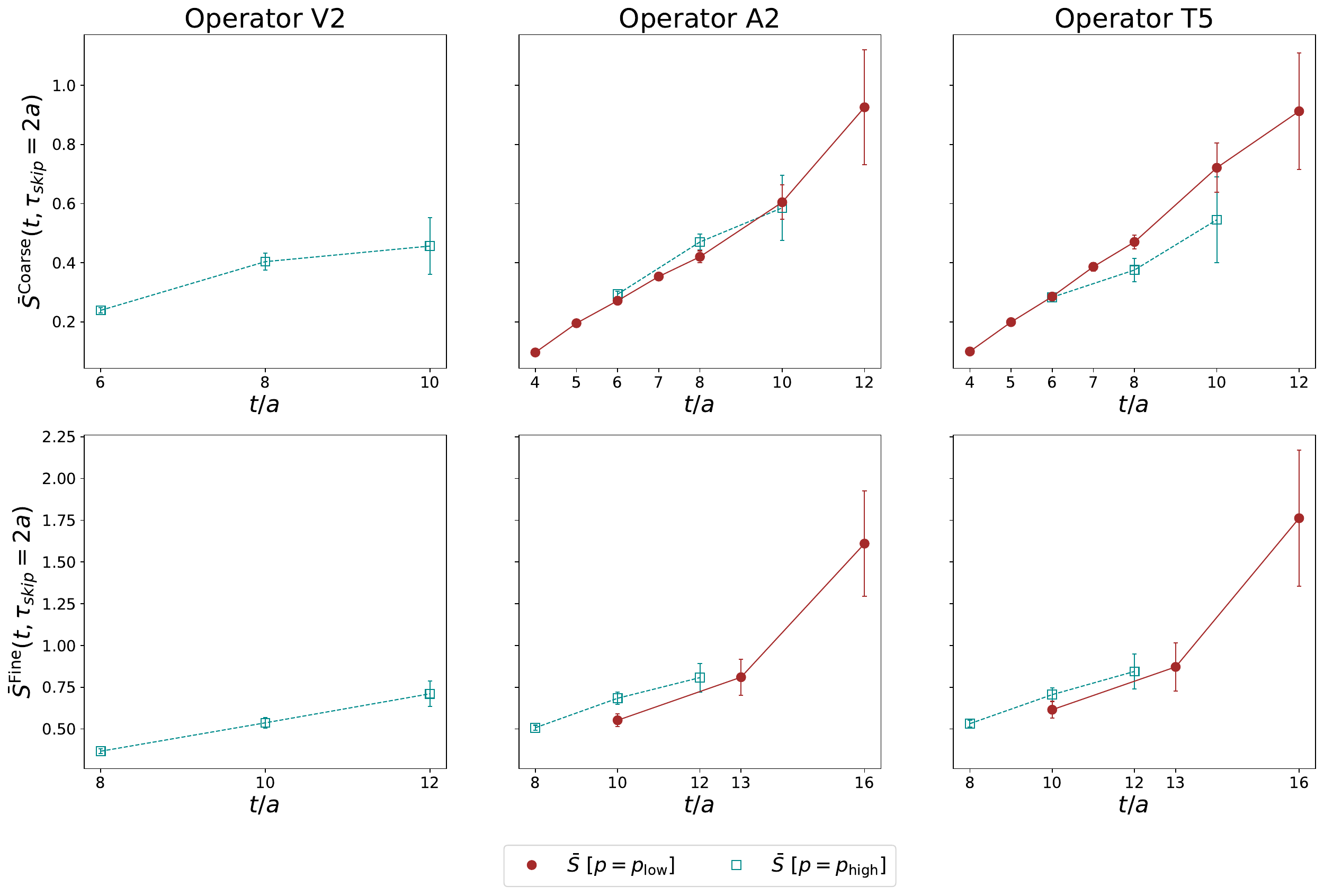}%
\caption{\label{fig:sumratio_highlight}%
Summed ratio data with $\tskip /a = 2$, normalized to the kinematic factor, for the operators $V2$, $A2$ and $T5$ of Eqs.~\eqref{eq:vector_axial_op} and \eqref{eq:tensor_op}.
The high- and low-momentum data are displayed using filled circles and empty squares, respectively, and are distinguished by different colors.
The three columns are organized by operator, while the upper and lower rows represent data from the coarse and fine lattices, respectively.}
\end{figure*}

The normalized summed ratio data, $ \bar S(t, \tskip) =  S(t, \tskip) / K$, where $\tskip /a$ is fixed at $2$, is presented in Figure \ref{fig:sumratio_highlight} for the same set of operators.
The normalization is such that to a value of the slope one can directly associate a value of the moment.
The summed ratio values are plotted as a function of the source-sink separation.
As previously indicated in Table \ref{table:sim_parms}, a total of eight distinct source-sink separations are available on the coarse lattice for the low momentum, whereas only three are available for all other cases.
The summed ratio value at $t/a=3$ is not displayed in the plot, as it cannot be computed for $\tskip/a \ge 2$.
Changing the value of $\tskip$ produced no significant difference in terms of moments.
Therefore, all summed ratios were calculated using $\tskip/a=2$\footnote{The same choice was made in our previous work \cite{Rodekamp:2023wpe}, where the value $\tskip/a=1$ was incorrectly specified.}.

Notably, in every case, the last available value of the source-sink separation is substantially noisier than the others.
Consequently, we have chosen to exclude it entirely from our analysis.
The smallest source-sink separation, $t/a=3$, cannot be used with the summed ratio method.
Thus, in order to ensure that the two extraction methods rely on the same data, also the source-sink separation $t/a=3$ has been excluded entirely from our analysis.

Using the remaining data points, we can perform a linear regression, and obtain, from the slope, one value of the bare moment.
If possible, we repeat the linear regression varying each time the initial and final source-sink separations considered.
The results of the various linear regressions are then averaged into one value using the Akaike information criterion.
This average yields a result which is consistent with the results of simple linear regressions;
more details are available in Appendix \ref{app:op_res}.

Thus, from this second method, we obtain a single value for the bare moment for each operator and momentum.
As for the other method, this implies that we obtain one value per lattice for the operator $V2$, whereas we obtain two values per lattice for both  $A2$ and $T5$.

The values of the moments derived from this second methodology are taken as the central value of the wide prior used in the Bayesian fitting procedure we perform on the ratio data, thus enabling us to rely on data for the prior estimation.
\begin{figure*}[ht!]
\includegraphics[width=\textwidth]{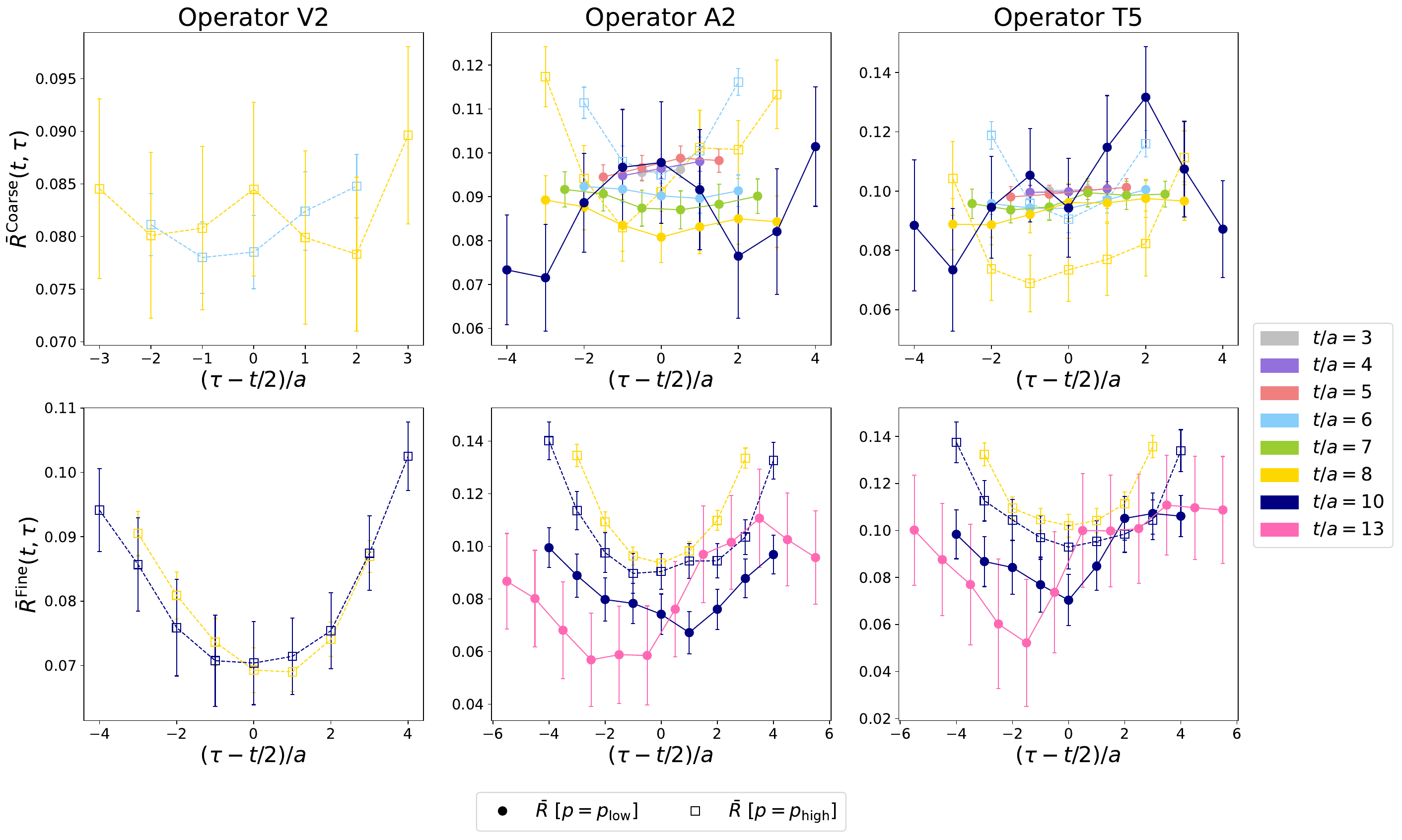}%
\caption{\label{fig:ratio_highlight}%
Ratio data, normalized to the kinematic factor, for the operators $V2$, $A2$ and $T5$ of Eqs.~\eqref{eq:vector_axial_op} and \eqref{eq:tensor_op}.
The data corresponding to high and low momentum are denoted by filled circles and empty squares, respectively.
Different values of the source-sink separation $t$, are represented by distinct colors.
The three columns are organized by operator, whereas the upper and lower rows display data from the coarse and fine lattice, respectively.}
\end{figure*}
%

\section{Results\label{sec:results}}

By employing the techniques explained in the preceding section, a substantial number of values for the renormalized moments $x^2_j$ are calculated.
Using Eq.~\eqref{eq:final_average}, these intermediate results are then combined into a final average, for which we compute the statistical uncertainty through bootstrap resampling.
Furthermore, the systematic uncertainty on the final average is computed using Eq.~\eqref{eq:final_systematic}.

\begin{figure*}[h!]
\includegraphics[width=\textwidth]{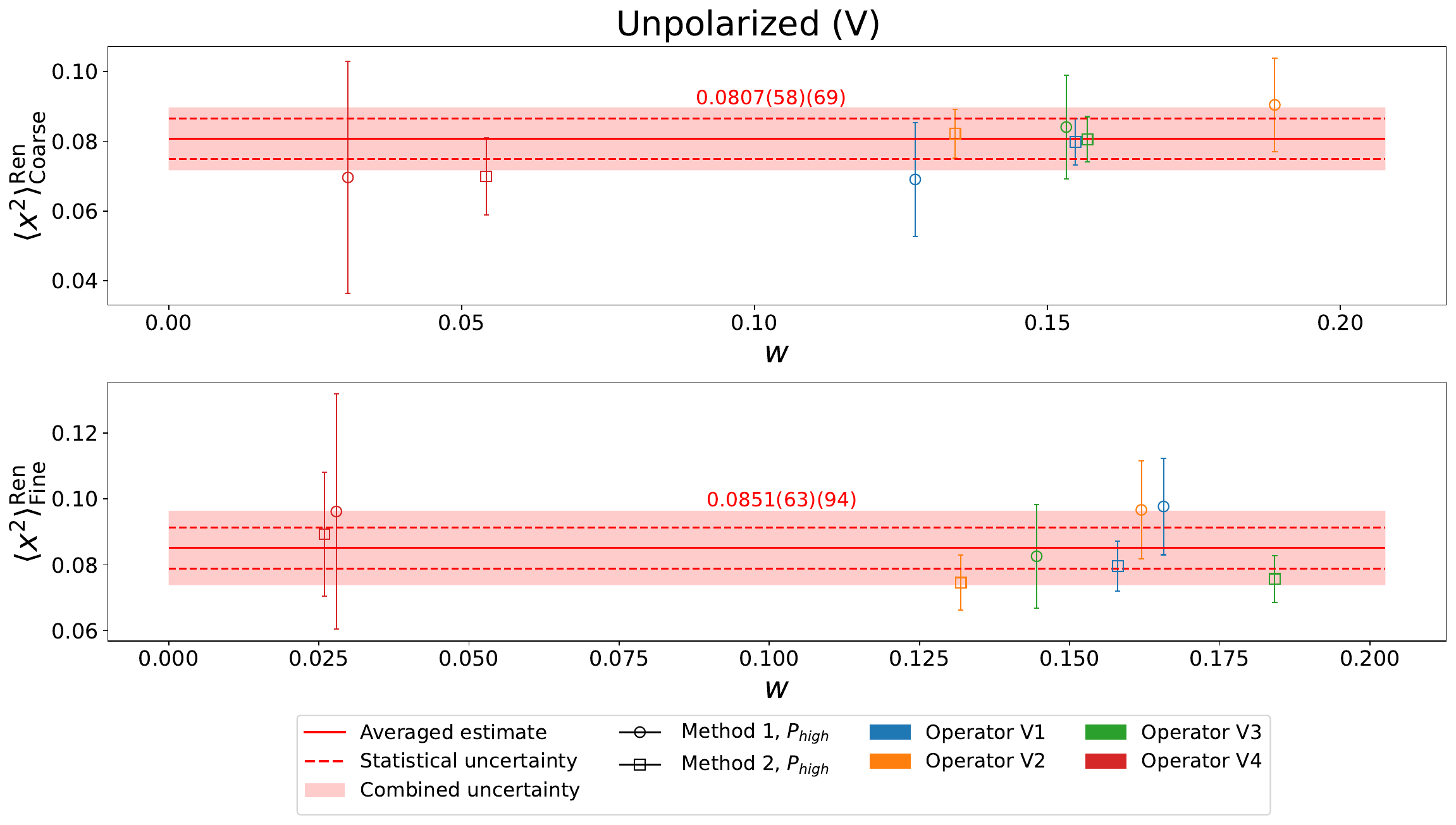}
\caption{\label{fig:result_vector}%
Estimates of unpolarized $x^2_j$ versus their weight in the final average.
Moments obtained from the coarse and fine lattice are displayed, respectively, on the top and bottom row.
They are resolved per operator, momentum and extraction method, which are represented by different colors and marker styles.
The final average is shown in red with a solid line, its statistical uncertainty with a dashed line, and the sum in quadrature of statistical and systematic uncertainty with a red band.
}
\end{figure*}

\begin{figure*}[h!]
\includegraphics[width=\textwidth]{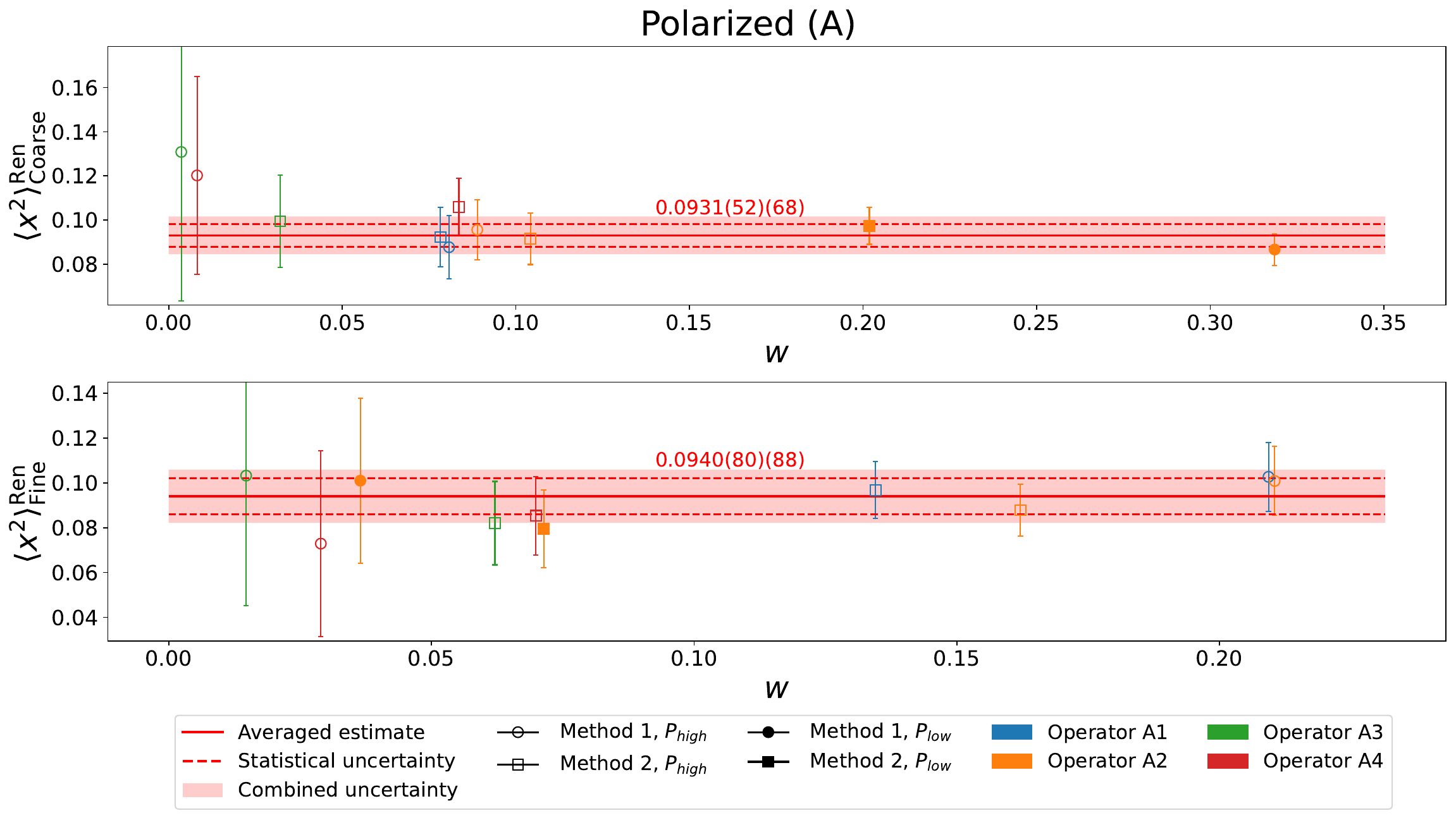}
\caption{\label{fig:result_axial}%
Estimates of polarized $x^2_j$ versus their weight in the final average, analogous to Figure \ref{fig:result_vector}.
}
\end{figure*}

\begin{figure*}[h!]
\includegraphics[width=\textwidth]{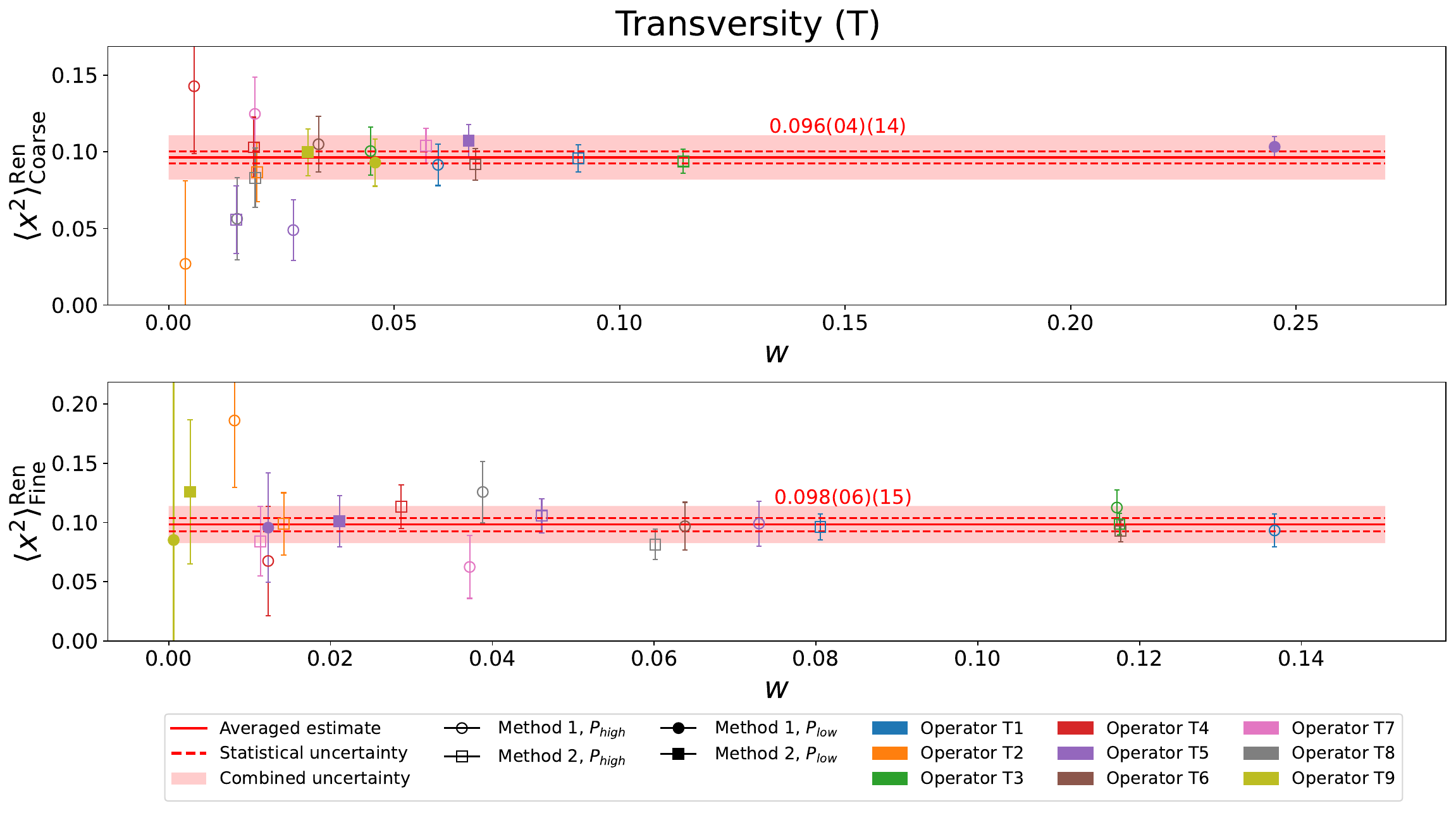}
\caption{\label{fig:result_tensor}%
Estimates of transversity $x^2_j$ versus their weight in the final average, analogous to Figure \ref{fig:result_vector}.
}
\end{figure*}

Figures \ref{fig:result_vector}, \ref{fig:result_axial} and \ref{fig:result_tensor} display the complete set of all the available renormalized moments, plotted as a function of their respective weights in the final average.
This is done separately for the unpolarized, polarized, and transversity channels.
The final average is also explicitly reported on the plots, along with its statistical and systematic uncertainties, enclosed respectively in the first and second parentheses.

\begin{figure*}[h!]
\includegraphics[width=\textwidth]{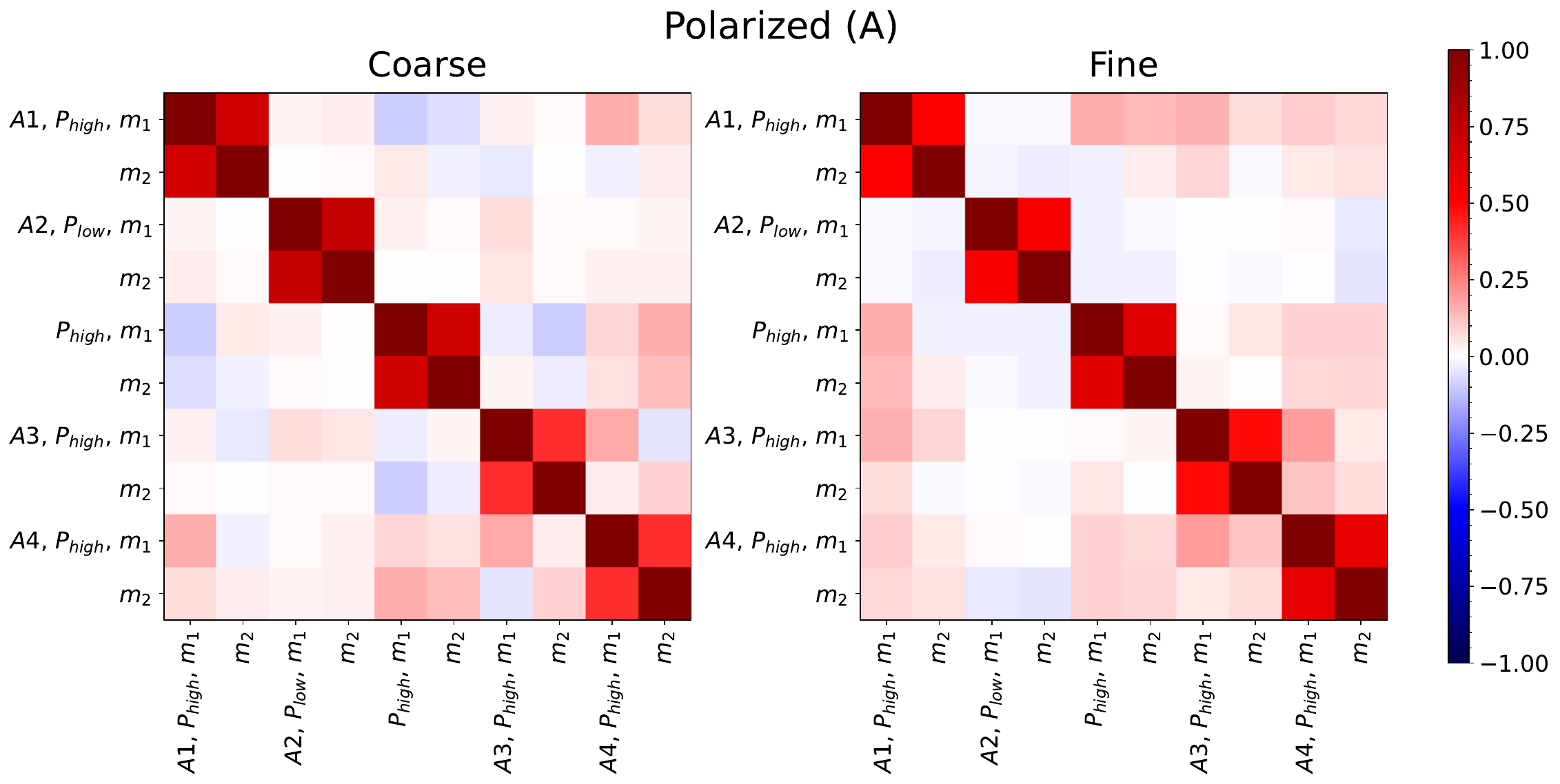}
\caption{\label{fig:correlation_axial}%
Correlation matrices of the moments shown in Figure \ref{fig:result_axial}.
The moments are labeled by operator, momentum, and extraction method.
A strong correlation exist only between results obtained from the same operator and momentum.
}
\end{figure*}

\begin{figure*}[ht!]
\includegraphics[width=\textwidth]{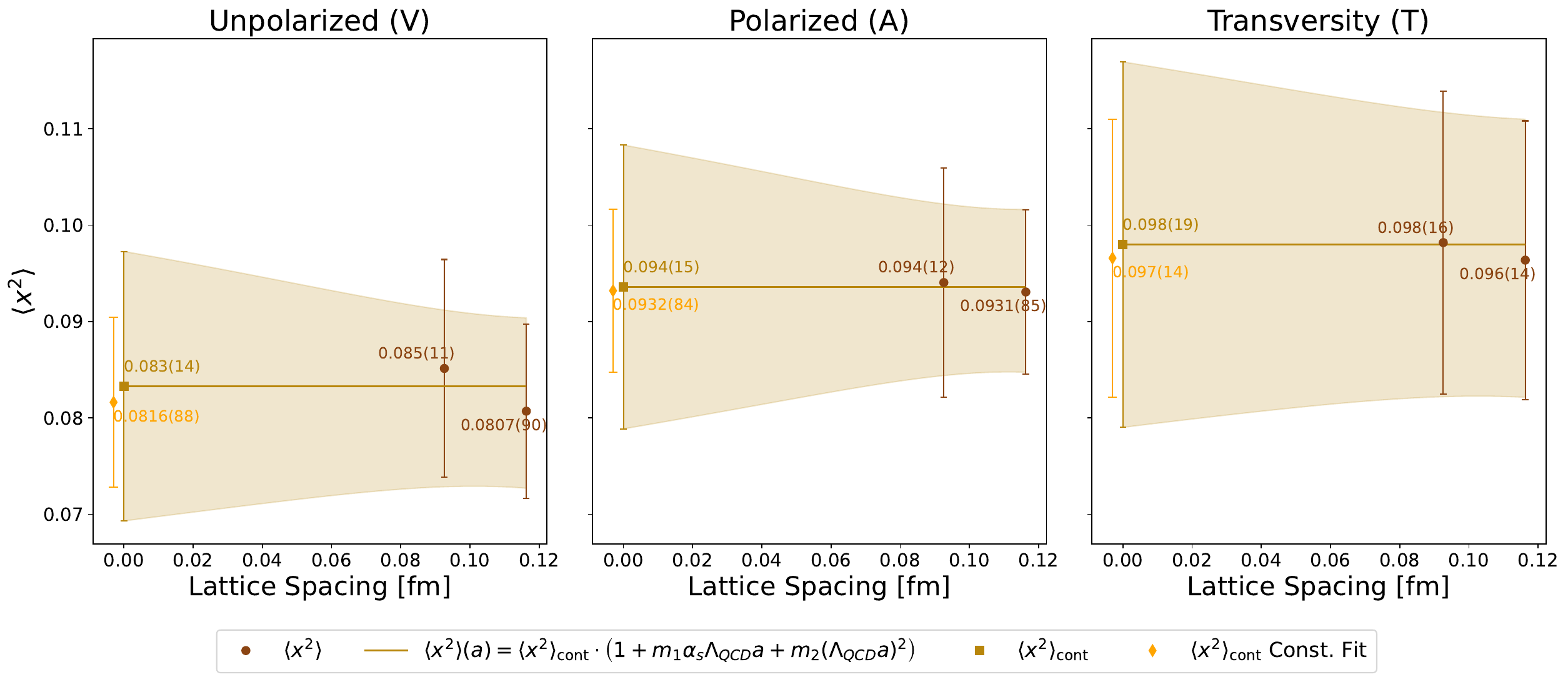}%
\caption{\label{fig:continuum_limit}%
Continuum extrapolation using a Bayesian fit with the model described in Eq.~\eqref{eq:continuum_limit}.
The limited amount of data makes this extrapolation dependent on the chosen priors for the coefficients $m_i$.
Further details are provided in the main text, and resulting estimates are listed in Table \ref{table:results}.
For comparison, the continuum value obtained from a constant fit is also shown on the plot, slightly displaced horizontally for visual clarity.}
\end{figure*}

For the computation of the final averages, correlations are automatically accounted for by our bootstrap analysis.
As an example, Figure \ref{fig:correlation_axial} shows the correlation matrices of the renormalized moments related to the polarized channel.
Looking at the correlation matrices, it is evident that a strong correlation exists only between results pertaining to the same operator and momentum.
This is to be expected, since in these cases the methodology employed to extract different values of the moments varies, but the underlying data remain unchanged.
The same behavior is observed for the unpolarized and transversity channels.

The numerical values obtained for the final estimate of the isovector third Mellin moments are summarized in Table \ref{table:results}.

The final step of the analysis is the continuum extrapolation.
With only two ensembles available, we acknowledge that a continuum extrapolation does not have enough degrees of freedom to judge the quality of the fit.
Nevertheless, in order to provide a continuum estimate, we perform a Bayesian fit, where priors on the free parameters are used to stabilize the fit.
The points from the two ensembles are interpreted as partially correlated Gaussian variables.
The correlation, we argue, is due to the use of the same operators, and hence to similar excited-state contamination in both cases.
Consequently, we conservatively assume that systematic uncertainties are fully correlated and construct a covariance matrix whose off‑diagonal entries are the products of single‑ensemble systematic uncertainties.
The functional form for the extrapolation is the following
\begin{equation}\label{eq:continuum_limit}
    \melmom \inCurve{a} = \melmom_{\text{cont}} \cdot \inCurve{1 + m_1 \inCurve{\alpha_s a \Lambda_{\text{QCD}} } + m_2 \inCurve{a \Lambda_{\text{QCD}} }^2 } \, .
\end{equation}
The combination $a \Lambda_{\text{QCD}}$ dictates the size of the discretization effects.
The fermion action we use is tree-level $O(a)$-improved, and the operators we consider are also expected to have no tree-level $O(a)$ errors; thus the above functional form.
We expect the dimensionless parameters $m_i$ to be of order 1; thus, we constrain them by the use of two equal Gaussian priors, of the form $\mathcal{N}\inCurve{0,2}$. We approximate $\alpha_s\approx0.3$.
A completely flat prior is used for the continuum value $\melmom_{\text{cont}}$.
The continuum value obtained from this procedure is a prior-constrained estimate.
To assess the robustness of our result, we repeat the extrapolation while varying the width of the prior on $m_i$ within the range $[0.5,32]$, and observe no sizable effect.
We emphasize that the priored extrapolation results in a more conservative continuum uncertainty.
The findings of our prior-width study are reported in Appendix~\ref{app:cont_est}.

The fit curves are shown in Figure \ref{fig:continuum_limit}.
For each channel, the continuum extrapolation curve is almost completely flat.
This is expected because, for each channel, the two available data points are compatible between uncertainties.
Indeed, performing a fit to a constant, we find a central value for $\melmom_{\text{cont}}$ which is completely compatible with the one already obtained.
However, the uncertainty obtained from a fit to a constant is sizably smaller.

The results of the continuum extrapolation are quoted in Table \ref{table:results}.
The total uncertainty on the continuum value is obtained using the Bayesian fit procedure described above.
To assess which parts of the uncertainty are statistical and which are systematic contributions, we compare the uncertainties obtained by varying the extrapolation procedure.
First, we set the single-ensemble systematic uncertainties to zero and repeat the fit to Eq.~\eqref{eq:continuum_limit}.
We obtain a smaller uncertainty on the continuum value, and interpret the difference in quadrature as the excited-state systematic uncertainty.
Additionally, keeping the single-ensemble systematic uncertainties to zero, we fit to a constant.
The result is again compatible, but the uncertainty is considerably lower.
This additional difference in uncertainty is interpreted as the systematic contribution resulting from discretization effects.
The uncertainty of the constant fit is taken as the statistical contribution.
As an example, we consider the polarized case, where the result of the Bayesian fit is $0.094(15)$.
We rewrite it as $0.094(04)(08)(12)$.
The statistical uncertainty is the value between the first set of parentheses.
Inside the second and third parentheses, we have the systematic uncertainties due to excited-state and discretization effects, respectively.
The systematic uncertainties appear to be the dominant contribution to the total uncertainty budget.
We reiterate that the continuum estimate we provide is not the result of an unconstrained continuum extrapolation.
Rather, it is a prior-constrained estimate, obtained with what we believe to be the best option available with the given data.

\begin{figure}[ht!]
\includegraphics[width=0.9\textwidth]{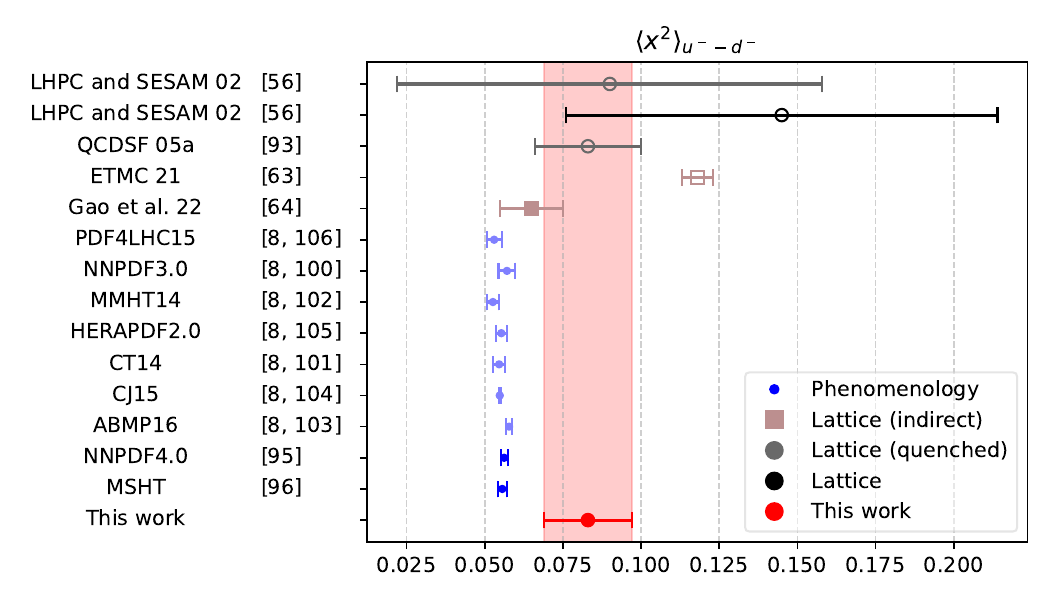}
\caption{\label{fig:comparison_vector}%
Comparison of unpolarized isovector third Mellin moments $\melmom_{u^- - d^-}$ computed by different collaborations.
The result obtained in this work is shown as a red circle;
the corresponding $1\sigma$ region is indicated by a red band.
Lattice computations are represented by circles and squares: black circles denote dynamical calculations performed using the same approach adopted here (based on local twist-two operators), gray circles denote quenched calculations using the same approach, and brown squares denote dynamical lattice calculations based on indirect determinations of the moments (e.g., Quasi- or Pseudo-PDF methods).
Empty markers are used for lattice calculations performed on a single lattice with a heavier-than-physical pion mass or with no pion mass below $500 \,\mathrm{MeV}$.
Phenomenological results are shown as blue dots.
A lighter shade of blue is used for values corresponding to older PDF ensembles.
Further details and references are provided in the main text.
}
\end{figure}

\begin{figure}[ht!]
\includegraphics[width=0.9\textwidth]{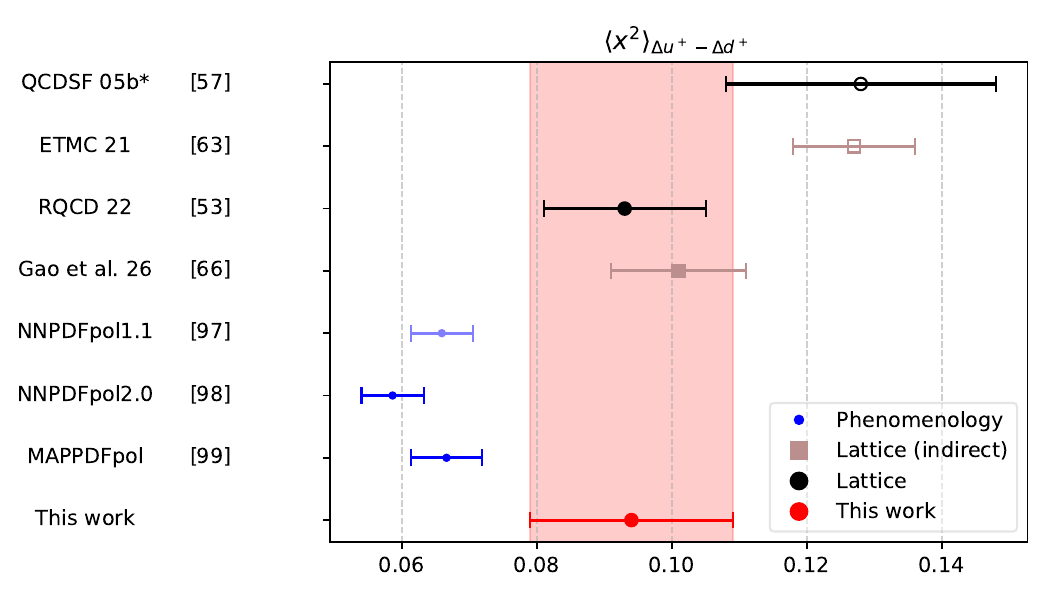}
\caption{\label{fig:comparison_axial}%
Comparison of polarized isovector third Mellin moments $\melmom_{\Delta u^+ - \Delta d^+}$ computed by different collaborations, analogous to Figure \ref{fig:comparison_vector}.
An asterisk next to a collaboration name indicates that the quoted result was derived by us using inputs from the corresponding reference, rather than taken directly from it.
Further details and references are provided in the main text.
}
\end{figure}

\begin{figure}[ht!]
\includegraphics[width=0.9\textwidth]{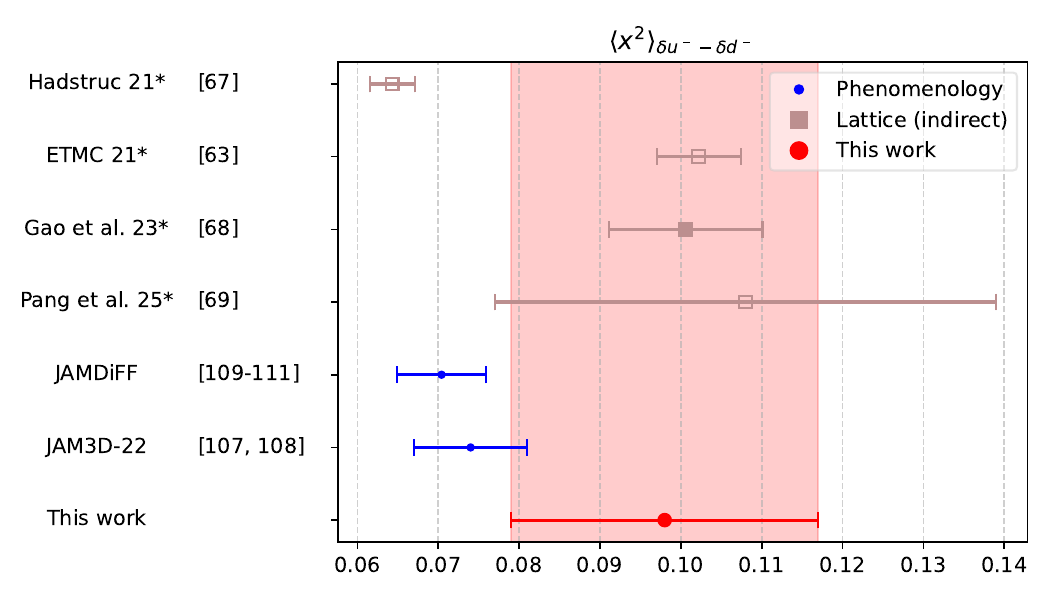}
\caption{\label{fig:comparison_tensor}%
Comparison of transversity isovector third Mellin moments $\melmom_{\delta u^- - \delta d^-}$ computed by different collaborations, analogous to Figure \ref{fig:comparison_vector}.
An asterisk next to a collaboration name indicates that the quoted result was derived by us using inputs from the corresponding reference, rather than taken directly from it.
Further details and references are provided in the main text.
}
\end{figure}

An overview of the comparison between our results and those of other collaborations is provided in Figures \ref{fig:comparison_vector}-\ref{fig:comparison_tensor}.
%
For the unpolarized, polarized, and transversity moment, the lattice calculations are taken from Refs.~\cite{LHPC:2002xzk,Gockeler:2004wp, Alexandrou:2021oih, Gao:2022uhg}, \cite{Gockeler:2005vw, Alexandrou:2021oih, Burger:2021knd, Gao:2026wlz}, and \cite{Alexandrou:2021oih, Gao:2023ktu, HadStruc:2021qdf, Pang:2024kza}, respectively.
Some of the quoted results have been derived by us using inputs from the corresponding references.
In Refs.~\cite{Gockeler:2005vw} and \cite{Burger:2021knd} the polarized moment was given in terms of the matrix element $a_2^{(u-d)}$, whose definition differs by a factor of $2$ from $\melmom_{\Delta u^+ - \Delta d^+}$\footnote{
For Ref.~\cite{Gockeler:2005vw} we considered the combination $3/2(a^{(p)}_2-a^{(n)}_2) Z$, where $a^{(p)}_2$ and $a^{(n)}_2$ are the matrix elements, in the $\overline{\mathrm{MS}}$ scheme at scale $\sqrt{5} \, \mathrm{GeV}$, taken from the reference, and $Z\approx1.038$ is the factor needed to evolve to our scale of $ 2 \, \mathrm{GeV}$.
}.
The transversity moment was either reported normalized to the tensor charge (Ref.~\cite{Gao:2023ktu}), or at a different renormalization scale (Ref.~\cite{Alexandrou:2021oih}) or both (Refs.~\cite{HadStruc:2021qdf} and \cite{Pang:2024kza});
thus, in order to compare with our result, we used the tensor charge from the FLAG review and computed the relevant numerical factors needed for a change of renormalization scale\footnote{
In order to evolve $\melmom_{\delta u^- - \delta d^-}$ and $\melmom_{\delta u^- - \delta d^-} / g_T^{u-d}$ from the $\overline{\mathrm{MS}}$ scheme at scale $\sqrt{2} \, \mathrm{GeV}$ to scale $2 \, \mathrm{GeV}$, we used the numerical factors $Z\approx0.865965$ and $Z\approx 0.906225$, respectively.
}.
If we exclude calculations performed on single lattice with a heavier-than-physical pion mass and calculations with no pion mass below $500 \,\mathrm{MeV}$, we find a good agreement with the results of other collaborations.
Ref.~\cite{Burger:2021knd} used our same methodology based on local twist-two operators, and the polarized moment they obtain is in perfect agreement with our result.
The authors of Refs.~\cite{Gao:2022uhg,Gao:2023ktu,Gao:2026wlz} compute the Mellin moments using the Pseudo-PDF approach.
For the polarized and transversity cases, their result lies well within one standard deviation of our value.
In the unpolarized case, the difference is slightly larger.

Results from phenomenology are also available.
For the unpolarized and polarized moment, we use the software library \texttt{LHAPDF} \cite{Buckley:2014ana}.
Using the unpolarized PDF sets \texttt{NNPDF4.0aN3LO} \cite{NNPDF:2024nan} and \texttt{MSHT20aN3LO} \cite{McGowan:2022nag} we obtain the values $\melmom_{u^- - d^-} = 0.0562(11)$ and $\melmom_{u^- - d^-} = 0.0556(15)$, respectively.
Using the polarized PDF sets \texttt{NNPDFpol1.1} \cite{Nocera:2014gqa}, \texttt{NNPDFpol2.0NNLO} \cite{Cruz-Martinez:2025ahf} and \texttt{MAPPDFpol1.0NNLO} \cite{Bertone:2024taw} we obtain the values $\melmom_{\Delta u^+ - \Delta d^+} = 0.0659(46)$, $\melmom_{\Delta u^+ - \Delta d^+} = 0.0586(46)$ and $\melmom_{\Delta u^+ - \Delta d^+} = 0.0666(53)$, respectively.
For the unpolarized moment, we also consider the values given in Table C.1 of Ref.~\cite{Lin:2017snn}, where they have been computed using the unpolarized PDF sets provided in Refs.~\cite{NNPDF:2017mvq, Dulat:2015mca, Harland-Lang:2014zoa, Alekhin:2017kpj, Accardi:2016qay, H1:2015ubc, Butterworth:2015oua}.
Finally, for the transversity moment, we use the transversity distribution provided by the global analysis \texttt{JAM3D-22} \cite{Gamberg:2022kdb}, and obtain the value\footnote{
The code used to compute this value is based on the \texttt{Google Colab} notebook \cite{Gamberg:jam3d_library} provided in Ref. \cite{Gamberg:2022kdb}.
} $\melmom_{\delta u^- - \delta d^-} = 0.074(7)$;
using the distribution by \texttt{JAMDiFF} \cite{Cocuzza:2023oam, Cocuzza:2023vqs}, we obtain the value\footnote{
The code used to compute this value is based on the \texttt{Google Colab} notebook \cite{Cocuzza:2023colab} provided in Ref. \cite{Cocuzza:2023oam}.
} $\melmom_{\delta u^- - \delta d^-} = 0.0704(55)$.
In the unpolarized and polarized case, our result exceeds the phenomenological values by approximately two standard deviations.
For the transversity moment the comparison is more favorable, with our result exceeding the values from phenomenology by less than 2 standard deviations.

\begin{table}
\caption{\label{table:results}%
Final averages for the isovector third Mellin Moments for the unpolarized, polarized and transversity channels.
For both the coarse and fine lattices, the central values have been obtained using Eq.~\eqref{eq:final_average};
the statistical uncertainty, denoted by the value within the first set of parentheses, was calculated through bootstrap resampling;
the systematic uncertainty, represented by the value within the second set of parentheses, was obtained using Eq.~\eqref{eq:final_systematic}.
The continuum values have been extrapolated from the two values at finite lattice spacing using a Bayesian fit.
In this case, the values between the three sets of parentheses represent, in order, the statistical uncertainty, the systematic uncertainty related to excited-state contamination, and the systematic uncertainty related to the size of discretization effects.
Further details are provided in the main text.}
\begin{ruledtabular}
\begin{tabular}{lcl}
                 & Ensemble & $\melmom$  \\
\hline
\hline
Unpolarized (V)  & Coarse        & 0.081(06)(07) \\
                 & Fine          & 0.085(06)(09) \\
                 & Continuum     & 0.083(04)(09)(10) \\
                 \hline
Polarized (A)    & Coarse        & 0.093(05)(07) \\
                 & Fine          & 0.094(08)(09) \\
                 & Continuum     & 0.094(04)(08)(12) \\
                 \hline
Transversity (T) & Coarse        & 0.096(04)(14) \\
                 & Fine          & 0.098(06)(15) \\
                 & Continuum     & 0.098(03)(15)(11)
\end{tabular}
\end{ruledtabular}
\end{table}
%

\section{Summary\label{sec:summary}}

In this work, the isovector third Mellin moments $\melmom$ of unpolarized, polarized, and transversity parton distribution functions have been computed using two lattice QCD ensembles at the physical pion mass and two different nonzero values of the momentum.

For the computation, we employed a large set of operators. 
For each kinematically allowed operator-momentum combination, we obtained two estimates of the Mellin moment.
For each ensemble, these intermediate estimates were then combined into a single average.
The variety of operators reflected in a variety of excited-state contamination.
By applying fit averaging, we were able to consistently extract Mellin moments across all operator choices using the same procedure.
The resulting values are mutually compatible, providing a nontrivial cross‑check of the robustness of our calculation.
Moreover, the large number of operators yields a correspondingly large set of intermediate results.
The spread of these results enables a reliable estimate of the systematic uncertainty associated with excited-state contamination.

The continuum limit cannot be reliably performed in an unconstrained way with only two ensembles.
To provide a continuum estimate, we performed a Bayesian fit in which the coefficients describing the lattice-spacing dependence were constrained to be natural-sized.
This procedure provides the following values: $\melmom_{u^- - d^-} = 0.083(14)$, $\melmom_{\Delta u^+ - \Delta d^+} = 0.094(15)$, $\melmom_{\delta u^- - \delta d^-} = 0.098(19)$.
These results are compatible with the ones at finite lattice spacing.

For both the unpolarized and polarized channels, our result deviates from the corresponding phenomenological estimate by approximately two standard deviations.
In the transversity channel, the agreement is slightly better, although the phenomenological value still lies more than one standard deviation away from our determination.
We find good agreement with other lattice results either obtained at or extrapolated to the physical pion mass.
However, a comprehensive comparison remains limited, as only a few such lattice calculations exist in the literature.
Moreover, this work provides the first direct determination of these observables at the physical pion mass.

\clearpage
\FloatBarrier
\appendix
%
%

\section{Choice of operators\label{app:op_gen}}

Operators of the form given in Eq.~\eqref{eq:op_definition1} can be directly linked to the third Mellin moments by means of Eq.~\eqref{eq:matele}.
However, these operators are not directly well suited for the computation because, in general, they mix under renormalization with lower- or equal-dimensional operators.
Lattice operators can mix if they have the same flavor structure, the same charge-conjugation parity, and the same  transformation properties under $H(4)$ (the group of lattice rotations and reflections).
For this reason, to avoid mixing, we first classify all irreps of $H(4)$ appearing in the decomposition of operators of the form \eqref{eq:op_definition1}, as well as in the decomposition of operators of equal or lower dimension.
We then rearrange the operators of Eq.~\eqref{eq:op_definition1} into irreps of $H(4)$, and focus only on those belonging to an irrep that appears only once (with a given charge-conjugation parity) in the above classification.

In general, all the possible operators of the form \eqref{eq:op_definition1} span reducible representations of $H(4)$ represented, respectively for the vector, axial and tensorial cases, by the tensor products $\tau^{(4)}_1 \otimes \tau^{(4)}_1 \otimes \tau^{(4)}_1$, $\,\,\,\, \tau^{(4)}_4 \otimes \tau^{(4)}_1 \otimes \tau^{(4)}_1\,\,$ and $\, \,\tau^{(6)}_1 \otimes \tau^{(4)}_1 \otimes \tau^{(4)}_1$, where $\tau^{(b)}_a$ is the standard symbol used to denote the $a$-th $b$-dimensional irrep of $H(4)$.
These tensor products can be easily decomposed into a sum of irreps by means of the character table of the group $H(4)$ (which one can find e.g. in Ref. \cite{Baake:1981qe}).
The decompositions are the following:
\begin{equation}
    \begin{alignedat}{1}
        \tau^{(4)}_1 \otimes \tau^{(4)}_1 \otimes \tau^{(4)}_1 &= 4\tau^{(4)}_1 \oplus \tau^{(4)}_2 \oplus \tau^{(4)}_4 \oplus 3\tau^{(8)}_1 \oplus 2\tau^{(8)}_2 ,\\
        \tau^{(4)}_4 \otimes \tau^{(4)}_1 \otimes \tau^{(4)}_1 &= \tau^{(4)}_1 \oplus \tau^{(4)}_3 \oplus 4\tau^{(4)}_4 \oplus 2\tau^{(8)}_1 \oplus 3\tau^{(8)}_2 ,\\
        \tau^{(6)}_1 \otimes \tau^{(4)}_1 \otimes \tau^{(4)}_1 &= \tau^{(1)}_1 \oplus \tau^{(1)}_4 \oplus \tau^{(2)}_1 \oplus \tau^{(2)}_2 \oplus 2\tau^{(3)}_1 \oplus \tau^{(3)}_2 \\
        &\quad \, \oplus \tau^{(3)}_3 \oplus 2\tau^{(3)}_4 \oplus 4\tau^{(6)}_1 \oplus 2\tau^{(6)}_2 \oplus 3\tau^{(6)}_3 \oplus 3\tau^{(6)}_4  .
    \end{alignedat}
\end{equation}
Comparing such a decomposition to the decomposition of lower and equal dimensional operators, we find the irreps that are suitable for a computation that does not involve mixing.
These are: $\tau^{(4)}_2$ for the vector case, $\tau^{(4)}_3$ for the axial case, and  $\tau^{(3)}_2$, $\tau^{(3)}_3$ and $\tau^{(6)}_2$ for the tensorial case.

Once these suitable irreps have been identified, we have to find an explicit expression for the operators belonging to them.
Such a task can be achieved using the recipe given in Ref. \cite{Sakata:1974hd}, which tells us how to compute the Clebsch-Gordan coefficients of a finite group.
In our case, the finite group is $H(4)$, and its Clebsch-Gordan coefficients tell us exactly how to rearrange the operators of the form \eqref{eq:op_definition1} into a combination which transforms, under $H(4)$, according to one of the selected irreps.
To proceed with this computation, one needs to know an explicit matrix representation of each one of the 384 elements of $H(4)$, and for each one of its 20 irreps.
These 384$\times$20 matrices can be constructed starting just from an explicit representation, for each irrep, of the 3 generators of $H(4)$; conveniently enough, these 3$\times$20 matrix representations are given in Table IX of Ref. \cite{Baake:1981qe}.

Putting everything together we are then able to obtain an explicit expression for our operators of interest, using as input only the explicit matrix representations of the generators of $H(4)$.
The results we obtain are shown in Eqs.~\eqref{eq:vector_axial_op} and \eqref{eq:tensor_op}.
We note here that for the tensorial case, from the three dimensional $\tau^{(3)}_2$ and $\tau^{(3)}_3$ irreps, we only report 1 and 2 operators respectively.
We omit three operators because their associated kinematic factors vanish for all available momenta.

Finally, we confirm that the bases we obtain for $H(4)$ using this methodology, coincide, up to a trivial rearrangement of operators inside the same irrep, with the bases of operators given in Ref. \cite{Gockeler:1996mu}, where the same results have been obtained in a different way.

\section{Results per operator\label{app:op_res}}

In this Appendix, we present the complete set of bare moments that enter our final averages.
The bare moments are estimated for each lattice, operator, and momentum.
Because the extraction of a Mellin moment requires a non‑zero kinematic factor, not every operator can be used in combination with every momentum.
Operators $A2$ and $T5$ can be used with both the high and low momentum. 
Operators $T9$ can be used only with the low momentum.
All remaining operators can be used solely with the high momentum.
For both the vector and axial channels, the employed operators belong to the same irrep.
Consequently, a single renormalization factor applies to all operators within each channel.
In the tensor channel this simplification does not hold, because the relevant operators transform according to different irreps.
Nevertheless, as can be seen from Table \ref{table:ren_factors}, the tensor renormalization factors are mutually compatible within uncertainties.
Hence, agreement among the unrenormalized results in a given channel implies agreement among the corresponding renormalized quantities. 

\begin{figure*}[ht!]
\makebox[\textwidth][c]{\includegraphics[width=\textwidth]{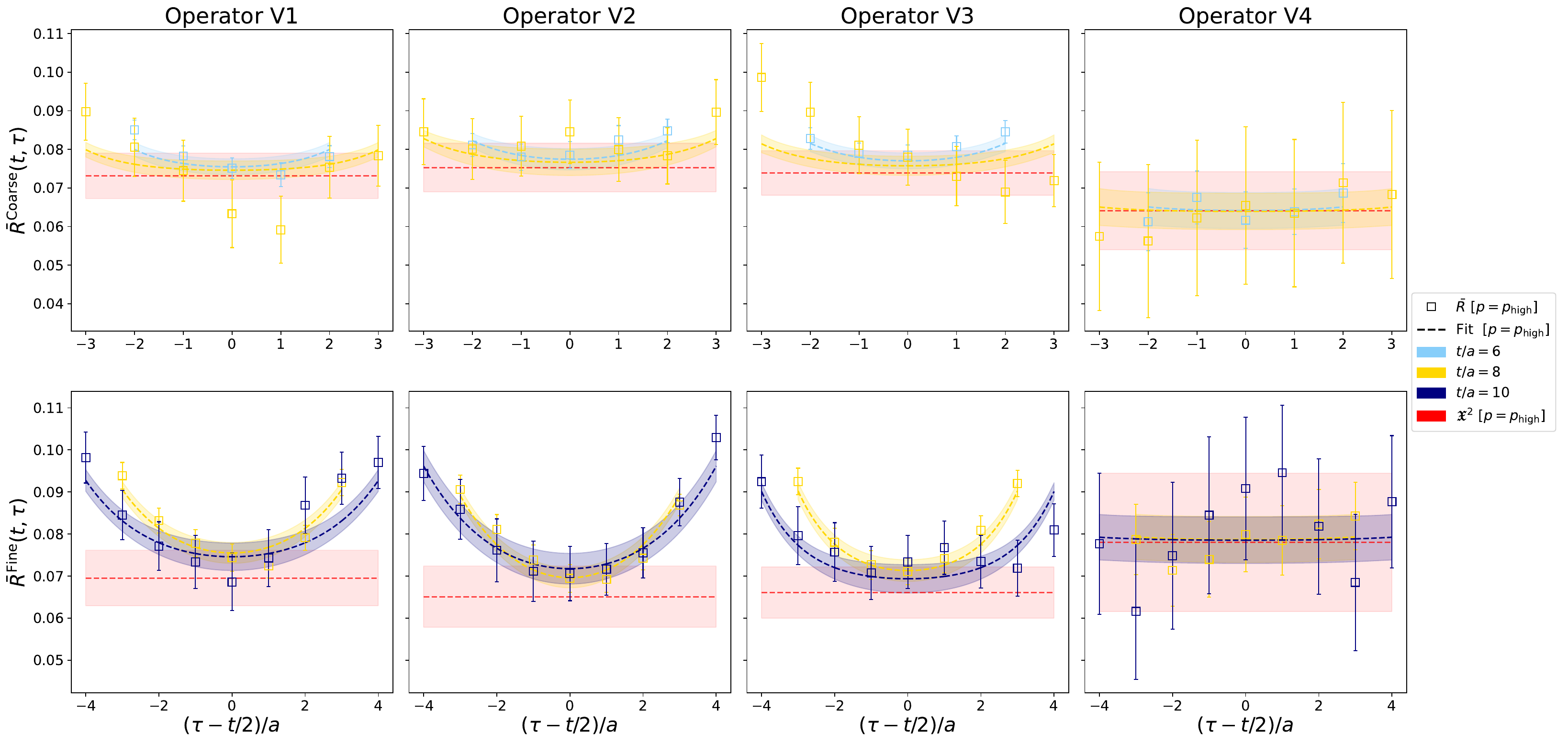}}%
\caption{\label{fig:fit_vector}%
Normalized ratio for the vector operators versus operator insertion time: different operators (columns) and ensembles (rows).
Different source-sink separations are distinguished by distinct colors.
The dashed curves illustrate the outcome of the model averaging procedure.
The shaded regions surrounding the curves indicate the $1\sigma$ uncertainty interval.
The red line denotes the value of the bare moment extracted from the fit average.
}
\end{figure*}

\begin{figure*}[ht!]
\makebox[\textwidth][c]{\includegraphics[width=\textwidth]{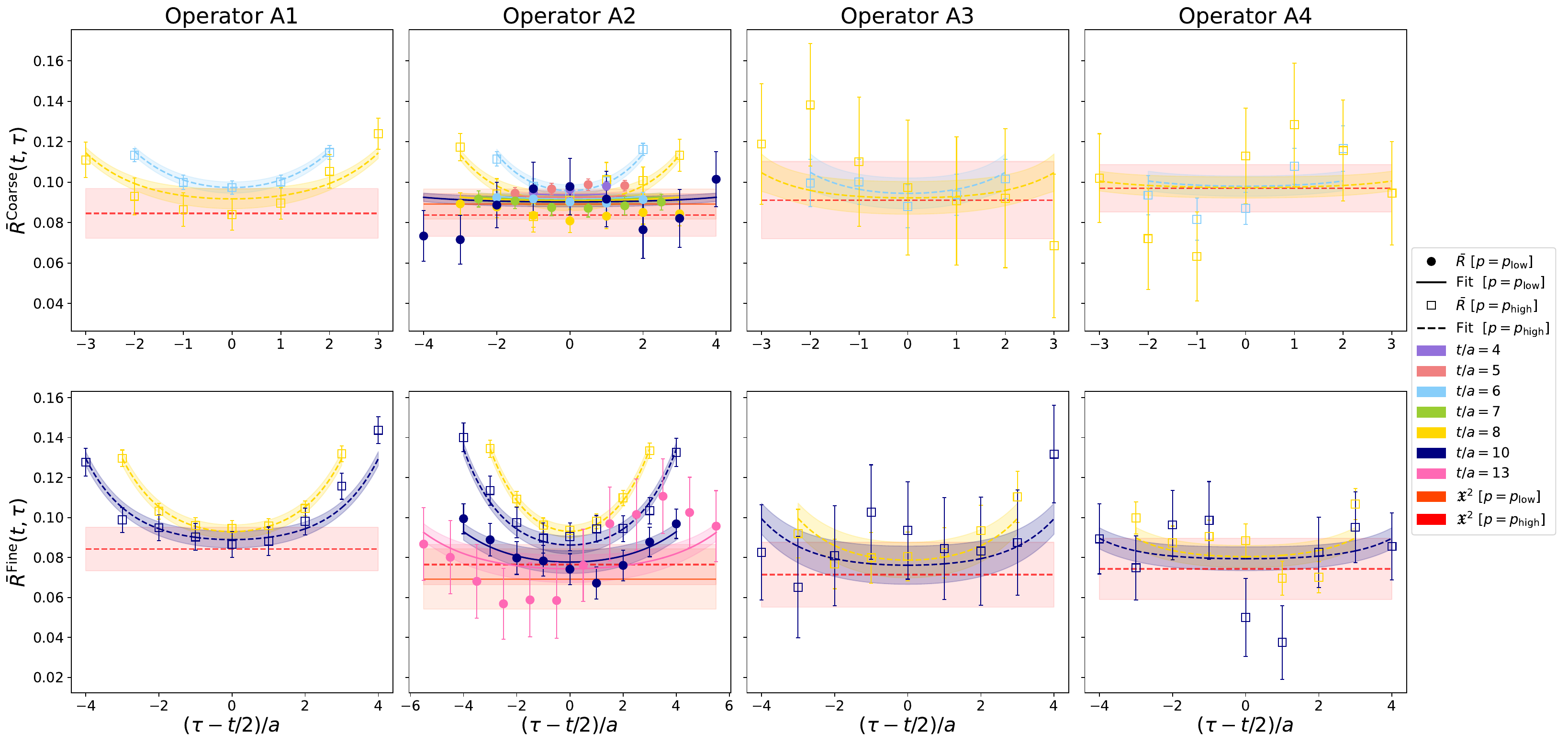}}%
\caption{\label{fig:fit_axial}%
Normalized ratio for the axial operators versus operator insertion time, analogous to Figure \ref{fig:fit_vector}.
In the axial case, one operator is also accessible at low momentum.
Different momenta are represented by different markers.
The outcome of the model averaging procedure is depicted using solid and dashed lines, respectively for the high- and low-momentum
cases.
The fit average over the low-momentum data yields additional values of the bare moment, which are depicted in orange.
}
\end{figure*}

\begin{figure*}[ht!]
\makebox[\textwidth][c]{\includegraphics[width=\textwidth]{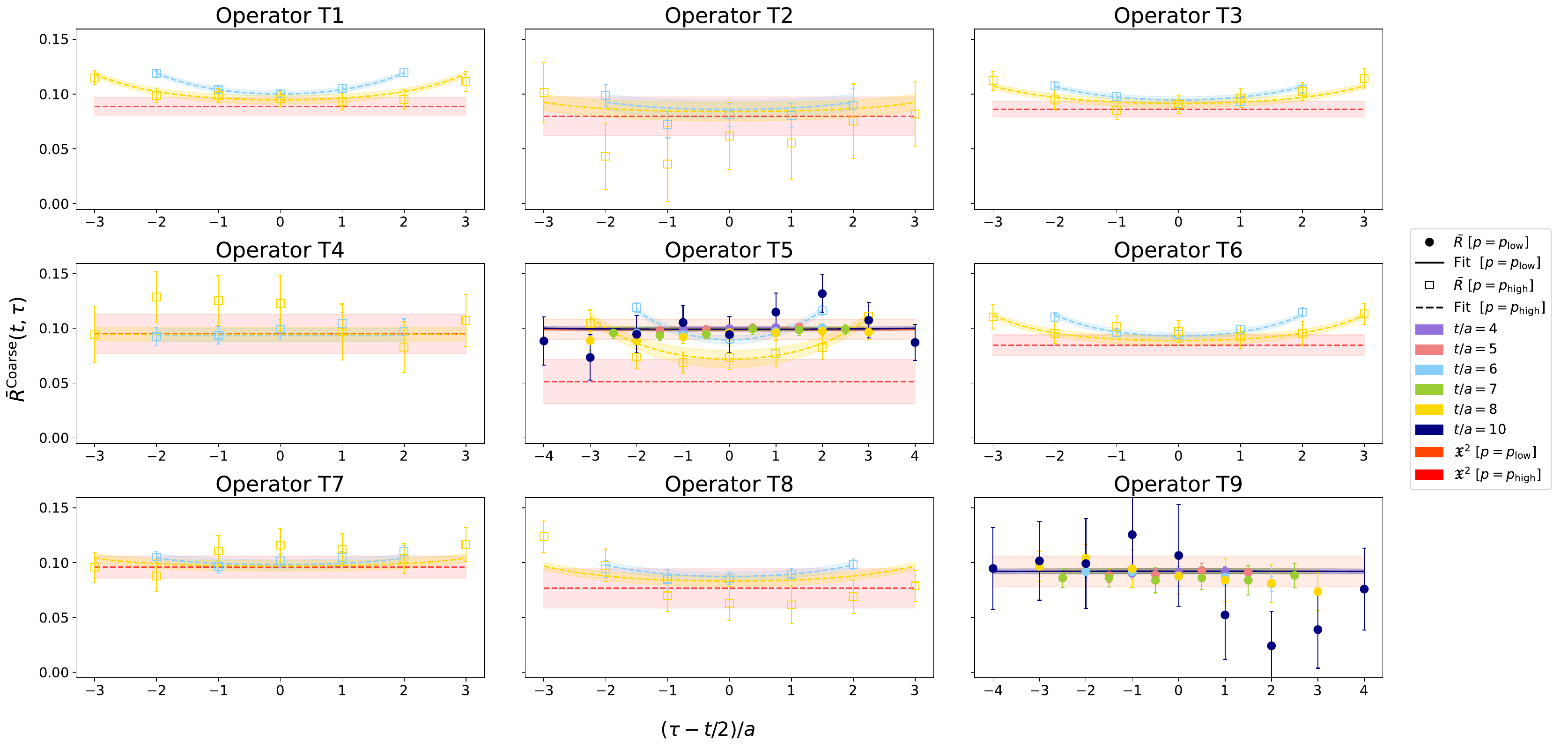}}%
\caption{\label{fig:fit_tensor_coarse}%
Normalized ratio, on the coarse lattice, for the tensor operators versus operator insertion time.
Analogous to Figure \ref{fig:fit_axial}, with two operators being also accessible at low momentum in the tensor case.
}
\end{figure*}

\begin{figure*}[ht!]
\makebox[\textwidth][c]{\includegraphics[width=\textwidth]{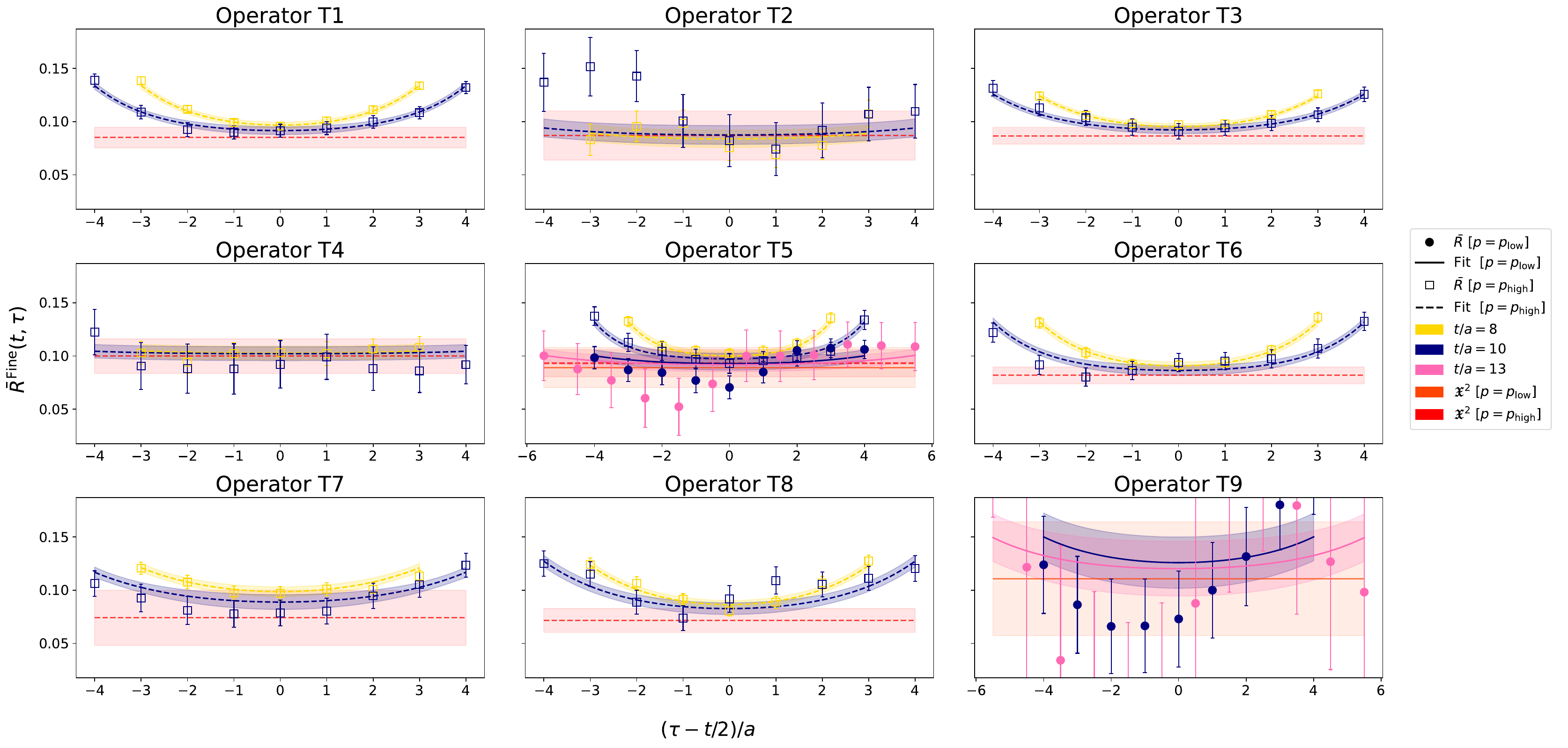}}%
\caption{\label{fig:fit_tensor_fine}%
Normalized ratio, on the fine lattice, for the tensor operators versus operator insertion time.
Analogous to Figure \ref{fig:fit_axial}, with two operators being also accessible at low momentum in the tensor case.
}
\end{figure*}

The results of the fits to the ratio data, together with the raw data, are displayed in Figures \ref{fig:fit_vector}–\ref{fig:fit_tensor_fine}.
For each operator, the plotted points have been normalized by the corresponding kinematic factor.
The values of the bare moments are indicated by the horizontal bands shown in the figures.
The curves that we overlay to the raw data points represent the outcome of the model averaging procedure.
They are not the outcome of a single fit;
instead, they have been obtained by averaging the outcomes of the different fits we consider.
The average is done, separately for each resample, using the AIC weights.
Then, averaging over the resamples, we obtain the central value and the surrounding $1\sigma$ region that we show in the plot.
For each channel and lattice, all the values of the bare moments obtained from the fit average procedure agree between uncertainties.
The only exception is the bare moment obtained on the coarse lattice from operator $T5$ with high momentum.
As is evident from Figure \ref{fig:fit_tensor_coarse}, this value is not compatible with the other ones obtained on the same ensemble.
However, on the fine ensemble this discrepancy disappears, as shown in Figure \ref{fig:fit_tensor_fine}.
This suggests that the outlier on the coarse ensemble may reflect a statistical fluctuation.

\begin{figure*}[ht!]
\makebox[\textwidth][c]{\includegraphics[width=\textwidth]{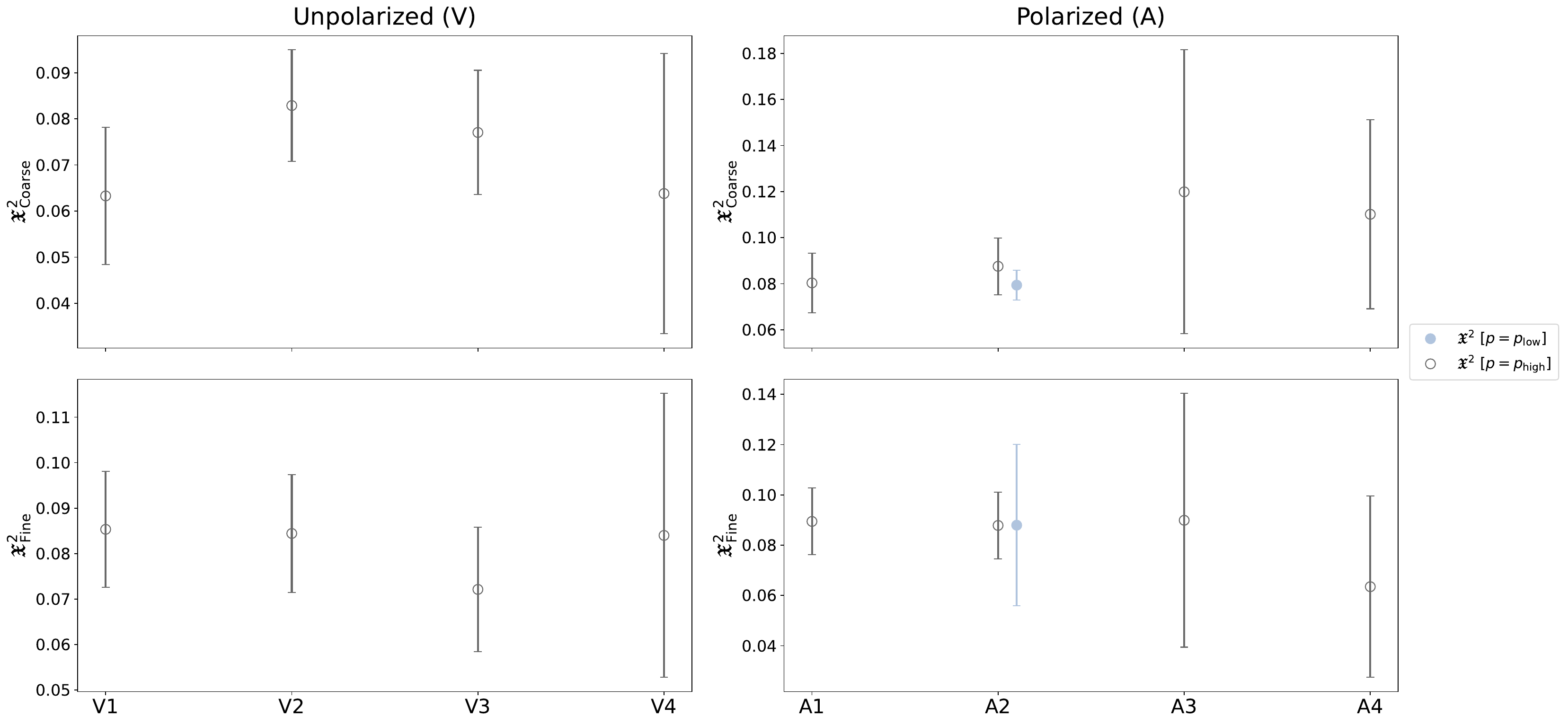}}%
\caption{\label{fig:sumratio_vector_axial}%
Overview of the unrenormalized summed ratio results for the vector and axial operators.
}
\end{figure*}

\begin{figure*}[ht!]
\makebox[\textwidth][c]{\includegraphics[width=\textwidth]{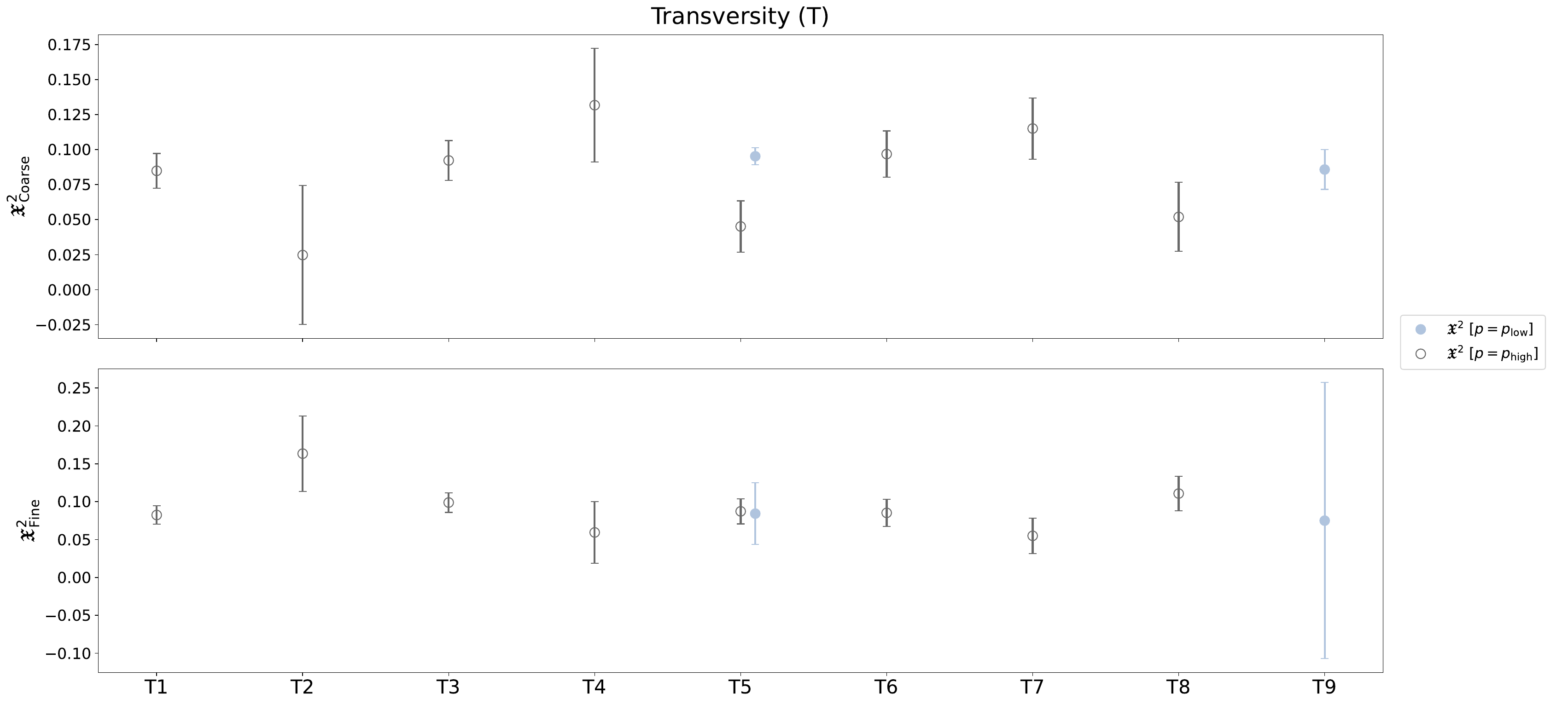}}%
\caption{\label{fig:sumratio_tensor}%
Overview of the unrenormalized summed ratio results for the tensor operators.
}
\end{figure*}

\begin{figure*}[ht!]
\makebox[\textwidth][c]{\includegraphics[width=\textwidth]{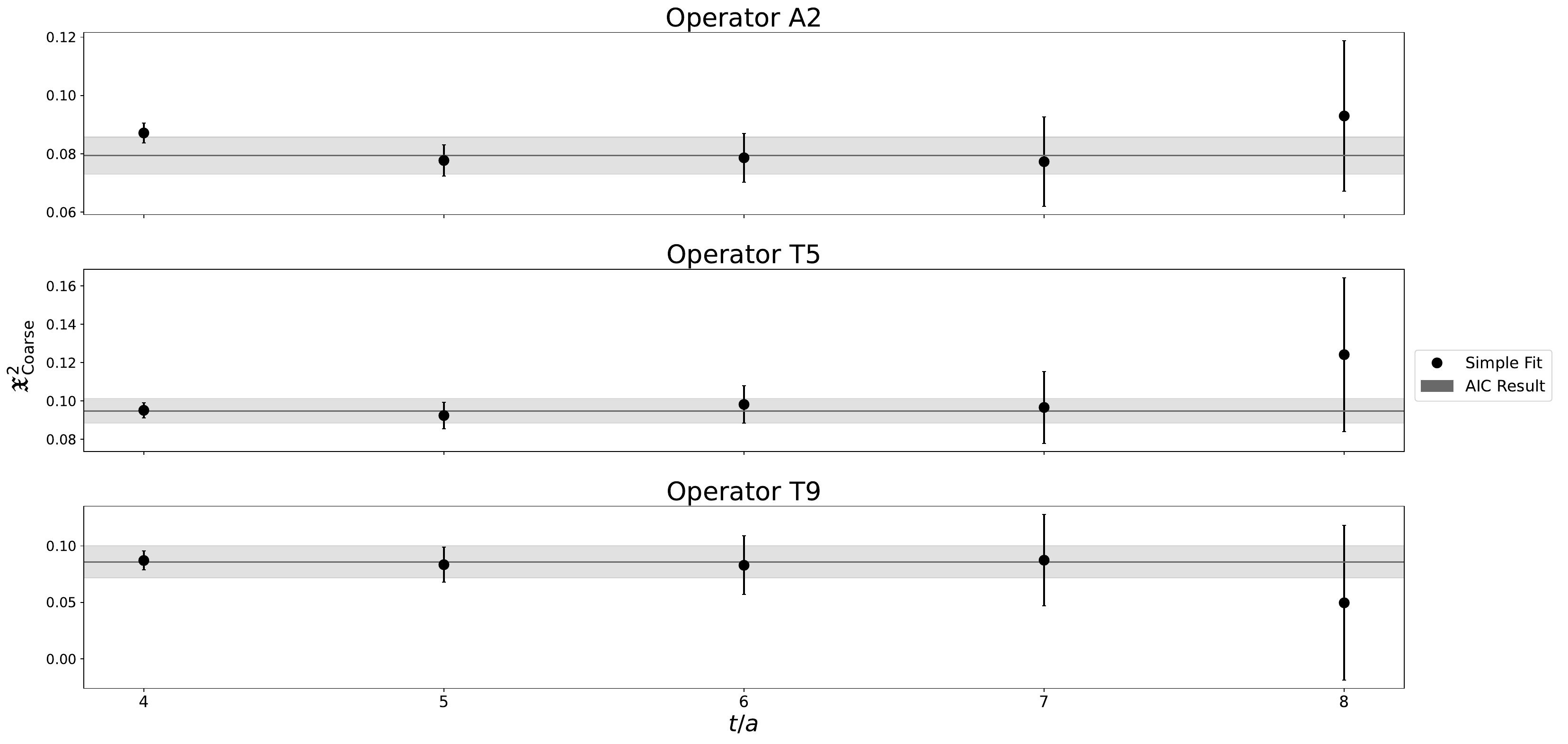}}%
\caption{\label{fig:sumratio_AIC}%
Comparison of the unrenormalized summed ratio results obtained with the AIC averaging procedure and results obtained from a simple fit.
The gray band indicates the result of the AIC averaging procedure.
The black dots are the result of a single linear regression;
they are displayed as a function of the lowest source-sink separation considered.
}
\end{figure*}

The bare moments extracted with the summed ratio method are shown in Figures \ref{fig:sumratio_vector_axial} and \ref{fig:sumratio_tensor}.
In the majority of cases only two distinct source–sink separations are available.
Consequently, the AIC averaging procedure we use for the summation method is trivial, and it does coincide with a simple linear regression.
For the low momentum on the coarse lattice, however, several source–sink separations are present.
In this case, the AIC averaging procedure is non trivial and takes into account multiple fits.
To illustrate the validity of the procedure, we compare the result we obtain to the results of simple linear regressions performed for different starting values of $t$.
The comparison is made in Figure \ref{fig:sumratio_AIC}.
Note that a simple linear regression starting at $t/a=6$ yields results in good agreement with those from the AIC averaging procedure, without underestimating the associated uncertainties.
This suggests that, for the high-momentum data, choosing $t/a=6$ as the smallest source–sink separation was a sensible choice.


\section{Model averaging\label{app:model_avg}}

To extract the Mellin moment directly from the ratio data, we perform multiple fits.
The different fits are distinguished by the choice of the functional form and by the data range considered.
Each fit yields a different estimate of the Mellin moment.
These values are combined into one by using the Akaike information criterion \cite{Jay:2020jkz, Neil:2022joj, Neil:2023pgt}.
The whole procedure is performed under a bootstrap analysis to extract the uncertainties. 
The fits have been performed on ratios normalized to the kinematic factors.
In this way, the fits yield moments instead of matrix elements. 

Three different functional forms have been employed to perform the fits
\begin{equation}\label{eq:ratio_functional_form}
\begin{alignedat}{1}
(i) \quad \bar R(t,\tau) &= \mathfrak{X}^2 , \\
(ii) \quad \bar R(t,\tau) &= \mathfrak{X}^2 + \bar R_{1} e^{-\tfrac{t}{2}\,\Delta E} \cosh \! \inSquare{ \left(\tfrac{t}{2} - \tau\right) \Delta E } , \\
(iii) \quad \bar R(t,\tau) &= \mathfrak{X}^2 + \bar R_{1} e^{-\tfrac{t}{2}\,\Delta E} \cosh \! \inSquare{ \left(\tfrac{t}{2} - \tau\right) \Delta E } + \bar R_{2} e^{-t \Delta E}.
\end{alignedat}
\end{equation}
The three functional forms differ in the number of terms retained from the spectral decomposition.
For each choice of operator, momentum, and ensemble, we perform one fit for each of three functional forms, repeated for several different data ranges.
Given that the ratio depends on two variables, the selection of the data ranges is not unique. 
We choose to prioritize points $(t,\tau)$ with larger values of $t$ and with $\tau$ close to $t/2$.
This choice is motivated by the expectation that the ratio converges to the matrix element in the limit $t-\tau,\tau \rightarrow\infty$, as stated in Eq.~\eqref{eq:ratio_limit}.
Furthermore, given the symmetry of the functional forms in Eq.~\eqref{eq:ratio_functional_form} about the midpoint $\tau=t/2$, we restrict to ranges which are symmetrical about this point.
Accordingly, the initial range is chosen as a subset of the points corresponding to the two largest source-sink separations.
For each value of $t$, the initial points are $(\tau-t/2)/a=0,\pm1$ if $t/a$ is even, and $\pm1/2,\pm3/2$ if $t/a$ is odd.
The range is then expanded iteratively, cycling continuously from the the largest to the smallest values of $t$ until all available points have been considered.
For a given $t$, we add the pair of points (or the point) with the smallest value of  $\left|\tau-t/2\right|$ not yet included.
Each time new points are added, a fit is performed.
The last fit always includes all the available points.
Using this procedure, we perform, for each functional form, a number of fits that goes from a minimum of 4 to a maximum of 16, depending on the number of data points available for each lattice and momentum combination.

The free parameters in the fits are $\mathfrak{X}^2$, $\Delta E$, $\bar R_1$ and $\bar R_2$.
All parameters exhibit an implicit dependence on the choice of the operator.
For $\bar R_1$ and $\bar R_2$ we used the prior $\mathcal{N}\inCurve{0,1}$, i.e. the normal distribution with mean 0 and standard deviation 1.
For $\Delta E$, in order to exclude negative energies from the fitting procedure, we used a log-normal prior and enforced a minimum energy gap $\Delta E_{\text{min}}=100\text{MeV}$.
To do so, we have rewritten the energy gap as $\Delta E = \Delta E_{\text{min}} + \Delta E'$, and used $\log{\Delta E'}$ as the free parameter of the fit.
The prior on $\log{\Delta E'}$ is modeled on the energy gap obtained from the fit of the two-point function.
Denoting the latter with $\Delta E_{2p}$, the chosen prior is such that $\log{\Delta E'} \sim \mathcal{N}\inCurve{\log{ \Delta E_{2p}'}, 1/2}$, with $\Delta E_{2p}' = \Delta E_{2p}-\Delta E_{\text{min}}$.
Depending on the choice of lattice and momentum, $\Delta E_{2p}$ ranges approximately from $1$ GeV to $1.6$ GeV.
Regarding $\mathfrak{X}^2$, the prior is modeled on the value obtained from the summed ratio method.
Denoting the central value and standard deviation on the latter with $\mathfrak{X}^2_S$ and $\sigma_S$ respectively, the chosen prior is a gaussian distribution with central value $\mathfrak{X}^2_S$ and with width $\text{max} ( \mathfrak{X}^2_S, \sigma_S )$.

For a given operator, depending on the amount of excited state contamination, certain kinds of fits, with a given functional form and a given data range, will work better than the others.
To asses which fit is the most relevant, we use the  Akaike information criterion \cite{Jay:2020jkz}
\begin{equation}
    \text{AIC} = \chi^2 + 2N_{\text{params}} - 2N_{\text{data}} .
\end{equation}
This allows us to combine the estimates we have for the free parameters of the fit, $p\in (\mathfrak{X}^2,\Delta E, \bar R_1, \bar R_2)$, into one single average
\begin{equation}\label{eq:AIC_average}
    \inBraket{p} = \frac{ \sum_i e^{-\frac{1}{2} \text{AIC}_i} p_i}{ \sum_i e^{-\frac{1}{2} \text{AIC}_i} },
\end{equation}
where the index $i$ runs over all available fits using the parameter $p$.
The weight $e^{-\frac{1}{2} \text{AIC}_i}$ takes into account the goodness of a fit by incorporating the $\chi^2$.
Furthermore, it penalizes fits with a great number of parameters and excluded data points.
A lengthy discussion about the validity of model averaging with the AIC, and how it compares to other model averaging procedures, can be found in \cite{Neil:2022joj}.
In the generalized form of Eq.~\eqref{eq:AIC_average} the single parameter $p$ is replaced by an arbitrary functions of the parameters $f(p)$.
Using this generalization, we can combine the fit curves from the various fits into a single averaged curve.
This is the curve we overlay on the data in Figures \ref{fig:fit_vector}–\ref{fig:fit_tensor_fine}.

\begin{figure*}[ht!]
\makebox[\textwidth][c]{\includegraphics[width=\textwidth]{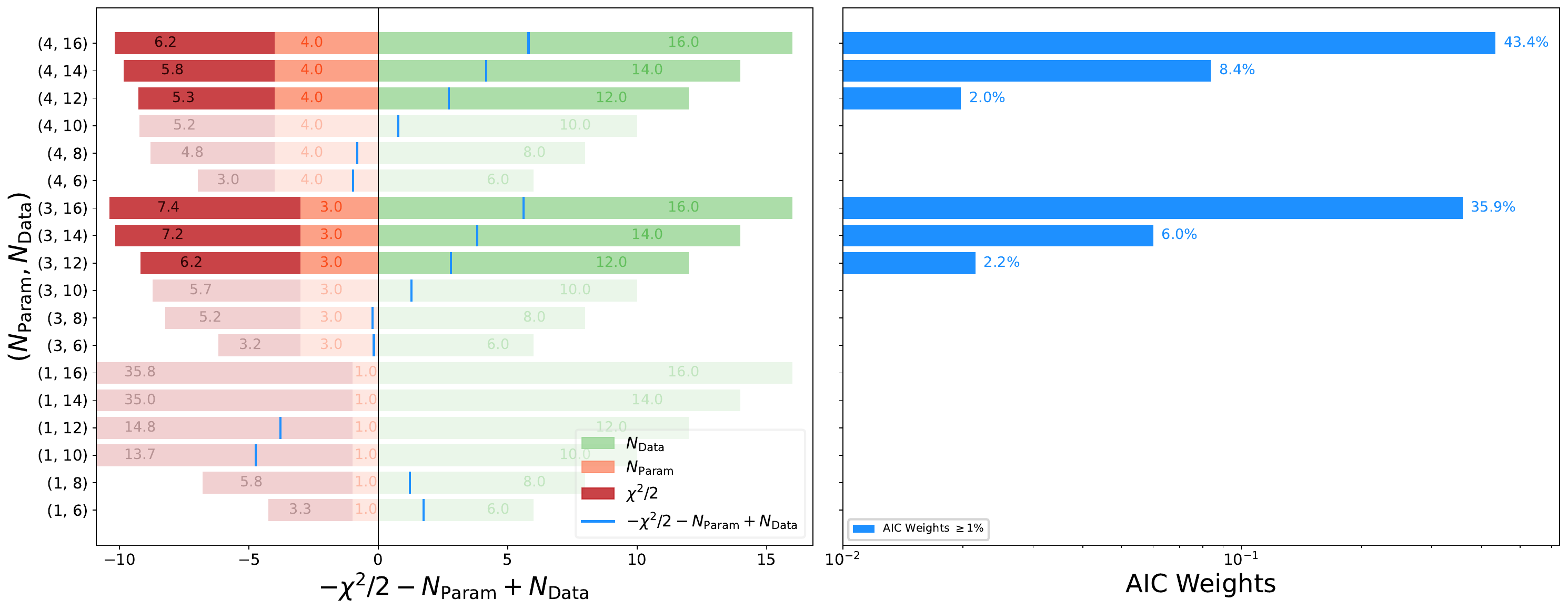}}%
\caption{\label{fig:AIC_operatorV1}%
Overview of the AIC weights for the model averaging procedure applied to the fine lattice data for operator $V1$.
Left: breakdown of the contributions to the AIC for each fit.
The fits are ordered according to increasing number of parameters and included data points.
The fits with 1, 3 and 4 parameters corresponds to the functional forms \emph{(i)}, \emph{(ii)} and \emph{(iii)} of Eq.~\eqref{eq:ratio_functional_form}, respectively.
Fits with AIC weight smaller than 1\% are shown in faint colors.
Right: AIC weights exceeding or equal to 1\%.
}
\end{figure*}

As a specific example, we now provide additional details regarding the model averaging procedure performed on the ratio data for operator $V1$ on the fine lattice.
The ratio data and the outcome of the model averaging procedure are shown on the bottom-left panel of Figure \ref{fig:fit_vector}.
In this case 
we perform 6 fits for each functional form, for a total of 18 different fits.
These fits differ in the number of parameters and in the number of data points employed.
An overview of the AIC weight associated to each fit is given in Figure \ref{fig:AIC_operatorV1}.
We find that the overall weight given to constant fits is completely negligible.
This is to be expected, as in this case the ratio data is manifestly not compatible to a constant and requires accounting for excited states.
In contrast, ratio data such as that on the bottom-right panel of Figure \ref{fig:fit_vector} are described well by a constant fit.
This reflects on the distribution of the AIC weights being significantly different from the one shown in Figure \ref{fig:AIC_operatorV1}.
The usefulness of the model averaging lies in the fact that we automatically take into account these differences.
We perform exactly the same fits in each case, and let any difference in the data reflect in a difference in the AIC weights.

The model averaging machinery that we just discussed has also been used for the two-point correlators and the summed ratios.
In these cases, the discussion is completely analogous, though greatly simplified.
For the two-point correlators, only two functional forms have been considered: a simple exponential, from which we extract only the ground state energy, and the sum of two exponentials, from which we extract also the energy gap of the first excited state.
Given that the data are one dimensional, the variation of the number of data points is straightforward.
We fix the last time slice and start with a total of 6 points.
We then increase the number of data points from below until we reach the first time slice.
For the summed ratios, the only functional form considered is a linear function.
The data is again one dimensional, and in this case multiple fits are performed by varying both the first and last value of $t$ considered in the fit.
This procedure differs from a simple fit only when there are more than two source-sink separations available.
Both for the two-point correlators and for the summed ratios, the estimates of the free parameters of the fits are combined into one single average using Eq.~\eqref{eq:AIC_average}, exactly as described above.

%

\section{Nonperturbative renormalization\label{app:renormalization}}

Using mostly the same methods as our earlier work for operators with one
derivative~\cite{Rodekamp:2023wpe}, we employ the nonperturbative
Rome-Southampton method~\cite{Martinelli:1994ty} to determine
renormalization factors $Z_{DDV}^\rho$, $Z_{DDA}^\rho$, and
$Z_{DDT}^\rho$ for the two-derivative isovector vector, axial, and
tensor operators. These depend on the irreducible representation
$\rho$ of the hypercubic group. As intermediate schemes, we use
variants of both RI$'$-MOM~\cite{Martinelli:1994ty, Gockeler:1998ye}
and RI-SMOM~\cite{Sturm:2009kb}, which we then perturbatively convert
and evolve to the $\overline{\mathrm{MS}}$ scheme at scale 2~GeV.

In Landau gauge, we compute the quark propagator,
\begin{equation}
  S(p) = \int d^4x \, e^{-ip\cdot x} \langle u(x) \bar u(0) \rangle,
\end{equation}
and the Green's functions for operator $\mathcal{O}$,
\begin{equation}
  G_\mathcal{O}(p',p) = \int d^4x'\, d^4x\, e^{-ip'\cdot x'} e^{ip\cdot x}
  \langle u(x') \mathcal{O}(0) \bar u(x) \rangle,
\end{equation}
yielding the amputated Green's functions
\begin{equation}
  \Lambda_\mathcal{O}(p',p) = S^{-1}(p')G_\mathcal{O}(p',p)S^{-1}(p).
\end{equation}
To obtain a precise signal, we employ plane wave sources to perform a
full volume average. Provided that the operator renormalizes
diagonally, so do the amputated Green's functions:
$\Lambda_\mathcal{O}^R = (Z_\mathcal{O}/Z_\psi)
\Lambda_\mathcal{O}$. As in our previous work, we avoid an explicit
determination of $Z_\psi$ and instead obtain $Z_\mathcal{O}/Z_V$ from
ratios, taking $Z_V$ from the determination using pion three-point
functions in Ref.~\cite{Hasan:2019noy}.

One complication is that our operators do not renormalize diagonally:
$\bar q \lrarrow{D}_{\{\mu}\lrarrow{D}_{\nu\}} \Gamma q$ can always
mix with $\partial_\mu\partial_\nu(\bar q \Gamma q)$. We avoid this
problem by choosing observables for which the mixing operator (a total
derivative) vanishes. This is trivial for the nucleon forward matrix
elements and for RI$'$-MOM kinematics; we will see below that it is
also achievable for RI-SMOM.

\subsection{Conditions and matching}

Our strategy starts with the decomposition, in the continuum, of
$\Lambda_{\shortOP_{\alpha\mu\nu}}(p',p)$ into products of
O(4)-invariant functions and kinematic tensors:
\begin{equation}
  \Lambda_{\shortOP_{\alpha\mu\nu}}(p',p) = \sum_i \Sigma_{\shortOP}^{(i)}(p^2)
  \Lambda^{(i)}_{\shortOP_{\alpha\mu\nu}}(p',p).
\end{equation}
Taking into account the reduced symmetry of the lattice, we then
decompose this into irreducible representations of the hypercubic
group, replacing $\alpha\mu\nu$ with $\rho n$, where $n$ ranges from 1
to the dimension of $\rho$. To isolate the O(4)-invariant functions,
we take the traces
\begin{equation}\label{eq:renorm_trace}
  \sum_n \Tr\left[ \Lambda^{(i)}_{\shortOP_{\rho n}}(p',p)
    \Lambda_{\shortOP_{\rho n}}(p',p) \right]
  = M^{ij}_\rho(p',p) \Sigma^{(j)}_{\mathcal{O}^X}(p^2),
\end{equation}
where $M$ is a computable kinematic matrix (see
\cite{Rodekamp:2023wpe}). Our tensor decompositions below imply that
at tree level, $\Sigma_{\mathcal{O}^X}^{(i)}=\delta^{i1}$, and our
preferred renormalization conditions are
$\Sigma^{(1)}_{\mathcal{O}^X_R,\rho}(\mu^2)=1$. However, in some
cases\footnote{This might be avoided by not summing over $n$, which
  could yield a larger system of equations constraining
  $\Sigma^{(i)}$.}  $M$ is such that Eq.~\eqref{eq:renorm_trace} can't
be solved for $\Sigma^{(1)}$; when necessary, we add a linear
combination of $\Sigma^{(j)}$, $j\neq 1$. For ratios of
renormalization factors between two different lattice irreps of the
same continuum operator, we take the same linear combination to ensure
the ratio is scale and scheme invariant and is unity under restoration
of $O(4)$ symmetry.

We now provide the tensor decompositions. For conciseness, only the
tensors that have possibly nonzero trace with the tree-level tensor
are considered. The two-derivative vector and axial operators are
\begin{align}
  \mathcal{O}^V_{\mu\nu\rho} &= -\mathcal{S}
   \bar\psi \tau_3 \gamma_{\mu} \lrarrow{D}_\nu \lrarrow{D}_\rho \psi,\\
  \mathcal{O}^A_{\mu\nu\rho} &= -\mathcal{S}
   \bar\psi \tau_3 \gamma_{\mu} \gamma_5 \lrarrow{D}_\nu \lrarrow{D}_\rho \psi,
\end{align}
where $\mathcal{S}$ takes the symmetric traceless part:
\begin{equation}
  \mathcal{S} T_{\mu\nu\rho} = \hat T_{\mu\nu\rho}
  - \frac{1}{2}\delta_{\{\mu\nu} \hat T_{\rho\}\alpha\alpha},
  \qquad \hat T_{\mu\nu\rho} = T_{\{\mu\nu\rho\}}.
\end{equation}
Here we have elected to include a factor of $(-i)^2$ in the operator
rather than a factor of $i^2$ in the tree-level tensor
below.\footnote{The corresponding factor of $-i$ was also included in
  calculation of Ref.~\cite{Rodekamp:2023wpe}, although it was missing
  from the text of the paper.} RI$'$-MOM kinematics correspond to $p'=p$ with
scale $\mu^2=p^2$; we obtain two tensors:
\begin{align}
  \Lambda^{(1)}_{\mathcal{O}^V_{\mu\nu\rho}}(p,p) &= \mathcal{S} \gamma_\mu p_\nu p_\rho,\\
  \Lambda^{(2)}_{\mathcal{O}^V_{\mu\nu\rho}}(p,p) &=
       \frac{\slashed{p}}{p^2}\mathcal{S} p_\mu p_\nu p_\rho.
\end{align}
For RI-SMOM, $p^2=(p')^2=(p'-p)^2=\mu^2$. Defining $q\equiv p'-p$ and
$\bar p\equiv (p'+p)/2$, we get six tensors:
\begin{align}
  \Lambda^{S(1)}_{\mathcal{O}^V_{\mu\nu\rho}}(p',p)
  &= \mathcal{S} \gamma_\mu \bar p_\nu \bar p_\rho, \\
  \Lambda^{S(2)}_{\mathcal{O}^V_{\mu\nu\rho}}(p',p)
  &= \mathcal{S} \gamma_\mu q_\nu q_\rho, \\
  \Lambda^{S(3)}_{\mathcal{O}^V_{\mu\nu\rho}}(p',p)
  &= \frac{\slashed{\bar p}}{\bar p^2}\mathcal{S} \bar p_\mu \bar p_\nu \bar p_\rho, \\
  \Lambda^{S(4)}_{\mathcal{O}^V_{\mu\nu\rho}}(p',p)
  &= \frac{\slashed{q}}{q^2}\mathcal{S} q_\mu \bar p_\nu \bar p_\rho, \\
  \Lambda^{S(5)}_{\mathcal{O}^V_{\mu\nu\rho}}(p',p)
  &= \frac{\slashed{\bar p}}{q^2}\mathcal{S} \bar p_\mu q_\nu q_\rho, \\
  \Lambda^{S(6)}_{\mathcal{O}^V_{\mu\nu\rho}}(p',p)
  &= \frac{\slashed{q}}{q^2}\mathcal{S} q_\mu q_\nu q_\rho.
\end{align}
Here, imposing $C$-symmetry results in considerably fewer tensors than
Gracey~\cite{Gracey:2010ci}. Since the mixing operator is a second
total derivative, its tensor structures must be of the form
$\mathcal{S} q_\mu q_\nu \Lambda_\rho$, where $\Lambda_\rho$ is a
tensor of the zero-derivative operator; we therefore conclude that
mixing can only occur in $\Sigma^{S(i)}$ for $i\in\{2,5,6\}$. For
$\mathcal{O}^A_{\mu\nu\rho}$, the tensors contain an additional factor
of $\gamma_5$.

The two-derivative tensor operator is
\begin{equation}
  \mathcal{O}^T_{\mu\nu\rho\sigma} = -\mathcal{S}
  \bar\psi \tau_3 \sigma_{\mu\nu} \lrarrow{D}_\rho \lrarrow{D}_\sigma \psi,
\end{equation}
where the symmetrization and trace subtraction is given by
\begin{equation}
  \begin{gathered}
    \mathcal{S}T_{\mu\nu\rho\sigma} = \hat T_{\mu\nu\rho\sigma}
    - \frac{5}{8} \hat T_{\mu\alpha\alpha\{\nu}\delta_{\rho\sigma\}}
    - \frac{1}{8} \delta_{\{\rho\sigma} \hat T_{\nu\}\mu\alpha\alpha}
    + \frac{6}{8} \delta_{\mu\{\nu} \hat T_{\rho\sigma\}\alpha\alpha},\\
    \hat T_{\mu\nu\rho\sigma} = \tilde T_{\mu\{\nu\rho\sigma\}},\qquad
    \tilde T_{\mu\nu\rho\sigma} = \frac{1}{2}\left(T_{\mu\nu\rho\sigma} - T_{\nu\mu\rho\sigma}\right).
  \end{gathered}
\end{equation}
In RI$'$-MOM kinematics, we again find two tensors:
\begin{align}
  \Lambda^{(1)}_{\mathcal{O}^T_{\mu\nu\rho\sigma}}(p,p)
  &= \mathcal{S} \sigma_{\mu\nu} p_\rho p_\sigma,\\
  \Lambda^{(2)}_{\mathcal{O}^T_{\mu\nu\rho\sigma}}(p,p)
  &= \frac{1}{p^2} \mathcal{S} \sigma_{\mu\alpha} p_\alpha p_\nu p_\rho p_\sigma.
\end{align}
As with the one-derivative case, we find that starting with the
antisymmetrization of $\mu$ and $\nu$ reduces Gracey's three
tensors~\cite{Gracey:2006zr} to two. For RI-SMOM,
Ref.~\cite{Braun:2016wnx} has 31 tensors within the category we
consider. The same antisymmetrization reduces the count to 18, then
$C$-symmetry leaves us with the following ten:
\begin{align}
  \Lambda^{S(1)}_{\mathcal{O}^T_{\mu\nu\rho\sigma}}(p',p)
  &= \mathcal{S} \sigma_{\mu\nu} \bar p_\rho \bar p_\sigma, \\
  \Lambda^{S(2)}_{\mathcal{O}^T_{\mu\nu\rho\sigma}}(p',p)
  &= \mathcal{S} \sigma_{\mu\nu} q_\rho q_\sigma, \\
  \Lambda^{S(3)}_{\mathcal{O}^T_{\mu\nu\rho\sigma}}(p',p)
  &= \frac{1}{\bar p^2} \mathcal{S} \sigma_{\mu\alpha} \bar p_\alpha \bar p_\nu \bar p_\rho \bar p_\sigma, \\
  \Lambda^{S(4)}_{\mathcal{O}^T_{\mu\nu\rho\sigma}}(p',p)
  &= \frac{1}{q^2} \mathcal{S} \sigma_{\mu\alpha} q_\alpha q_\nu \bar p_\rho \bar p_\sigma, \\
  \Lambda^{S(5)}_{\mathcal{O}^T_{\mu\nu\rho\sigma}}(p',p)
  &= \frac{1}{q^2} \mathcal{S} \sigma_{\mu\alpha} q_\alpha \bar p_\nu q_\rho \bar p_\sigma, \\
  \Lambda^{S(6)}_{\mathcal{O}^T_{\mu\nu\rho\sigma}}(p',p)
  &= \frac{1}{q^2} \mathcal{S} \sigma_{\mu\alpha} \bar p_\alpha q_\nu q_\rho \bar p_\sigma, \\
  \Lambda^{S(7)}_{\mathcal{O}^T_{\mu\nu\rho\sigma}}(p',p)
  &= \frac{1}{q^2} \mathcal{S} \sigma_{\mu\alpha} \bar p_\alpha \bar p_\nu q_\rho q_\sigma, \\
  \Lambda^{S(8)}_{\mathcal{O}^T_{\mu\nu\rho\sigma}}(p',p)
  &= \frac{1}{q^2} \mathcal{S} \sigma_{\mu\alpha} q_\alpha q_\nu q_\rho q_\sigma, \\
  \Lambda^{S(9)}_{\mathcal{O}^T_{\mu\nu\rho\sigma}}(p',p)
  &= \frac{1}{q^2\bar p^2} \sigma_{\alpha\beta} \bar p_\alpha q_\beta \mathcal{S} \bar p_\mu q_\nu \bar p_\rho \bar p_\sigma, \\
  \Lambda^{S(10)}_{\mathcal{O}^T_{\mu\nu\rho\sigma}}(p',p)
  &= \frac{1}{q^2\bar p^2} \sigma_{\alpha\beta} \bar p_\alpha q_\beta \mathcal{S} \bar p_\mu q_\nu q_\rho q_\sigma,
\end{align}
and mixing can only occur in $\Sigma^{S(i)}$ for $i\in\{2,7,8,10\}$.

After converting to $\overline{\mathrm{MS}}$, we evolve using the
four-loop anomalous dimension for the vector and axial
cases~\cite{Velizhanin:2014fua, Baikov:2015tea} and the three-loop
anomalous dimension for the tensor case~\cite{Gracey:2006zr}. For
RI$'$-MOM, we use three-loop matching~\cite{Gracey:2006zr}, whereas
for RI-SMOM, matching is done at three-loop order for the vector and
axial operators~\cite{Kniehl:2020nhw} and two-loop order for the
tensor~\cite{Braun:2016wnx}.

\subsection{Calculation}

We again follow Refs.~\cite{Hasan:2019noy, Rodekamp:2023wpe}. Using
periodic boundary conditions in space and time for the quarks (this is
partially twisted, since the sea quarks are antiperiodic in time), we
construct plane-wave sources with momenta either along the
four-dimensional diagonal
$p^{(\prime)} \in \{ \frac{2\pi}{L}(k,k,k,k),
\frac{2\pi}{L}(k,k,k,-k)\}$ or along the two-dimensional diagonal
$p^{(\prime)} \in \{\frac{2\pi}{L}(k,k,0,0),
\frac{2\pi}{L}(k,0,k,0)\}$, where
$k\in\{2,3,\dots,\frac{L}{4a}\}$. This allows both RI$'$-MOM and
RI-SMOM kinematics for each of the diagonals. On the fine ensemble, we
used 54 gauge configurations. Likewise on the coarse ensemble, we
initially chose 54 configurations; however, on one configuration, the
twisted boundary condition made the Dirac operator near-singular and
the multigrid solver was unable to converge. Omitting this
configuration, we used the remaining 53.

After perturbatively matching to $\overline{\mathrm{MS}}$ and evolving
to the scale 2~GeV, there remains residual dependence on the scale
$\mu$. This is caused by the window problem of being unable to neglect
effects of both orders $a^2\mu^2$ and $\Lambda_\text{QCD}^2/\mu^2$. We
first apply a tree-level correction factor\footnote{The square root of
  this was also used as a correction in Ref.~\cite{Rodekamp:2023wpe},
  although not mentioned in the article.}
$[(\sin \tfrac{2\pi a}{L} k)/(\tfrac{2\pi a}{L} k)]^2$. Then following
Ref.~\cite{Boucaud:2005rm}, we fit the data using a polynomial in
$\mu^2$ plus sometimes a pole term, i.e.\ with the form
$A + B\mu^2 + C\mu^4 + D/\mu^2$; the constant term $A$ is our estimate
of the ratio $Z_\mathcal{O}/Z_V$.

\begin{figure*}
  \centering
  \includegraphics[width=0.495\textwidth]{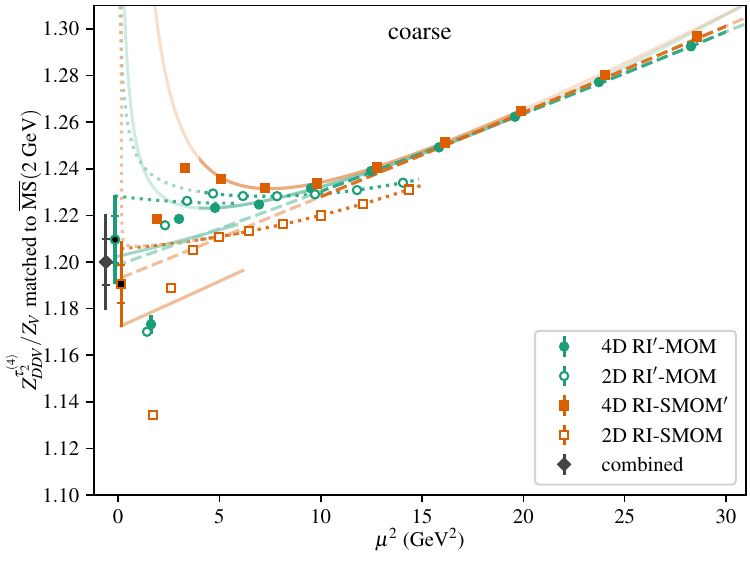}
  \includegraphics[width=0.495\textwidth]{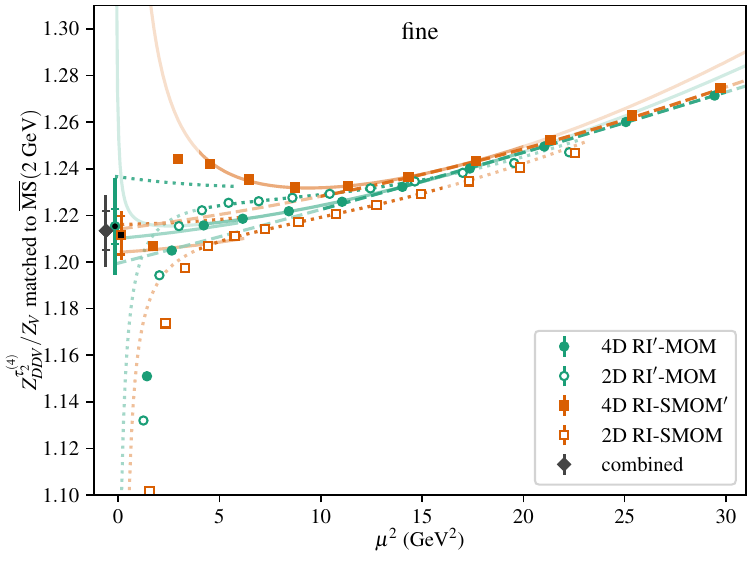}
  \includegraphics[width=0.495\textwidth]{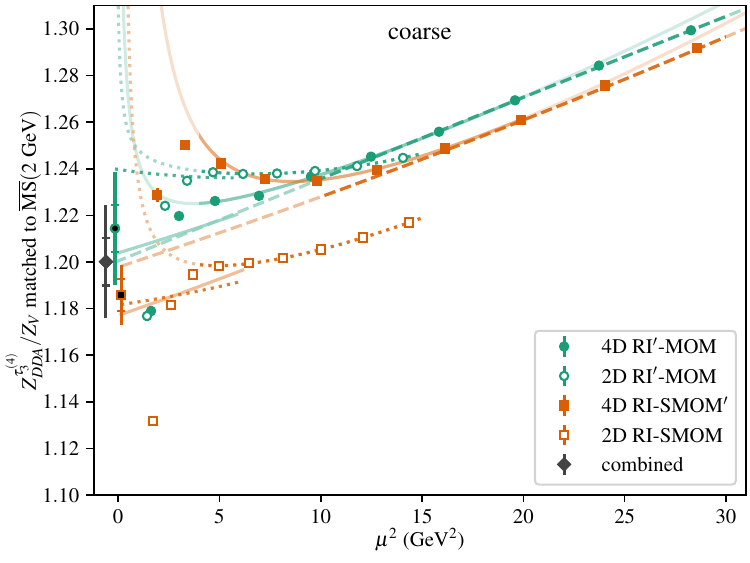}
  \includegraphics[width=0.495\textwidth]{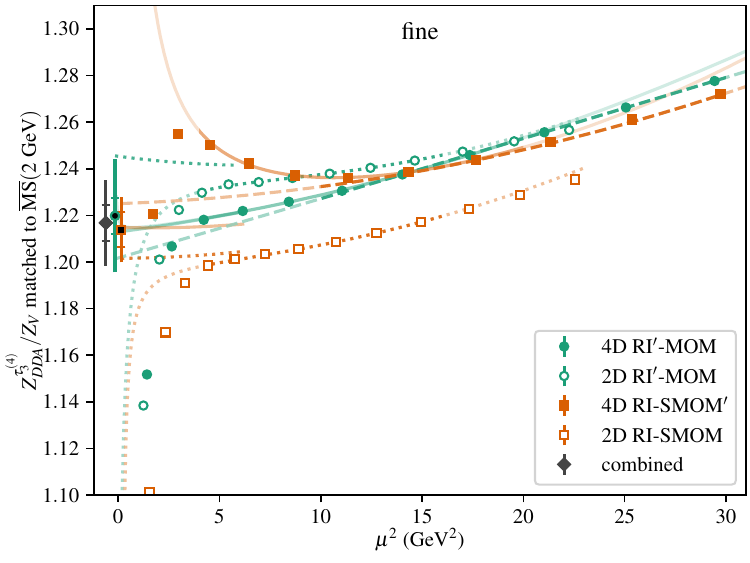}
  \includegraphics[width=0.495\textwidth]{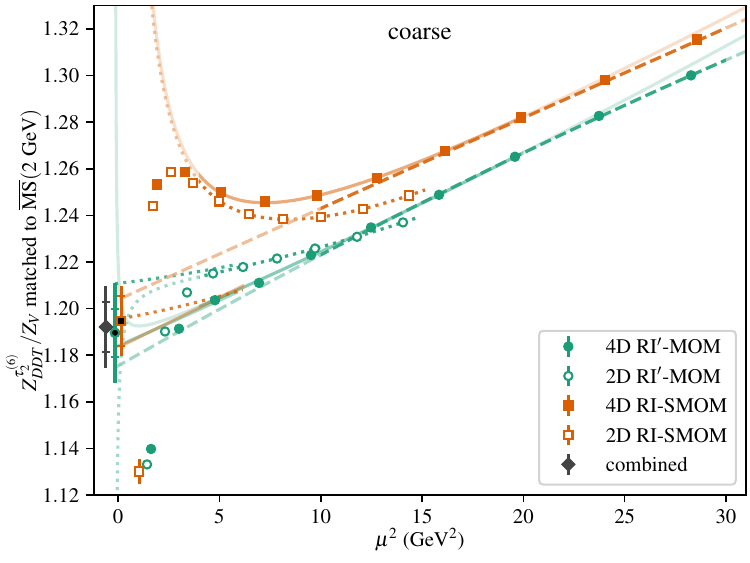}
  \includegraphics[width=0.495\textwidth]{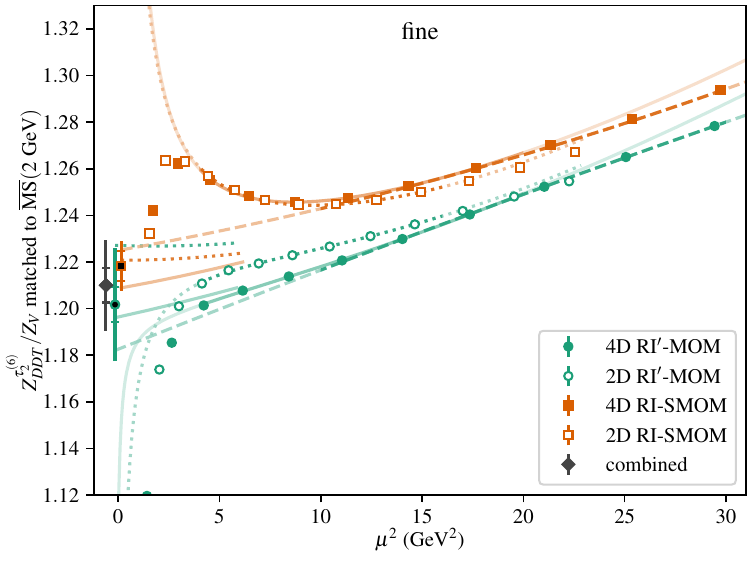}
  \caption{Ratios of renormalization factors
    $Z_{DDV}^{\tau^{(4)}_2}/Z_V$ (top), $Z_{DDA}^{\tau^{(4)}_3}/Z_V$
    (middle), and $Z_{DDT}^{\tau^{(6)}_2}/Z_V$ (bottom) on the coarse
    (left) and fine (right) ensembles, determined using RI$'$-MOM
    (green circles) and RI-SMOM (orange squares) kinematics and then
    matched to $\overline{\mathrm{MS}}$ at scale 2~GeV. The open and
    filled symbols correspond to momenta along the two-dimensional and
    four-dimensional diagonals, respectively. For most points, the
    statistical error is smaller than the symbol. In most cases, the
    renormalization condition is $\Sigma^{(1)}(\mu^2)=1$ or
    $\Sigma^{S(1)}(\mu^2)=1$; RI-SMOM$'$ indicates the condition
    $\Sigma^{S(1)}(\mu^2) + \Sigma^{S(3)}(\mu^2) = 1$; and the tensor
    RI-SMOM condition is
    $\Sigma^{S(1)}(\mu^2) + \frac{1}{2}\Sigma^{S(4)}(\mu^2) +
    \frac{1}{4}\Sigma^{S(5)}(\mu^2)=1$. Fits (a) are shown with solid
    curves, (b) are shown with dotted curves, and (c) are shown with
    dashed curves; they are also shown with the pole term set to zero
    in the range $\mu^2\in [0,6]$~GeV$^2$. The points near $\mu^2=0$
    show the combined estimates with statistical and systematic
    uncertainties.}
  \label{fig:renormalization}
\end{figure*}

Estimates of the ratio $Z_\mathcal{O}/Z_V$ for one irrep of the
two-derivative vector, axial, and tensor operators are shown in
Fig.~\ref{fig:renormalization}. Systematic uncertainties are explored
by performing three fits for each operator and renormalization scheme:
(a) the 4D data in the range $\mu^2\in[4,20]$~GeV$^2$, (b) the 2D data
in the range $[4,15]$~GeV$^2$, and (c) setting $D=0$, the 4D data in
the range $[10,30]$~GeV$^2$, which has limited sensitivity to a
$\mu^{-1}$ term. The fit quality to these sometimes very precise data
can be poor, so we scale the statistical uncertainty by
$\sqrt{\chi^2/\text{dof}}$ whenever this is greater than one. The
systematic uncertainty, which we estimate from the spread of results,
is dominant, and we conservatively take the maximum of statistical
uncertainties among the fits.

\begin{figure*}
  \centering
  \includegraphics[width=0.495\textwidth]{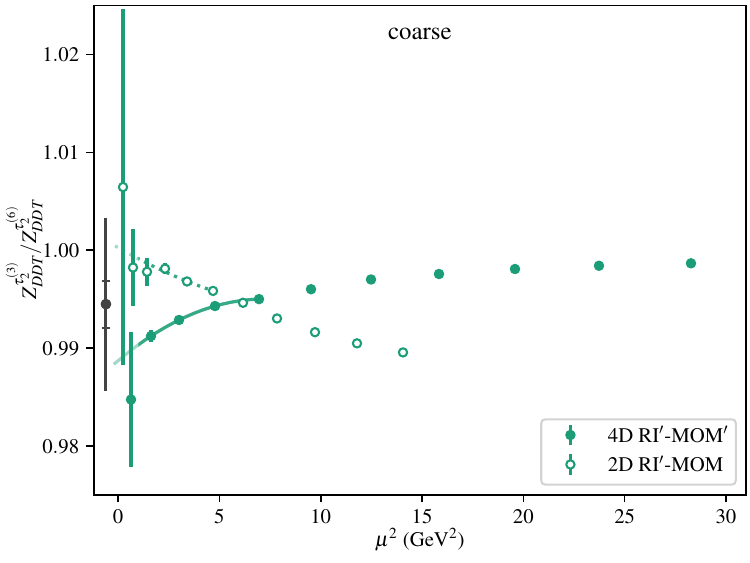}
  \includegraphics[width=0.495\textwidth]{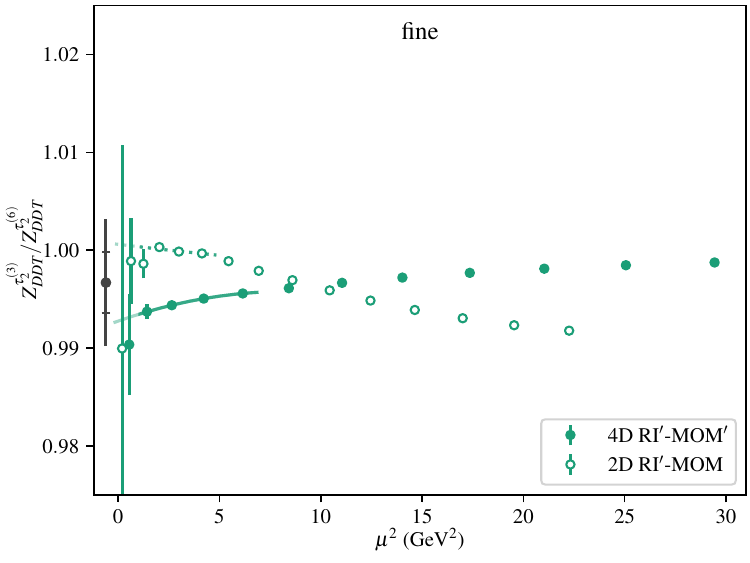}
  \includegraphics[width=0.495\textwidth]{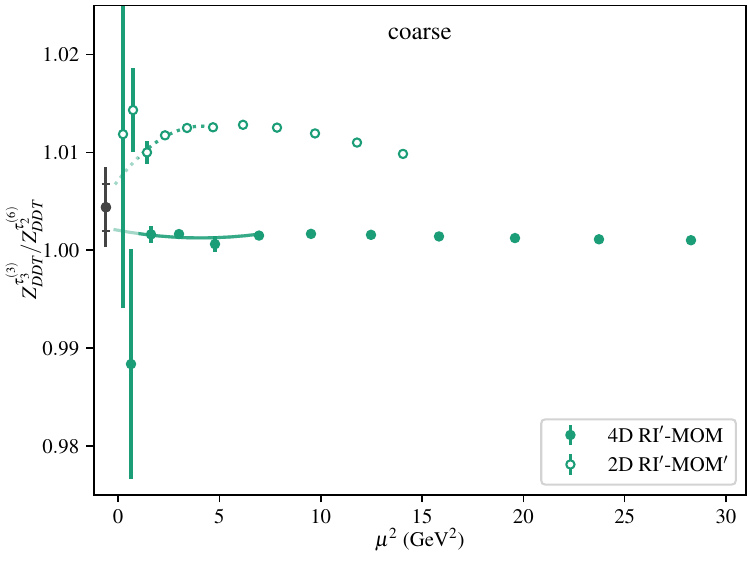}
  \includegraphics[width=0.495\textwidth]{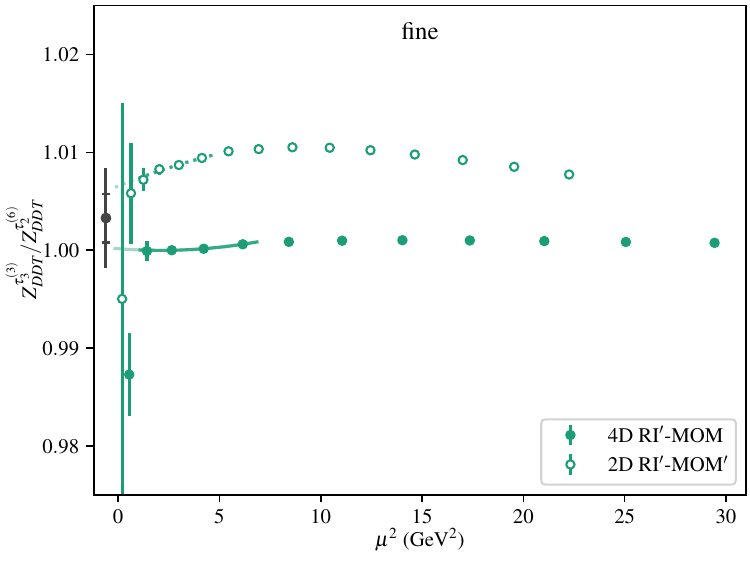}
  \caption{Scale and scheme-invariant ratios of renormalization
    factors $Z_{DDT}^{\tau^{(3)}_2}/Z_{DDT}^{\tau^{(6)}_2}$ (top) and
    $Z_{DDT}^{\tau^{(3)}_3}/Z_{DDT}^{\tau^{(6)}_2}$ (bottom) on the
    coarse (left) and fine (right) ensembles, determined using the
    condition $\Sigma^{(1)}(\mu^2)=1$ (RI$'$-MOM) or
    $\Sigma^{(1)}(\mu^2)+\frac{1}{2}\Sigma^{(2)}(\mu^2)=1$
    (RI$'$-MOM$'$). Fits (a) are shown with solid curves and fits (b)
    are shown with dotted curves.}
  \label{fig:renormalization_ratio}
\end{figure*}

For the two-derivative tensor operator, we computed matrix elements
belonging to three different irreps. The two relevant ratios of
renormalization factors in different irreps,
$Z_{DDT}^{\rho'}/Z_{DDT}^\rho$ are shown in
Fig.~\ref{fig:renormalization_ratio}. We were only able to identify
the same $O(4)$-invariant function in two different irreps using
RI$'$-MOM kinematics. As perturbation theory is not required here, we
can use much smaller momenta, performing fits with $D=0$ to (a) the 4D
data in the range $\mu^2\in[1,8]$~GeV$^2$ and (b) the 2D data in the
range $[1,5]$~GeV$^2$. These ratios are mostly within 1\% of unity.

Our final ratios of renormalization factors are given in
Table~\ref{table:ren_factors}.

\begin{table}
\begin{threeparttable}
\caption{\label{table:ren_factors}%
Vector current renormalization factor from Ref. \cite{Hasan:2019noy} and ratios of renormalization factors computed in this work.}
\begin{ruledtabular}
\begin{tabular}{lll}
                                                             & Coarse             & Fine              \\
\hline
$Z_V$                                                        & $0.9094(36)$       & $0.9438(1)$       \\
$Z^{\tau^{(4)}_2}_{DDV} / Z_V \tnote{a}$                     & $1.2000(100)(181)$ & $1.2134(83)(131)$ \\
$Z^{\tau^{(4)}_3}_{DDA} / Z_V \tnote{a}$                     & $1.2001(102)(221)$ & $1.2168(77)(167)$ \\
$Z^{\tau^{(6)}_2}_{DDT} / Z_V \tnote{a}$                     & $1.1921(108)(138)$ & $1.2017(74)(180)$ \\
$Z^{\tau^{(3)}_2}_{DDT} / Z^{\tau^{(6)}_2}_{DDT}$            & $0.9945(24)(85)$   & $0.9967(31)(57)$  \\
$Z^{\tau^{(3)}_3}_{DDT} / Z^{\tau^{(6)}_2}_{DDT}$            & $1.0044(24)(33)$   & $1.0033(25)(44)$
\end{tabular}
\end{ruledtabular}
\begin{tablenotes}
\footnotesize
\item[a] $\overline{\mathrm{MS}}$ scheme at scale 2~GeV.
\end{tablenotes}
\end{threeparttable}
\end{table}

%
%
%
%
\FloatBarrier

\section{Continuum estimate\label{app:cont_est}}

In this Appendix, we provide a functional-form and prior-width study motivating the choices we made for the continuum estimate in Sec.~\ref{sec:results}.
For the continuum estimate, we tested the following ansätze
\begin{equation}\label{eq:cont_est_functional_form}
\begin{alignedat}{1}
(i)   \quad \melmom \inCurve{a} &= \melmom_{\text{cont}} \, , \\
(ii)  \quad \melmom \inCurve{a} &= \melmom_{\text{cont}} \cdot \inCurve{1 + m_1 \, \alpha_s a \Lambda_{\text{QCD}}  } \, , \\
(iii) \quad \melmom \inCurve{a} &= \melmom_{\text{cont}} \cdot \inCurve{1 + m_2 \inCurve{a \Lambda_{\text{QCD}} }^2 } \, , \\
(iv)  \quad \melmom \inCurve{a} &= \melmom_{\text{cont}} \cdot \inCurve{1 + m_1 \inCurve{\alpha_s a \Lambda_{\text{QCD}} } + m_2 \inCurve{a \Lambda_{\text{QCD}} }^2 } \, .
\end{alignedat}
\end{equation}
The functional form \emph{(i)} represents a fit to a constant.
The functional forms \emph{(ii)} and \emph{(iii)} are two-parameter fits, and are respectively linear and quadratic in the lattice spacing.
The functional form \emph{(iv)} is that of Eq.~\eqref{eq:continuum_limit}, and, as stated in Sec.~\ref{sec:results}, is the functional form we expect in our case.
However, the functional form \emph{(iv)} has three free parameters.
Having only two lattice spacings, a fit to this functional form requires the use of priors.
We constrained the parameters $m_1$ and $m_2$ with the Gaussian prior $\mathcal{N}(0,\sigma)$, and performed the fit to the form \emph{(iv)} multiple times, varying $\sigma$ from 0.5 to 32.

\begin{figure*}[ht!]
\makebox[\textwidth][c]{\includegraphics[width=\textwidth]{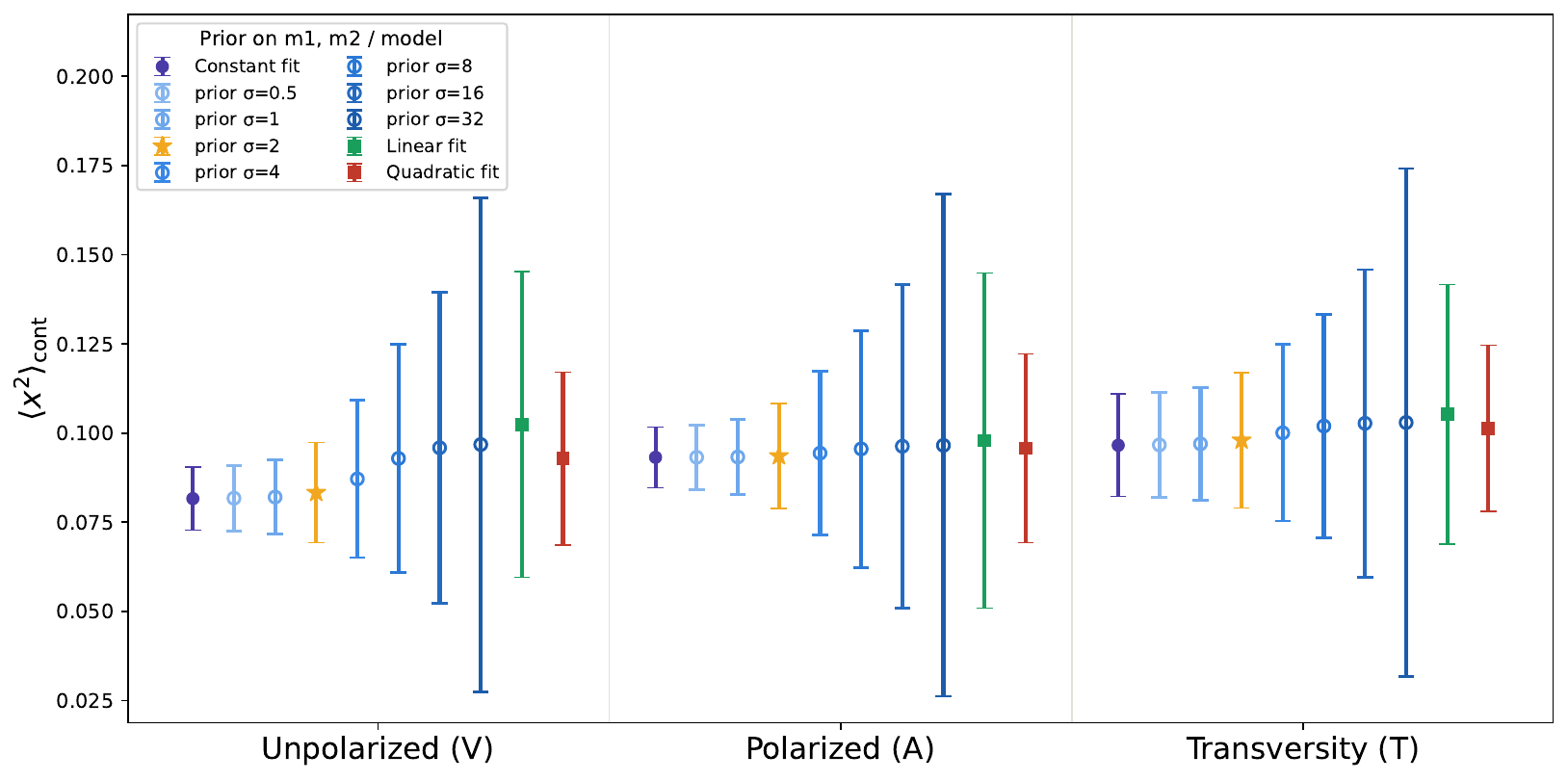}}%
\caption{\label{fig:cont_est_study}%
Prior-width and functional-form study of the continuum estimate.
For the three channels, the obtained value $\melmom_{\rm cont}$ is plotted for different choices of functional form and prior width on the parameters.
A filled circle is used for the outcome of the fit to a constant.
Empty circles are used for the three-parameter fits, where $m_1$ and $m_2$ are constrained by the Gaussian prior $\mathcal{N}(0,\sigma)$;
from left to right in the plot, $\sigma$ increases from $0.5$ to $32$.
The value corresponding to $\sigma=2$ is highlighted with a yellow star.
The outcomes of the linear and quadratic two-parameter fits are plotted with a green and red square, respectively.
}
\end{figure*}

The results we obtain are presented in Fig.~\ref{fig:cont_est_study}.
In the figure, the continuum value $\melmom_{\text{cont}}$ is plotted for each of the functional forms employed for the extrapolation, and, in the case of form \emph{(iv)}, for each choice of prior width.
The value that we choose as our final result is highlighted with a yellow star in the figure.
As is evident, for each channel, our choice is completely compatible with the outcome of a constant fit, but it exhibits a larger uncertainty.
This fact justifies our preference for a prior-constrained estimate over a constant fit.
In the case of functional form \emph{(iv)}, an increasing value of the prior width $\sigma$ leads, as expected, to results characterized by an increasing uncertainty.
More interestingly, the central values of the results remain stable;
we observe a slight upward trend in central values for increasing values of $\sigma$, but these values are in any case compatible with ours.
Analogously, the results of the linear and quadratic two-parameter fits are compatible with our results, but they exhibit a larger uncertainty.

Given the functional-form and prior-width study presented here, we can conclude that our choice for the continuum estimates is the best one available with the given data.
Although obtained from a prior-constrained extrapolation, our results show no significant difference in central values with respect to the results of a fit to a constant.
Moreover, they exhibit a smaller uncertainty compared to that of the values obtained from the other models studied.

%

\begin{acknowledgments}
%
We thank the Budapest-Marseille-Wuppertal Collaboration for making their configurations available to us and Nesreen Hasan for calculating some of the correlation functions analyzed here during the course of a different project.
%
Calculations for this project were done using the Qlua software suite \cite{Pochinsky:QLua}, and some of them made use of the QOPQDP adaptive multigrid solver \cite{Babich:2010qb, Osborn:QOPQDP}.
%
We gratefully acknowledge the computing time granted by the JARA Vergabegremium and provided on the JARA Partition part of the supercomputer JURECA \cite{Juelich:2018JURECA} at Jülich Supercomputing Centre (JSC);
%
computing time granted by the John von Neumann Institute for Computing (NIC) on the supercomputers JUQUEEN \cite{Juelich:2015JUQUEEN}, JURECA, and JUWELS \cite{Juelich:2019JUWELS} at JSC;
%
and computing time granted by the HLRS Steering Committee on Hazel Hen at the High Performance Computing Centre Stuttgart (HLRS).
%
This work is supported by the Deutsche Forschungsgemeinschaft (DFG, German Research Foundation) as part of the EXC 3701 "Color meets Flavor" – project no. 533766364, by the MKW NRW under funding code NW21-024-A (NRW-FAIR), and by the Helmholtz Association through the AIDAS laboratory and through EXNET-01-07\_CmF.
%
M.E.~is supported by the U.S.~DOE Office of Science, Office of Nuclear Physics, through grant DE-FG02-96ER40965.
%
S.M.~is supported by the U.S. Department of Energy, Office of Science, Office of High Energy Physics under Award Number DE-SC0009913.
\end{acknowledgments}
%
%

\bibliography{bibliography}

@article{Rodekamp:2023wpe,
    author = "Rodekamp, Marcel and Engelhardt, Michael and Green, Jeremy R. and Krieg, Stefan and Liuti, Simonetta and Meinel, Stefan and Negele, John W. and Pochinsky, Andrew and Syritsyn, Sergey",
    title = "{Moments of nucleon unpolarized, polarized, and transversity parton distribution functions from lattice QCD at the physical point}",
    eprint = "2401.05360",
    archivePrefix = "arXiv",
    primaryClass = "hep-lat",
    reportNumber = "DESY-23-212",
    doi = "10.1103/PhysRevD.109.074508",
    journal = "Phys. Rev. D",
    volume = "109",
    number = "7",
    pages = "074508",
    year = "2024"
}

@article{Spiegel:2024dec,
    author = "Spiegel, Sebastian and Lehner, Christoph",
    title = "{High-precision continuum limit study of the HVP short-distance window}",
    eprint = "2410.17053",
    archivePrefix = "arXiv",
    primaryClass = "hep-lat",
    doi = "10.1103/mj3d-yq87",
    journal = "Phys. Rev. D",
    volume = "111",
    number = "11",
    pages = "114517",
    year = "2025"
}

@article{Blum:2012uh,
    author = "Blum, Thomas and Izubuchi, Taku and Shintani, Eigo",
    title = "{New class of variance-reduction techniques using lattice symmetries}",
    eprint = "1208.4349",
    archivePrefix = "arXiv",
    primaryClass = "hep-lat",
    reportNumber = "RBRC-967",
    doi = "10.1103/PhysRevD.88.094503",
    journal = "Phys. Rev. D",
    volume = "88",
    number = "9",
    pages = "094503",
    year = "2013"
}

@article{Shintani:2014vja,
    author = "Shintani, Eigo and Arthur, Rudy and Blum, Thomas and Izubuchi, Taku and Jung, Chulwoo and Lehner, Christoph",
    title = "{Covariant approximation averaging}",
    eprint = "1402.0244",
    archivePrefix = "arXiv",
    primaryClass = "hep-lat",
    doi = "10.1103/PhysRevD.91.114511",
    journal = "Phys. Rev. D",
    volume = "91",
    number = "11",
    pages = "114511",
    year = "2015"
}

@article{NNPDF:2024nan,
    author = "Ball, Richard D. and others",
    collaboration = "NNPDF",
    title = "{The path to $\hbox {N}^3\hbox {LO}$ parton distributions}",
    eprint = "2402.18635",
    archivePrefix = "arXiv",
    primaryClass = "hep-ph",
    reportNumber = "Nikhef-2023-020, TIF-UNIMI-2023-23, Edinburgh 2023/29, CERN-TH-2024-033",
    doi = "10.1140/epjc/s10052-024-12891-7",
    journal = "Eur. Phys. J. C",
    volume = "84",
    number = "7",
    pages = "659",
    year = "2024"
}

@article{McGowan:2022nag,
    author = "McGowan, J. and Cridge, T. and Harland-Lang, L. A. and Thorne, R. S.",
    title = "{Approximate N$^{3}$LO parton distribution functions with theoretical uncertainties: MSHT20aN$^3$LO PDFs}",
    eprint = "2207.04739",
    archivePrefix = "arXiv",
    primaryClass = "hep-ph",
    doi = "10.1140/epjc/s10052-023-11236-0",
    journal = "Eur. Phys. J. C",
    volume = "83",
    number = "3",
    pages = "185",
    year = "2023",
    note = "[Erratum: Eur.Phys.J.C 83, 302 (2023)]"
}

@article{Cruz-Martinez:2025ahf,
    author = "Cruz-Martinez, Juan and Hasenack, Toon and Hekhorn, Felix and Magni, Giacomo and Nocera, Emanuele R. and Rabemananjara, Tanjona R. and Rojo, Juan and Sharma, Tanishq and van Seeventer, Gijs",
    title = "{NNPDFpol2.0: a global determination of polarised PDFs and their uncertainties at next-to-next-to-leading order}",
    eprint = "2503.11814",
    archivePrefix = "arXiv",
    primaryClass = "hep-ph",
    reportNumber = "CERN-TH-2025-054",
    doi = "10.1007/JHEP07(2025)168",
    journal = "JHEP",
    volume = "07",
    pages = "168",
    year = "2025"
}

@article{Bertone:2024taw,
    author = "Bertone, Valerio and Chiefa, Amedeo and Nocera, Emanuele R.",
    collaboration = "MAP (Multi-dimensional Analyses of Partonic distributions)",
    title = "{Helicity-dependent parton distribution functions at next-to-next-to-leading order accuracy from inclusive and semi-inclusive deep-inelastic scattering data}",
    eprint = "2404.04712",
    archivePrefix = "arXiv",
    primaryClass = "hep-ph",
    doi = "10.1016/j.physletb.2025.139497",
    journal = "Phys. Lett. B",
    volume = "865",
    pages = "139497",
    year = "2025"
}

@article{DeRoeck:2011na,
    author = "De Roeck, A. and Thorne, R. S.",
    title = "{Structure Functions}",
    eprint = "1103.0555",
    archivePrefix = "arXiv",
    primaryClass = "hep-ph",
    doi = "10.1016/j.ppnp.2011.06.001",
    journal = "Prog. Part. Nucl. Phys.",
    volume = "66",
    pages = "727--781",
    year = "2011"
}

@article{Forte:2013wc,
    author = "Forte, Stefano and Watt, Graeme",
    title = "{Progress in the Determination of the Partonic Structure of the Proton}",
    eprint = "1301.6754",
    archivePrefix = "arXiv",
    primaryClass = "hep-ph",
    reportNumber = "IFUM-1005-FT, ZU-TH-02-13",
    doi = "10.1146/annurev-nucl-102212-170607",
    journal = "Ann. Rev. Nucl. Part. Sci.",
    volume = "63",
    pages = "291--328",
    year = "2013"
}

@article{Blumlein:2012bf,
    author = "Blümlein, Johannes",
    title = "{The Theory of Deeply Inelastic Scattering}",
    eprint = "1208.6087",
    archivePrefix = "arXiv",
    primaryClass = "hep-ph",
    reportNumber = "DESY-12-096, DO-TH-12-19, SFB-CPP-12-38, LPN-12-056",
    doi = "10.1016/j.ppnp.2012.09.006",
    journal = "Prog. Part. Nucl. Phys.",
    volume = "69",
    pages = "28--84",
    year = "2013"
}

@article{Perez:2012um,
    author = "Perez, E. and Rizvi, E.",
    title = "{The Quark and Gluon Structure of the Proton}",
    eprint = "1208.1178",
    archivePrefix = "arXiv",
    primaryClass = "hep-ex",
    doi = "10.1088/0034-4885/76/4/046201",
    journal = "Rept. Prog. Phys.",
    volume = "76",
    pages = "046201",
    year = "2013"
}

@article{Ball:2012wy,
    author = "Ball, Richard D. and others",
    title = "{Parton Distribution Benchmarking with LHC Data}",
    eprint = "1211.5142",
    archivePrefix = "arXiv",
    primaryClass = "hep-ph",
    reportNumber = "CERN-PH-TH-2012-263",
    doi = "10.1007/JHEP04(2013)125",
    journal = "JHEP",
    volume = "04",
    pages = "125",
    year = "2013"
}

@article{Gao:2017yyd,
    author = "Gao, Jun and Harland-Lang, Lucian and Rojo, Juan",
    title = "{The Structure of the Proton in the LHC Precision Era}",
    eprint = "1709.04922",
    archivePrefix = "arXiv",
    primaryClass = "hep-ph",
    doi = "10.1016/j.physrep.2018.03.002",
    journal = "Phys. Rept.",
    volume = "742",
    pages = "1--121",
    year = "2018"
}

@article{Ethier:2020way,
    author = "Ethier, Jacob J. and Nocera, Emanuele R.",
    title = "{Parton Distributions in Nucleons and Nuclei}",
    eprint = "2001.07722",
    archivePrefix = "arXiv",
    primaryClass = "hep-ph",
    reportNumber = "Nikhef/2020-003",
    doi = "10.1146/annurev-nucl-011720-042725",
    journal = "Ann. Rev. Nucl. Part. Sci.",
    volume = "70",
    pages = "43--76",
    year = "2020"
}

@article{Constantinou:2022yye,
    author = "Constantinou, Martha and others",
    title = "{Lattice QCD Calculations of Parton Physics}",
    eprint = "2202.07193",
    archivePrefix = "arXiv",
    primaryClass = "hep-lat",
    reportNumber = "JLAB-THY-22-3564,MIT-CTP/5408,MSUHEP-22-004",
    month = "2",
    year = "2022",
    journal = ""
}

@article{Lin:2025hka,
    author = "Lin, Huey-Wen",
    title = "{Mapping parton distributions of hadrons with lattice QCD}",
    eprint = "2506.05025",
    archivePrefix = "arXiv",
    primaryClass = "hep-lat",
    reportNumber = "MSUHEP-24-012",
    doi = "10.1016/j.ppnp.2025.104177",
    journal = "Prog. Part. Nucl. Phys.",
    volume = "144",
    pages = "104177",
    year = "2025"
}

@article{Cichy:2018mum,
    author = "Cichy, Krzysztof and Constantinou, Martha",
    title = "{A guide to light-cone PDFs from Lattice QCD: an overview of approaches, techniques and results}",
    eprint = "1811.07248",
    archivePrefix = "arXiv",
    primaryClass = "hep-lat",
    doi = "10.1155/2019/3036904",
    journal = "Adv. High Energy Phys.",
    volume = "2019",
    pages = "3036904",
    year = "2019"
}

@article{Gao:2022uhg,
    author = "Gao, Xiang and Hanlon, Andrew D. and Holligan, Jack and Karthik, Nikhil and Mukherjee, Swagato and Petreczky, Peter and Syritsyn, Sergey and Zhao, Yong",
    title = "{Unpolarized proton PDF at NNLO from lattice QCD with physical quark masses}",
    eprint = "2212.12569",
    archivePrefix = "arXiv",
    primaryClass = "hep-lat",
    doi = "10.1103/PhysRevD.107.074509",
    journal = "Phys. Rev. D",
    volume = "107",
    number = "7",
    pages = "074509",
    year = "2023"
}

@article{Detmold:2025lyb,
    author = "Detmold, William and Grebe, Anthony V. and Kanamori, Issaku and Lin, C. -J. David and Perry, Robert J. and Zhao, Yong",
    collaboration = "HOPE",
    title = "{Parton physics from a heavy-quark operator product expansion: Lattice QCD calculation of the fourth moment of the pion distribution amplitude}",
    eprint = "2509.04799",
    archivePrefix = "arXiv",
    primaryClass = "hep-lat",
    reportNumber = "MIT-CTP/5913, FERMILAB-PUB-25-0615-T",
    doi = "10.1103/xs4z-nblq",
    journal = "Phys. Rev. D",
    volume = "113",
    number = "1",
    pages = "014510",
    year = "2026"
}

@article{Gockeler:1996mu,
    author = "Göckeler, M. and Horsley, R. and Ilgenfritz, Ernst-Michael and Perlt, H. and Rakow, Paul E. L. and Schierholz, G. and Schiller, A.",
    title = "{Lattice operators for moments of the structure functions and their transformation under the hypercubic group}",
    eprint = "hep-lat/9602029",
    archivePrefix = "arXiv",
    reportNumber = "DESY-96-031, HLRZ-96-11, HUB-EP-96-4",
    doi = "10.1103/PhysRevD.54.5705",
    journal = "Phys. Rev. D",
    volume = "54",
    pages = "5705--5714",
    year = "1996"
}

@article{Sakata:1974hd,
    author = "Sakata, I.",
    title = "{A general method for obtaining Clebsch-Gordan coefficients of finite groups. I. Its application to point and space groups}",
    doi = "10.1063/1.1666528",
    journal = "J. Math. Phys.",
    volume = "15",
    pages = "1702--1709",
    year = "1974"
}

@article{Baake:1981qe,
    author = "Baake, M. and Gemünden, B. and Oedingen, R.",
    title = "{Structure and Representations of the Symmetry Group of the Four-dimensional Cube}",
    reportNumber = "BONN-HE-81-13",
    doi = "10.1063/1.525461",
    journal = "J. Math. Phys.",
    volume = "23",
    pages = "944",
    year = "1982",
    note = "[Erratum: J.Math.Phys. 23, 2595 (1982)]"
}

@article{Francis:2025pgf,
    author = "Francis, Anthony and Fritzsch, Patrick and Karur, Rohith and Kim, Jangho and Pederiva, Giovanni and Pefkou, Dimitra A. and Rago, Antonio and Shindler, Andrea and Walker-Loud, Andr{\'e} and Zafeiropoulos, Savvas",
    title = "{Moments of parton distributions functions of the pion from lattice QCD using gradient flow}",
    eprint = "2510.26738",
    archivePrefix = "arXiv",
    primaryClass = "hep-lat",
    month = "10",
    year = "2025",
    journal=""
}

@article{Francis:2025rya,
    author = "Francis, Anthony and others",
    title = "{Gradient Flow for Parton Distribution Functions: First Application to the Pion}",
    eprint = "2509.02472",
    archivePrefix = "arXiv",
    primaryClass = "hep-lat",
    doi = "10.1103/z3wr-zk8n",
    journal = "Phys. Rev. Lett.",
    volume = "136",
    number = "17",
    pages = "171903",
    year = "2026"
}

@article{Gusken:1989qx,
    author = "Güsken, S.",
    editor = "Cabibbo, N. and Marinari, E. and Parisi, G. and Petronzio, Roberto and Maiani, L. and Martinelli, G. and Pettorino, Roberto",
    title = "{A Study of smearing techniques for hadron correlation functions}",
    doi = "10.1016/0920-5632(90)90273-W",
    journal = "Nucl. Phys. B Proc. Suppl.",
    volume = "17",
    pages = "361--364",
    year = "1990"
}

@article{Bali:2016lva,
    author = {Bali, Gunnar S. and Lang, Bernhard and Musch, Bernhard U. and Sch{\"a}fer, Andreas},
    title = "{Novel quark smearing for hadrons with high momenta in lattice QCD}",
    eprint = "1602.05525",
    archivePrefix = "arXiv",
    primaryClass = "hep-lat",
    doi = "10.1103/PhysRevD.93.094515",
    journal = "Phys. Rev. D",
    volume = "93",
    number = "9",
    pages = "094515",
    year = "2016"
}

@article{Bruno:2014jqa,
    author = "Bruno, Mattia and others",
    title = "{Simulation of QCD with N$_{f} =$ 2 $+$ 1 flavors of non-perturbatively improved Wilson fermions}",
    eprint = "1411.3982",
    archivePrefix = "arXiv",
    primaryClass = "hep-lat",
    reportNumber = "DESY-14-216, FTUAM-14-48, HIM-2014-01, HU-EP-14-51, MITP-14-091, SFB-CPP-14-89, IFT-UAM-CSIC-14-117",
    doi = "10.1007/JHEP02(2015)043",
    journal = "JHEP",
    volume = "02",
    pages = "043",
    year = "2015"
}

@article{BMW:2010skj,
    author = "Dürr, S. and Fodor, Z. and Hoelbling, C. and Katz, S. D. and Krieg, S. and Kurth, T. and Lellouch, L. and Lippert, T. and Szabó, K. K. and Vulvert, G.",
    collaboration = "BMW",
    title = "{Lattice QCD at the physical point: Simulation and analysis details}",
    eprint = "1011.2711",
    archivePrefix = "arXiv",
    primaryClass = "hep-lat",
    reportNumber = "WUB-10-34, CPT-P055-2010",
    doi = "10.1007/JHEP08(2011)148",
    journal = "JHEP",
    volume = "08",
    pages = "148",
    year = "2011"
}

@article{BMW:2010ucx,
    author = "Dürr, S. and Fodor, Z. and Hoelbling, C. and Katz, S. D. and Krieg, S. and Kurth, T. and Lellouch, L. and Lippert, T. and Szabó, K. K. and Vulvert, G.",
    collaboration = "BMW",
    title = "{Lattice QCD at the physical point: light quark masses}",
    eprint = "1011.2403",
    archivePrefix = "arXiv",
    primaryClass = "hep-lat",
    reportNumber = "WUB-10-33, CPT-P054-2010",
    doi = "10.1016/j.physletb.2011.05.053",
    journal = "Phys. Lett. B",
    volume = "701",
    pages = "265--268",
    year = "2011"
}

@article{Martinelli:1994ty,
    author = "Martinelli, G. and Pittori, C. and Sachrajda, Christopher T. and Testa, M. and Vladikas, A.",
    title = "{A general method for non-perturbative renormalization of lattice operators}",
    eprint = "hep-lat/9411010",
    archivePrefix = "arXiv",
    reportNumber = "CERN-TH-7342-94, LPTHE-ORSAY-94-52, ROME-1022-1994, SHEP-94-95-03",
    doi = "10.1016/0550-3213(95)00126-D",
    journal = "Nucl. Phys. B",
    volume = "445",
    pages = "81--108",
    year = "1995"
}

@article{Gockeler:1998ye,
    author = "Göckeler, M. and Horsley, R. and Oelrich, H. and Perlt, H. and Petters, D. and Rakow, Paul E. L. and Schäfer, A. and Schierholz, G. and Schiller, A.",
    title = "{Non-perturbative renormalization of composite operators in lattice QCD}",
    eprint = "hep-lat/9807044",
    archivePrefix = "arXiv",
    reportNumber = "DESY-98-097, TPR-98-19, HUB-EP-98-45",
    doi = "10.1016/S0550-3213(99)00036-X",
    journal = "Nucl. Phys. B",
    volume = "544",
    pages = "699--733",
    year = "1999"
}

@article{Sturm:2009kb,
    author = "Sturm, C. and Aoki, Y. and Christ, N. H. and Izubuchi, T. and Sachrajda, C. T. C. and Soni, A.",
    title = "{Renormalization of quark bilinear operators in a momentum-subtraction scheme with a nonexceptional subtraction point}",
    eprint = "0901.2599",
    archivePrefix = "arXiv",
    primaryClass = "hep-ph",
    reportNumber = "CU-TP-1186, KANAZAWA-09-01, RBRC-771, SHEP-09-02",
    doi = "10.1103/PhysRevD.80.014501",
    journal = "Phys. Rev. D",
    volume = "80",
    pages = "014501",
    year = "2009"
}

@article{Gracey:2010ci,
    author = "Gracey, J. A.",
    title = "{RI$'$/SMOM scheme amplitudes for deep inelastic scattering operators at one loop in QCD}",
    eprint = "1009.3895",
    archivePrefix = "arXiv",
    primaryClass = "hep-ph",
    reportNumber = "LTH-885",
    doi = "10.1103/PhysRevD.83.054024",
    journal = "Phys. Rev. D",
    volume = "83",
    pages = "054024",
    year = "2011"
}

@article{Gracey:2006zr,
    author = "Gracey, J. A.",
    title = "{Three loop anomalous dimensions of higher moments of the non-singlet twist-2 Wilson and transversity operators in the $\overline{\text{MS}}$ and RI$'$ schemes}",
    eprint = "hep-ph/0609231",
    archivePrefix = "arXiv",
    reportNumber = "LTH-718",
    doi = "10.1088/1126-6708/2006/10/040",
    journal = "JHEP",
    volume = "10",
    pages = "040",
    year = "2006"
}

@article{Braun:2016wnx,
    author = "Braun, Vladimir M. and others",
    title = "{The {\ensuremath{\rho}}-meson light-cone distribution amplitudes from lattice QCD}",
    eprint = "1612.02955",
    archivePrefix = "arXiv",
    primaryClass = "hep-lat",
    reportNumber = "LTH-1108",
    doi = "10.1007/JHEP04(2017)082",
    journal = "JHEP",
    volume = "04",
    pages = "082",
    year = "2017"
}

@article{Velizhanin:2014fua,
    author = "Velizhanin, V. N.",
    title = "{Four-loop anomalous dimension of the third and fourth moments of the nonsinglet twist-2 operator in QCD}",
    eprint = "1411.1331",
    archivePrefix = "arXiv",
    primaryClass = "hep-ph",
    reportNumber = "HU-MATHEMATIK-2014-31, HU-EP-14-47",
    doi = "10.1142/S0217751X20501997",
    journal = "Int. J. Mod. Phys. A",
    volume = "35",
    number = "32",
    pages = "2050199",
    year = "2020"
}

@article{Baikov:2015tea,
    author = {Baikov, P. A. and Chetyrkin, K. G. and K{\"u}hn, J. H.},
    editor = {Bl{\"u}mlein, Johannes and Jansen, Karl and Kr{\"a}mer, Michael and K{\"u}hn, Johann H.},
    title = "{Massless Propagators, $R(s)$ and Multiloop QCD}",
    eprint = "1501.06739",
    archivePrefix = "arXiv",
    primaryClass = "hep-ph",
    reportNumber = "TTP15-002",
    doi = "10.1016/j.nuclphysbps.2015.03.002",
    journal = "Nucl. Part. Phys. Proc.",
    volume = "261-262",
    pages = "3--18",
    year = "2015"
}

@article{Kniehl:2020nhw,
    author = "Kniehl, Bernd A. and Veretin, Oleg L.",
    title = "{Moments $n=2$ and $n=3$ of the Wilson twist-two operators at three loops in the RI${}'$/SMOM scheme}",
    eprint = "2009.11325",
    archivePrefix = "arXiv",
    primaryClass = "hep-ph",
    reportNumber = "DESY 20-156, DESY-20-156, 0418-9833",
    doi = "10.1016/j.nuclphysb.2020.115229",
    journal = "Nucl. Phys. B",
    volume = "961",
    pages = "115229",
    year = "2020"
}

@article{Boucaud:2005rm,
    author = "Boucaud, Philippe and de Soto, F. and Leroy, J. P. and Le Yaouanc, A. and Micheli, J. and Moutarde, H. and Pène, O. and Rodriguez-Quintero, J.",
    title = "{Artifacts and $\langle A^2 \rangle$ power corrections: Reexamining $Z_\psi(p^2)$ and $Z_V$ in the momentum-subtraction scheme}",
    eprint = "hep-lat/0504017",
    archivePrefix = "arXiv",
    reportNumber = "LPT-ORSAY-02-99, UHU-FT-05-09",
    doi = "10.1103/PhysRevD.74.034505",
    journal = "Phys. Rev. D",
    volume = "74",
    pages = "034505",
    year = "2006"
}

@article{Jay:2020jkz,
    author = "Jay, William I. and Neil, Ethan T.",
    title = "{Bayesian model averaging for analysis of lattice field theory results}",
    eprint = "2008.01069",
    archivePrefix = "arXiv",
    primaryClass = "stat.ME",
    reportNumber = "FERMILAB-PUB-20-374-T",
    doi = "10.1103/PhysRevD.103.114502",
    journal = "Phys. Rev. D",
    volume = "103",
    pages = "114502",
    year = "2021"
}

@article{Neil:2023pgt,
    author = "Neil, Ethan T. and Sitison, Jacob W.",
    title = "{Model averaging approaches to data subset selection}",
    eprint = "2305.19417",
    archivePrefix = "arXiv",
    primaryClass = "stat.ME",
    doi = "10.1103/PhysRevE.108.045308",
    journal = "Phys. Rev. E",
    volume = "108",
    number = "4",
    pages = "045308",
    year = "2023"
}

@article{Neil:2022joj,
    author = "Neil, Ethan T. and Sitison, Jacob W.",
    title = "{Improved information criteria for Bayesian model averaging in lattice field theory}",
    eprint = "2208.14983",
    archivePrefix = "arXiv",
    primaryClass = "stat.ME",
    doi = "10.1103/PhysRevD.109.014510",
    journal = "Phys. Rev. D",
    volume = "109",
    number = "1",
    pages = "014510",
    year = "2024"
}

@article{Shindler:2023xpd,
    author = "Shindler, Andrea",
    title = "{Moments of parton distribution functions of any order from lattice QCD}",
    eprint = "2311.18704",
    archivePrefix = "arXiv",
    primaryClass = "hep-lat",
    reportNumber = "TTK-23-31",
    doi = "10.1103/PhysRevD.110.L051503",
    journal = "Phys. Rev. D",
    volume = "110",
    number = "5",
    pages = "L051503",
    year = "2024"
}

@article{Bali:2014gha,
    author = {Bali, Gunnar S. and Collins, Sara and Gl{\"a}{\ss}le, Benjamin and G{\"o}ckeler, Meinulf and Najjar, Johannes and R{\"o}dl, Rudolf H. and Sch{\"a}fer, Andreas and Schiel, Rainer W. and Sternbeck, Andr{\'e} and S{\"o}ldner, Wolfgang},
    title = "{The moment $\langle x\rangle_{u-d}$ of the nucleon from $N_f=2$ lattice QCD down to nearly physical quark masses}",
    eprint = "1408.6850",
    archivePrefix = "arXiv",
    primaryClass = "hep-lat",
    doi = "10.1103/PhysRevD.90.074510",
    journal = "Phys. Rev. D",
    volume = "90",
    number = "7",
    pages = "074510",
    year = "2014"
}

@article{Knippschild:20118n,
  author = "Knippschild, Bastian  and  Wittig, Hartmut  and  Capitani, Stefano  and  Della Morte, Michele",
  title = "{Systematic errors in extracting nucleon properties from lattice QCD}",
  doi = "10.22323/1.105.0147",
  journal = "PoS",
  year = 2011,
  volume = "Lattice 2010",
  pages = "147"
}

@article{Donnellan:2011/a,
  author = "Donnellan, Michael  and  Bulava, John  and  Sommer, Rainer",
    collaboration = "ALPHA",
    title = "{The $B^*B\pi$ Coupling in the Static Limit}",
    eprint = "1011.4393",
    archivePrefix = "arXiv",
    primaryClass = "hep-lat",
  doi = "10.22323/1.105.0303",
  journal = "PoS",
  year = 2011,
  volume = "Lattice 2010",
  pages = "303"
}

@article{Buckley:2014ana,
    author = {Buckley, Andy and Ferrando, James and Lloyd, Stephen and Nordstr{\"o}m, Karl and Page, Ben and R{\"u}fenacht, Martin and Sch{\"o}nherr, Marek and Watt, Graeme},
    title = "{LHAPDF6: parton density access in the LHC precision era}",
    eprint = "1412.7420",
    archivePrefix = "arXiv",
    primaryClass = "hep-ph",
    reportNumber = "GLAS-PPE-2014-05, MCNET-14-29, IPPP-14-111, DCPT-14-222",
    doi = "10.1140/epjc/s10052-015-3318-8",
    journal = "Eur. Phys. J. C",
    volume = "75",
    pages = "132",
    year = "2015"
}

@article{Nocera:2014gqa,
    author = "Nocera, Emanuele R. and Ball, Richard D. and Forte, Stefano and Ridolfi, Giovanni and Rojo, Juan",
    collaboration = "NNPDF",
    title = "{A first unbiased global determination of polarized PDFs and their uncertainties}",
    eprint = "1406.5539",
    archivePrefix = "arXiv",
    primaryClass = "hep-ph",
    reportNumber = "CERN-PH-TH-2014-106, IFUN-1028-FT, EDINBURGH-14-11, OUTP-14-06P",
    doi = "10.1016/j.nuclphysb.2014.08.008",
    journal = "Nucl. Phys. B",
    volume = "887",
    pages = "276--308",
    year = "2014"
}

@article{Gockeler:2005cj,
    author = "Göckeler, M. and Hägler, Ph. and Horsley, R. and Pleiter, D. and Rakow, Paul E. L. and Schäfer, A. and Schierholz, G. and Zanotti, J. M.",
    collaboration = "QCDSF, UKQCD",
    title = "{Quark helicity flip generalized parton distributions from two-flavor lattice QCD}",
    eprint = "hep-lat/0507001",
    archivePrefix = "arXiv",
    reportNumber = "DESY-05-107, EDINBURGH-2005-05, MPP-2005-60",
    doi = "10.1016/j.physletb.2005.09.002",
    journal = "Phys. Lett. B",
    volume = "627",
    pages = "113--123",
    year = "2005"
}

@article{Alexandrou:2016tjj,
    author = "Alexandrou, Constantia",
    editor = "Pasquini, Barbara and Salm{\`e}, Giovanni",
    title = "{Parton distribution functions from Lattice QCD}",
    eprint = "1602.08726",
    archivePrefix = "arXiv",
    primaryClass = "hep-lat",
    doi = "10.1007/s00601-016-1073-5",
    journal = "Few Body Syst.",
    volume = "57",
    number = "8",
    pages = "621--626",
    year = "2016"
}

@article{Wandzura:1977qf,
    author = "Wandzura, S. and Wilczek, Frank",
    title = "{Sum Rules for Spin Dependent Electroproduction: Test of Relativistic Constituent Quarks}",
    reportNumber = "Print-77-0325 (PRINCETON)",
    doi = "10.1016/0370-2693(77)90700-6",
    journal = "Phys. Lett. B",
    volume = "72",
    pages = "195--198",
    year = "1977"
}

@article{Jaffe:1989xx,
    author = "Jaffe, R. L.",
    title = "{$g_{2}$---The Nucleon's Other Spin-Dependent Structure Function}",
    reportNumber = "MIT-CTP-1798",
    journal = "Comments Nucl. Part. Phys.",
    volume = "19",
    number = "5",
    pages = "239--257",
    year = "1990"
}

@article{Jaffe:1991ra,
    author = "Jaffe, R. L. and Ji, Xiang-Dong",
    title = "{Chiral odd parton distributions and Drell-Yan processes}",
    reportNumber = "MIT-CTP-2005",
    doi = "10.1016/0550-3213(92)90110-W",
    journal = "Nucl. Phys. B",
    volume = "375",
    pages = "527--560",
    year = "1992"
}

@article{Capitani:1994qn,
    author = "Capitani, Stefano and Rossi, Giancarlo",
    title = "{Deep inelastic scattering in improved lattice QCD. 1. The First moment of structure functions}",
    eprint = "hep-lat/9401014",
    archivePrefix = "arXiv",
    reportNumber = "ROM2F-93-38, ROME-978-1993",
    doi = "10.1016/0550-3213(94)00428-H",
    journal = "Nucl. Phys. B",
    volume = "433",
    pages = "351--389",
    year = "1995"
}

@article{Gockeler:1995wg,
    author = "Göckeler, M. and Horsley, R. and Ilgenfritz, Ernst-Michael and Perlt, H. and Rakow, Paul E. L. and Schierholz, G. and Schiller, A.",
    title = "{Polarized and unpolarized nucleon structure functions from lattice QCD}",
    eprint = "hep-lat/9508004",
    archivePrefix = "arXiv",
    reportNumber = "DESY-95-128, HLRZ-95-36, HUB-EP-95-9",
    doi = "10.1103/PhysRevD.53.2317",
    journal = "Phys. Rev. D",
    volume = "53",
    pages = "2317--2325",
    year = "1996"
}

@article{Gockeler:1996hg,
    author = "Göckeler, M. and Horsley, R. and Ilgenfritz, Ernst-Michael and Perlt, H. and Rakow, Paul E. L. and Schierholz, G. and Schiller, A.",
    title = "{Perturbative renormalization of lattice bilinear quark operators}",
    eprint = "hep-lat/9603006",
    archivePrefix = "arXiv",
    reportNumber = "DESY-96-034, HLRZ-96-13, HUB-EP-96-5",
    doi = "10.1016/0550-3213(96)00217-9",
    journal = "Nucl. Phys. B",
    volume = "472",
    pages = "309--333",
    year = "1996"
}

@article{FlavourLatticeAveragingGroupFLAG:2024oxs,
    author = "Aoki, Y. and others",
    collaboration = "Flavour Lattice Averaging Group (FLAG)",
    title = "{FLAG review 2024}",
    eprint = "2411.04268",
    archivePrefix = "arXiv",
    primaryClass = "hep-lat",
    reportNumber = "CERN-TH-2024-192, FERMILAB-PUB-24-0785-T",
    doi = "10.1103/nfzp-p5dn",
    journal = "Phys. Rev. D",
    volume = "113",
    number = "1",
    pages = "014508",
    year = "2026"
}

@article{Alexandrou:2023qbg,
    author = "Alexandrou, Constantia and Bacchio, Simone and Constantinou, Martha and Finkenrath, Jacob and Frezzotti, Roberto and Kostrzewa, Bartosz and Koutsou, Giannis and Spanoudes, Gregoris and Urbach, Carsten",
    collaboration = "Extended Twisted Mass",
    title = "{Nucleon axial and pseudoscalar form factors using twisted-mass fermion ensembles at the physical point}",
    eprint = "2309.05774",
    archivePrefix = "arXiv",
    primaryClass = "hep-lat",
    doi = "10.1103/PhysRevD.109.034503",
    journal = "Phys. Rev. D",
    volume = "109",
    number = "3",
    pages = "034503",
    year = "2024"
}

@article{Jang:2023zts,
    author = "Jang, Yong-Chull and Gupta, Rajan and Bhattacharya, Tanmoy and Yoon, Boram and Lin, Huey-Wen",
    collaboration = "Precision Neutron Decay Matrix Elements (PNDME)",
    title = "{Nucleon isovector axial form factors}",
    eprint = "2305.11330",
    archivePrefix = "arXiv",
    primaryClass = "hep-lat",
    reportNumber = "Los Alamos LA-UR-23-25225",
    doi = "10.1103/PhysRevD.109.014503",
    journal = "Phys. Rev. D",
    volume = "109",
    number = "1",
    pages = "014503",
    year = "2024"
}

@article{Walker-Loud:2019cif,
    author = "Walker-Loud, Andr{\'e} and others",
    title = "{Lattice QCD Determination of $g_A$}",
    eprint = "1912.08321",
    archivePrefix = "arXiv",
    primaryClass = "hep-lat",
    reportNumber = "RIKEN-iTHEMS-Report-19, LLNL-PROC-800060",
    doi = "10.22323/1.317.0020",
    journal = "PoS",
    volume = "CD2018",
    pages = "020",
    year = "2020"
}

@article{Alexandrou:2019brg,
    author = "Alexandrou, C. and Bacchio, S. and Constantinou, M. and Finkenrath, J. and Hadjiyiannakou, K. and Jansen, K. and Koutsou, G. and Vaquero Aviles-Casco, A.",
    title = "{Nucleon axial, tensor, and scalar charges and $\sigma$-terms in lattice QCD}",
    eprint = "1909.00485",
    archivePrefix = "arXiv",
    primaryClass = "hep-lat",
    doi = "10.1103/PhysRevD.102.054517",
    journal = "Phys. Rev. D",
    volume = "102",
    number = "5",
    pages = "054517",
    year = "2020"
}

@article{Gupta:2018qil,
    author = "Gupta, Rajan and Jang, Yong-Chull and Yoon, Boram and Lin, Huey-Wen and Cirigliano, Vincenzo and Bhattacharya, Tanmoy",
    title = "{Isovector Charges of the Nucleon from 2+1+1-flavor Lattice QCD}",
    eprint = "1806.09006",
    archivePrefix = "arXiv",
    primaryClass = "hep-lat",
    reportNumber = "LA-UR-18-25335, MSUHEP-18-011",
    doi = "10.1103/PhysRevD.98.034503",
    journal = "Phys. Rev. D",
    volume = "98",
    pages = "034503",
    year = "2018"
}

@article{Chang:2018uxx,
    author = "Chang, C. C. and others",
    title = "{A per-cent-level determination of the nucleon axial coupling from quantum chromodynamics}",
    eprint = "1805.12130",
    archivePrefix = "arXiv",
    primaryClass = "hep-lat",
    reportNumber = "BNL-203631-2018-JAAM, INT-PUB-18-021, LLNL-JRNL-747003, RBRC-1283, LTH-1166, RIKEN-ITHEMS-REPORT-18",
    doi = "10.1038/s41586-018-0161-8",
    journal = "Nature",
    volume = "558",
    number = "7708",
    pages = "91--94",
    year = "2018"
}

@article{Berkowitz:2017gql,
    author = "Berkowitz, Evan and others",
    title = "{An accurate calculation of the nucleon axial charge with lattice QCD}",
    eprint = "1704.01114",
    archivePrefix = "arXiv",
    primaryClass = "hep-lat",
    reportNumber = "JLAB-THY-17-2485",
    month = "4",
    year = "2017",
    journal = "",
}

@article{Bhattacharya:2016zcn,
    author = "Bhattacharya, Tanmoy and Cirigliano, Vincenzo and Cohen, Saul and Gupta, Rajan and Lin, Huey-Wen and Yoon, Boram",
    title = "{Axial, Scalar and Tensor Charges of the Nucleon from 2+1+1-flavor Lattice QCD}",
    eprint = "1606.07049",
    archivePrefix = "arXiv",
    primaryClass = "hep-lat",
    reportNumber = "LA-UR-16-20522",
    doi = "10.1103/PhysRevD.94.054508",
    journal = "Phys. Rev. D",
    volume = "94",
    number = "5",
    pages = "054508",
    year = "2016"
}

@article{Djukanovic:2024krw,
    author = "Djukanovic, Dalibor and von Hippel, Georg and Meyer, Harvey B. and Ottnad, Konstantin and Wittig, Hartmut",
    title = "{Improved analysis of isovector nucleon matrix elements with Nf=2+1 flavors of O(a) improved Wilson fermions}",
    eprint = "2402.03024",
    archivePrefix = "arXiv",
    primaryClass = "hep-lat",
    reportNumber = "MITP-24-014",
    doi = "10.1103/PhysRevD.109.074507",
    journal = "Phys. Rev. D",
    volume = "109",
    number = "7",
    pages = "074507",
    year = "2024"
}

@article{Tsuji:2023llh,
    author = "Tsuji, Ryutaro and Aoki, Yasumichi and Ishikawa, Ken-Ichi and Kuramashi, Yoshinobu and Sasaki, Shoichi and Sato, Kohei and Shintani, Eigo and Watanabe, Hiromasa and Yamazaki, Takeshi",
    collaboration = "PACS",
    title = "{Nucleon form factors in Nf=2+1 lattice QCD at the physical point: Finite lattice spacing effect on the root-mean-square radii}",
    eprint = "2311.10345",
    archivePrefix = "arXiv",
    primaryClass = "hep-lat",
    reportNumber = "UTHEP-783, UTCCS-P-149, HUPD-2307, YITP-23-143",
    doi = "10.1103/PhysRevD.109.094505",
    journal = "Phys. Rev. D",
    volume = "109",
    number = "9",
    pages = "094505",
    year = "2024"
}

@article{Bali:2023sdi,
    author = {Bali, Gunnar S. and Collins, Sara and Heybrock, Simon and L{\"o}ffler, Marius and R{\"o}dl, Rudolf and S{\"o}ldner, Wolfgang and Weish{\"a}upl, Simon},
    collaboration = "RQCD",
    title = "{Octet baryon isovector charges from Nf=2+1 lattice QCD}",
    eprint = "2305.04717",
    archivePrefix = "arXiv",
    primaryClass = "hep-lat",
    doi = "10.1103/PhysRevD.108.034512",
    journal = "Phys. Rev. D",
    volume = "108",
    number = "3",
    pages = "034512",
    year = "2023"
}

@article{QCDSFUKQCDCSSM:2023qlx,
    author = "Smail, R. E. and others",
    collaboration = "QCDSF/UKQCD/CSSM",
    title = "{Constraining beyond the standard model nucleon isovector charges}",
    eprint = "2304.02866",
    archivePrefix = "arXiv",
    primaryClass = "hep-lat",
    reportNumber = "ADP-23-10/T1219, DESY-23-047, LTH 1336, MIT-CTP/5581",
    doi = "10.1103/PhysRevD.108.094511",
    journal = "Phys. Rev. D",
    volume = "108",
    number = "9",
    pages = "094511",
    year = "2023"
}

@article{Tsuji:2022ric,
    author = "Tsuji, Ryutaro and Tsukamoto, Natsuki and Aoki, Yasumichi and Ishikawa, Ken-Ichi and Kuramashi, Yoshinobu and Sasaki, Shoichi and Shintani, Eigo and Yamazaki, Takeshi",
    collaboration = "PACS",
    title = "{Nucleon isovector couplings in Nf=2+1 lattice QCD at the physical point}",
    eprint = "2207.11914",
    archivePrefix = "arXiv",
    primaryClass = "hep-lat",
    reportNumber = "UTHEP-771, UTCCS-P-145, HUPD-2209",
    doi = "10.1103/PhysRevD.106.094505",
    journal = "Phys. Rev. D",
    volume = "106",
    number = "9",
    pages = "094505",
    year = "2022"
}

@article{Djukanovic:2022wru,
    author = "Djukanovic, Dalibor and von Hippel, Georg and Koponen, Jonna and Meyer, Harvey B. and Ottnad, Konstantin and Schulz, Tobias and Wittig, Hartmut",
    title = "{Isovector axial form factor of the nucleon from lattice QCD}",
    eprint = "2207.03440",
    archivePrefix = "arXiv",
    primaryClass = "hep-lat",
    reportNumber = "MITP-22-053",
    doi = "10.1103/PhysRevD.106.074503",
    journal = "Phys. Rev. D",
    volume = "106",
    number = "7",
    pages = "074503",
    year = "2022"
}

@article{Park:2021ypf,
    author = "Park, Sungwoo and Gupta, Rajan and Yoon, Boram and Mondal, Santanu and Bhattacharya, Tanmoy and Jang, Yong-Chull and Jo{\'o}, B{\'a}lint and Winter, Frank",
    collaboration = "Nucleon Matrix Elements (NME)",
    title = "{Precision nucleon charges and form factors using (2+1)-flavor lattice QCD}",
    eprint = "2103.05599",
    archivePrefix = "arXiv",
    primaryClass = "hep-lat",
    reportNumber = "LA-UR-21-20526, JLAB-THY-22-3583",
    doi = "10.1103/PhysRevD.105.054505",
    journal = "Phys. Rev. D",
    volume = "105",
    number = "5",
    pages = "054505",
    year = "2022"
}

@article{Blossier:2013ioa,
    author = "Blossier, B. and Boucaud, Ph. and Brinet, M. and De Soto, F. and Morenas, V. and Pène, O. and Petrov, K. and Rodríguez-Quintero, J.",
    collaboration = "ETM",
    title = "{High statistics determination of the strong coupling constant in Taylor scheme and its OPE Wilson coefficient from lattice QCD with a dynamical charm}",
    eprint = "1310.3763",
    archivePrefix = "arXiv",
    primaryClass = "hep-ph",
    reportNumber = "LPT-Orsay-13-77, UHU-FT-13-09",
    doi = "10.1103/PhysRevD.89.014507",
    journal = "Phys. Rev. D",
    volume = "89",
    number = "1",
    pages = "014507",
    year = "2014"
}

@article{Hasan:2019noy,
    author = "Hasan, Nesreen and Green, Jeremy and Meinel, Stefan and Engelhardt, Michael and Krieg, Stefan and Negele, John and Pochinsky, Andrew and Syritsyn, Sergey",
    title = "{Nucleon axial, scalar, and tensor charges using lattice QCD at the physical pion mass}",
    eprint = "1903.06487",
    archivePrefix = "arXiv",
    primaryClass = "hep-lat",
    reportNumber = "DESY-19-044, DESY 19-044",
    doi = "10.1103/PhysRevD.99.114505",
    journal = "Phys. Rev. D",
    volume = "99",
    number = "11",
    pages = "114505",
    year = "2019"
}

@article{Harris:2019bih,
    author = "Harris, Tim and von Hippel, Georg and Junnarkar, Parikshit and Meyer, Harvey B. and Ottnad, Konstantin and Wilhelm, Jonas and Wittig, Hartmut and Wrang, Linus",
    title = "{Nucleon isovector charges and twist-2 matrix elements with $N_f=2+1$ dynamical Wilson quarks}",
    eprint = "1905.01291",
    archivePrefix = "arXiv",
    primaryClass = "hep-lat",
    doi = "10.1103/PhysRevD.100.034513",
    journal = "Phys. Rev. D",
    volume = "100",
    number = "3",
    pages = "034513",
    year = "2019"
}

@article{Shintani:2018ozy,
    author = "Shintani, Eigo and Ishikawa, Ken-Ichi and Kuramashi, Yoshinobu and Sasaki, Shoichi and Yamazaki, Takeshi",
    title = "{Nucleon form factors and root-mean-square radii on a (10.8  fm)$^4$ lattice at the physical point}",
    eprint = "1811.07292",
    archivePrefix = "arXiv",
    primaryClass = "hep-lat",
    doi = "10.1103/PhysRevD.99.014510",
    journal = "Phys. Rev. D",
    volume = "99",
    number = "1",
    pages = "014510",
    year = "2019",
    note = "[Erratum: Phys.Rev.D 102, 019902 (2020)]"
}

@article{Ishikawa:2018rew,
    author = "Ishikawa, Ken-Ichi and Kuramashi, Yoshinobu and Sasaki, Shoichi and Tsukamoto, Natsuki and Ukawa, Akira and Yamazaki, Takeshi",
    collaboration = "PACS",
    title = "{Nucleon form factors on a large volume lattice near the physical point in 2+1 flavor QCD}",
    eprint = "1807.03974",
    archivePrefix = "arXiv",
    primaryClass = "hep-lat",
    reportNumber = "UTHEP-721, UTCCS-P-113, HUPD-1806",
    doi = "10.1103/PhysRevD.98.074510",
    journal = "Phys. Rev. D",
    volume = "98",
    number = "7",
    pages = "074510",
    year = "2018"
}

@article{Liang:2018pis,
    author = "Liang, Jian and Yang, Yi-Bo and Draper, Terrence and Gong, Ming and Liu, Keh-Fei",
    title = "{Quark spins and Anomalous Ward Identity}",
    eprint = "1806.08366",
    archivePrefix = "arXiv",
    primaryClass = "hep-ph",
    doi = "10.1103/PhysRevD.98.074505",
    journal = "Phys. Rev. D",
    volume = "98",
    number = "7",
    pages = "074505",
    year = "2018"
}

@article{Yamanaka:2018uud,
    author = "Yamanaka, Nodoka and Hashimoto, Shoji and Kaneko, Takashi and Ohki, Hiroshi",
    collaboration = "JLQCD",
    title = "{Nucleon charges with dynamical overlap fermions}",
    eprint = "1805.10507",
    archivePrefix = "arXiv",
    primaryClass = "hep-lat",
    reportNumber = "KEK-CP-365",
    doi = "10.1103/PhysRevD.98.054516",
    journal = "Phys. Rev. D",
    volume = "98",
    number = "5",
    pages = "054516",
    year = "2018"
}

@article{Liu:2021irg,
    author = "Liu, Liuming and Chen, Ting and Draper, Terrence and Liang, Jian and Liu, Keh-Fei and Wang, Gen and Yang, Yi-Bo",
    collaboration = "{\ensuremath{\chi}}QCD",
    title = "{Nucleon isovector scalar charge from overlap fermions}",
    eprint = "2103.12933",
    archivePrefix = "arXiv",
    primaryClass = "hep-lat",
    doi = "10.1103/PhysRevD.104.094503",
    journal = "Phys. Rev. D",
    volume = "104",
    number = "9",
    pages = "094503",
    year = "2021"
}

@article{Abramczyk:2019fnf,
    author = "Abramczyk, Michael and Blum, Thomas and Izubuchi, Taku and Jung, Chulwoo and Lin, Meifeng and Lytle, Andrew and Ohta, Shigemi and Shintani, Eigo",
    title = "{Nucleon mass and isovector couplings in 2+1-flavor dynamical domain-wall lattice QCD near physical mass}",
    eprint = "1911.03524",
    archivePrefix = "arXiv",
    primaryClass = "hep-lat",
    reportNumber = "KEK-TH-2167, RBRC-1320",
    doi = "10.1103/PhysRevD.101.034510",
    journal = "Phys. Rev. D",
    volume = "101",
    number = "3",
    pages = "034510",
    year = "2020"
}

@article{Alexandrou:2022dtc,
    author = "Alexandrou, C. and others",
    title = "{Moments of the nucleon transverse quark spin densities using lattice QCD}",
    eprint = "2202.09871",
    archivePrefix = "arXiv",
    primaryClass = "hep-lat",
    doi = "10.1103/PhysRevD.107.054504",
    journal = "Phys. Rev. D",
    volume = "107",
    number = "5",
    pages = "054504",
    year = "2023"
}

@article{Bhattacharya:2015wna,
    author = "Bhattacharya, Tanmoy and Cirigliano, Vincenzo and Cohen, Saul and Gupta, Rajan and Joseph, Anosh and Lin, Huey-Wen and Yoon, Boram",
    collaboration = "PNDME",
    title = "{Iso-vector and Iso-scalar Tensor Charges of the Nucleon from Lattice QCD}",
    eprint = "1506.06411",
    archivePrefix = "arXiv",
    primaryClass = "hep-lat",
    reportNumber = "DESY-15-128, LA-UR-15-23801",
    doi = "10.1103/PhysRevD.92.094511",
    journal = "Phys. Rev. D",
    volume = "92",
    number = "9",
    pages = "094511",
    year = "2015"
}

@article{Bhattacharya:2015esa,
    author = "Bhattacharya, Tanmoy and Cirigliano, Vincenzo and Gupta, Rajan and Lin, Huey-Wen and Yoon, Boram",
    title = "{Neutron Electric Dipole Moment and Tensor Charges from Lattice QCD}",
    eprint = "1506.04196",
    archivePrefix = "arXiv",
    primaryClass = "hep-lat",
    reportNumber = "LA-UR-15-24210",
    doi = "10.1103/PhysRevLett.115.212002",
    journal = "Phys. Rev. Lett.",
    volume = "115",
    number = "21",
    pages = "212002",
    year = "2015"
}

@article{Mondal:2020cmt,
    author = "Mondal, Santanu and Gupta, Rajan and Park, Sungwoo and Yoon, Boram and Bhattacharya, Tanmoy and Lin, Huey-Wen",
    title = "{Moments of nucleon isovector structure functions in $2+1+1$-flavor QCD}",
    eprint = "2005.13779",
    archivePrefix = "arXiv",
    primaryClass = "hep-lat",
    reportNumber = "LA-UR-20-23801",
    doi = "10.1103/PhysRevD.102.054512",
    journal = "Phys. Rev. D",
    volume = "102",
    number = "5",
    pages = "054512",
    year = "2020"
}

@article{Alexandrou:2020sml,
    author = "Alexandrou, C. and Bacchio, S. and Constantinou, M. and Finkenrath, J. and Hadjiyiannakou, K. and Jansen, K. and Koutsou, G. and Panagopoulos, H. and Spanoudes, G.",
    title = "{Complete flavor decomposition of the spin and momentum fraction of the proton using lattice QCD simulations at physical pion mass}",
    eprint = "2003.08486",
    archivePrefix = "arXiv",
    primaryClass = "hep-lat",
    doi = "10.1103/PhysRevD.101.094513",
    journal = "Phys. Rev. D",
    volume = "101",
    number = "9",
    pages = "094513",
    year = "2020"
}

@article{Alexandrou:2019ali,
    author = "Alexandrou, C. and others",
    title = "{Moments of nucleon generalized parton distributions from lattice QCD simulations at physical pion mass}",
    eprint = "1908.10706",
    archivePrefix = "arXiv",
    primaryClass = "hep-lat",
    doi = "10.1103/PhysRevD.101.034519",
    journal = "Phys. Rev. D",
    volume = "101",
    number = "3",
    pages = "034519",
    year = "2020"
}

@article{Mondal:2021oot,
    author = "Mondal, Santanu and Bhattacharya, Tanmoy and Gupta, Rajan and Jo{\'o}, B{\'a}lint and Lin, Huey-Wen and Park, Sungwoo and Winter, Frank and Yoon, Boram",
    title = "{Nucleon isovector momentum fraction, helicity and transversity moment using Lattice QCD}",
    eprint = "2201.00067",
    archivePrefix = "arXiv",
    primaryClass = "hep-lat",
    reportNumber = "LA-UR-21-32143, MSUHEP-21-036, JLAB-CST-21-3504",
    doi = "10.22323/1.396.0513",
    journal = "PoS",
    volume = "LATTICE2021",
    pages = "513",
    year = "2021"
}

@article{Mondal:2020ela,
    author = "Mondal, Santanu and Gupta, Rajan and Park, Sungwoo and Yoon, Boram and Bhattacharya, Tanmoy and Jo{\'o}, B{\'a}lint and Winter, Frank",
    collaboration = "Nucleon Matrix Elements (NME)",
    title = "{Nucleon momentum fraction, helicity and transversity from 2+1-flavor lattice QCD}",
    eprint = "2011.12787",
    archivePrefix = "arXiv",
    primaryClass = "hep-lat",
    reportNumber = "LA-UR-20-28586",
    doi = "10.1007/JHEP04(2021)044",
    journal = "JHEP",
    volume = "21",
    pages = "004",
    year = "2020"
}

@article{Yang:2018nqn,
    author = "Yang, Yi-Bo and Liang, Jian and Bi, Yu-Jiang and Chen, Ying and Draper, Terrence and Liu, Keh-Fei and Liu, Zhaofeng",
    title = "{Proton Mass Decomposition from the QCD Energy Momentum Tensor}",
    eprint = "1808.08677",
    archivePrefix = "arXiv",
    primaryClass = "hep-lat",
    doi = "10.1103/PhysRevLett.121.212001",
    journal = "Phys. Rev. Lett.",
    volume = "121",
    number = "21",
    pages = "212001",
    year = "2018"
}

@article{Green:2012ud,
    author = "Green, J. R. and Engelhardt, M. and Krieg, S. and Negele, J. W. and Pochinsky, A. V. and Syritsyn, S. N.",
    title = "{Nucleon Structure from Lattice QCD Using a Nearly Physical Pion Mass}",
    eprint = "1209.1687",
    archivePrefix = "arXiv",
    primaryClass = "hep-lat",
    reportNumber = "MIT-CTP-4399, NT-LBL-12-015, UCB-NPAT-12-013, WUB-12-18",
    doi = "10.1016/j.physletb.2014.05.075",
    journal = "Phys. Lett. B",
    volume = "734",
    pages = "290--295",
    year = "2014"
}

@article{LHPC:2010jcs,
    author = "Bratt, J. D. and others",
    collaboration = "LHPC",
    title = "{Nucleon structure from mixed action calculations using 2+1 flavors of asqtad sea and domain wall valence fermions}",
    eprint = "1001.3620",
    archivePrefix = "arXiv",
    primaryClass = "hep-lat",
    reportNumber = "TUM-T39-10-01, TUM-EFT-6-10, CERN-PH-TH-2010-005, JLAB-THY-10-1128",
    doi = "10.1103/PhysRevD.82.094502",
    journal = "Phys. Rev. D",
    volume = "82",
    pages = "094502",
    year = "2010"
}

@article{Aoki:2010xg,
    author = "Aoki, Yasumichi and Blum, Tom and Lin, Huey-Wen and Ohta, Shigemi and Sasaki, Shoichi and Tweedie, Robert and Zanotti, James and Yamazaki, Takeshi",
    title = "{Nucleon isovector structure functions in (2+1)-flavor QCD with domain wall fermions}",
    eprint = "1003.3387",
    archivePrefix = "arXiv",
    primaryClass = "hep-lat",
    doi = "10.1103/PhysRevD.82.014501",
    journal = "Phys. Rev. D",
    volume = "82",
    pages = "014501",
    year = "2010"
}

@article{Bali:2018zgl,
    author = {Bali, Gunnar S. and Collins, Sara and G{\"o}ckeler, Meinulf and R{\"o}dl, Rudolf and Sch{\"a}fer, Andreas and Sternbeck, Andr{\'e}},
    title = "{Nucleon generalized form factors from two-flavor lattice QCD}",
    eprint = "1812.08256",
    archivePrefix = "arXiv",
    primaryClass = "hep-lat",
    doi = "10.1103/PhysRevD.100.014507",
    journal = "Phys. Rev. D",
    volume = "100",
    number = "1",
    pages = "014507",
    year = "2019"
}

@article{Alexandrou:2017oeh,
    author = "Alexandrou, C. and Constantinou, M. and Hadjiyiannakou, K. and Jansen, K. and Kallidonis, C. and Koutsou, G. and Vaquero Avil{\'e}s-Casco, A. and Wiese, C.",
    title = "{Nucleon Spin and Momentum Decomposition Using Lattice QCD Simulations}",
    eprint = "1706.02973",
    archivePrefix = "arXiv",
    primaryClass = "hep-lat",
    reportNumber = "DESY-17-086",
    doi = "10.1103/PhysRevLett.119.142002",
    journal = "Phys. Rev. Lett.",
    volume = "119",
    number = "14",
    pages = "142002",
    year = "2017"
}

@article{Abdel-Rehim:2015owa,
    author = "Abdel-Rehim, A. and others",
    title = "{Nucleon and pion structure with lattice QCD simulations at physical value of the pion mass}",
    eprint = "1507.04936",
    archivePrefix = "arXiv",
    primaryClass = "hep-lat",
    doi = "10.1103/PhysRevD.92.114513",
    journal = "Phys. Rev. D",
    volume = "92",
    number = "11",
    pages = "114513",
    year = "2015",
    note = "[Erratum: Phys.Rev.D 93, 039904 (2016)]"
}

@article{Lin:2017snn,
    author = "Lin, Huey-Wen and others",
    title = "{Parton distributions and lattice QCD calculations: a community white paper}",
    eprint = "1711.07916",
    archivePrefix = "arXiv",
    primaryClass = "hep-ph",
    reportNumber = "DESY-17-185, IFJPAN-IV-2017-19, INT-PUB-17-042, JLAB-THY-17-2604, MSUHEP-17-017, OUTP-17-15P, SMU-HEP-17-08, NIKHEF-2017-047",
    doi = "10.1016/j.ppnp.2018.01.007",
    journal = "Prog. Part. Nucl. Phys.",
    volume = "100",
    pages = "107--160",
    year = "2018"
}

@article{Gockeler:2005vw,
    author = "Göckeler, M. and Horsley, R. and Pleiter, D. and Rakow, Paul E. L. and Schäfer, A. and Schierholz, G. and Stüben, H. and Zanotti, J. M.",
    title = "{Investigation of the second moment of the nucleon's $g_1$ and $g_2$ structure functions in two-flavor lattice QCD}",
    eprint = "hep-lat/0506017",
    archivePrefix = "arXiv",
    reportNumber = "DESY-05-076, EDINBURGH-2005-04, MPP-2005-52",
    doi = "10.1103/PhysRevD.72.054507",
    journal = "Phys. Rev. D",
    volume = "72",
    pages = "054507",
    year = "2005"
}

@article{Gockeler:2004vx,
    author = "Göckeler, M. and Hägler, P. and Horsley, R. and Pleiter, D. and Rakow, Paul E. L. and Schäfer, A. and Schierholz, G. and Zanotti, J. M.",
    editor = "Bodwin, Geoffrey T. and Sinclair, D. K. and Eichten, E. and Holmgren, D. and Kronfeld, Andreas S. and Mackenzie, P. and Okamoto, M. and Simone, J. and El-Khadra, Aida X.",
    collaboration = "QCDSF",
    title = "{Generalized parton distributions and structure functions from full lattice QCD}",
    eprint = "hep-lat/0409162",
    archivePrefix = "arXiv",
    reportNumber = "DESY-04-189, LU-ITP-2004-029, EDINBURGH-2004-21",
    doi = "10.1016/j.nuclphysbps.2004.11.141",
    journal = "Nucl. Phys. B Proc. Suppl.",
    volume = "140",
    pages = "399--404",
    year = "2005"
}

@article{LHPC:2002xzk,
    author = "Dolgov, D. and others",
    collaboration = "LHPC, TXL",
    title = "{Moments of nucleon light cone quark distributions calculated in full lattice QCD}",
    eprint = "hep-lat/0201021",
    archivePrefix = "arXiv",
    reportNumber = "MIT-CTP-3219, JLAB-THY-02-06",
    doi = "10.1103/PhysRevD.66.034506",
    journal = "Phys. Rev. D",
    volume = "66",
    pages = "034506",
    year = "2002"
}

@article{Fan:2020nzz,
    author = "Fan, Zhouyou and Gao, Xiang and Li, Ruizi and Lin, Huey-Wen and Karthik, Nikhil and Mukherjee, Swagato and Petreczky, Peter and Syritsyn, Sergey and Yang, Yi-Bo and Zhang, Rui",
    title = "{Isovector parton distribution functions of the proton on a superfine lattice}",
    eprint = "2005.12015",
    archivePrefix = "arXiv",
    primaryClass = "hep-lat",
    doi = "10.1103/PhysRevD.102.074504",
    journal = "Phys. Rev. D",
    volume = "102",
    number = "7",
    pages = "074504",
    year = "2020"
}

@article{Alexandrou:2021oih,
    author = "Alexandrou, Constantia and Constantinou, Martha and Hadjiyiannakou, Kyriakos and Jansen, Karl and Manigrasso, Floriano",
    title = "{Flavor decomposition of the nucleon unpolarized, helicity, and transversity parton distribution functions from lattice QCD simulations}",
    eprint = "2106.16065",
    archivePrefix = "arXiv",
    primaryClass = "hep-lat",
    doi = "10.1103/PhysRevD.104.054503",
    journal = "Phys. Rev. D",
    volume = "104",
    number = "5",
    pages = "054503",
    year = "2021"
}

@article{Gao:2023ktu,
    author = "Gao, Xiang and Hanlon, Andrew D. and Mukherjee, Swagato and Petreczky, Peter and Shi, Qi and Syritsyn, Sergey and Zhao, Yong",
    title = "{Transversity PDFs of the proton from lattice QCD with physical quark masses}",
    eprint = "2310.19047",
    archivePrefix = "arXiv",
    primaryClass = "hep-lat",
    doi = "10.1103/PhysRevD.109.054506",
    journal = "Phys. Rev. D",
    volume = "109",
    number = "5",
    pages = "054506",
    year = "2024"
}

@article{HadStruc:2021qdf,
    author = "Egerer, Colin and others",
    collaboration = "HadStruc",
    title = "{Transversity parton distribution function of the nucleon using the pseudodistribution approach}",
    eprint = "2111.01808",
    archivePrefix = "arXiv",
    primaryClass = "hep-lat",
    reportNumber = "JLAB-THY-21-3521",
    doi = "10.1103/PhysRevD.105.034507",
    journal = "Phys. Rev. D",
    volume = "105",
    number = "3",
    pages = "034507",
    year = "2022"
}

@article{Gamberg:2022kdb,
    author = "Gamberg, Leonard and Malda, Michel and Miller, Joshua A. and Pitonyak, Daniel and Prokudin, Alexei and Sato, Nobuo",
    collaboration = "JAM",
    title = "{Updated QCD global analysis of single transverse-spin asymmetries: Extracting $\tilde H$, and the role of the Soffer bound and lattice QCD}",
    eprint = "2205.00999",
    archivePrefix = "arXiv",
    primaryClass = "hep-ph",
    reportNumber = "JLAB-THY-22-3604",
    doi = "10.1103/PhysRevD.106.034014",
    journal = "Phys. Rev. D",
    volume = "106",
    number = "3",
    pages = "034014",
    year = "2022"
}

@misc{Gamberg:jam3d_library,
  author       = {Gamberg, Leonard and Malda, M. and Miller, J. and Pitonyak, Daniel and Prokudin, Alexei and Sato, Nobuo},
  title        = {JAM3D Library Notebook},
  howpublished = {\url{https://colab.research.google.com/github/pitonyak25/jam3d_dev_lib/blob/main/JAM3D_Library.ipynb}},
  note         = {Accessed: 2026-04-13}
}

@article{Gockeler:2004wp,
    author = "Göckeler, M. and Horsley, R. and Pleiter, D. and Rakow, Paul E. L. and Schierholz, G.",
    collaboration = "QCDSF",
    title = "{A Lattice determination of moments of unpolarised nucleon structure functions using improved Wilson fermions}",
    eprint = "hep-ph/0410187",
    archivePrefix = "arXiv",
    reportNumber = "DESY-04-194, EDINBURGH-2004-24, LTH-638, LU-ITP-2004-039",
    doi = "10.1103/PhysRevD.71.114511",
    journal = "Phys. Rev. D",
    volume = "71",
    pages = "114511",
    year = "2005"
}

@article{NNPDF:2017mvq,
    author = "Ball, Richard D. and others",
    collaboration = "NNPDF",
    title = "{Parton distributions from high-precision collider data}",
    eprint = "1706.00428",
    archivePrefix = "arXiv",
    primaryClass = "hep-ph",
    reportNumber = "EDINBURGH-2017-08, NIKHEF-2017-006, OUTP-17-04P, TIF-UNIMI-2017-3, CAVENDISH-HEP-17-06, CERN-TH-2017-077, Edinburgh 2017/08,
  Nikhef/2017-006, OUTP-17-04P,TIF-UNIMI-2017-3",
    doi = "10.1140/epjc/s10052-017-5199-5",
    journal = "Eur. Phys. J. C",
    volume = "77",
    number = "10",
    pages = "663",
    year = "2017"
}

@article{Dulat:2015mca,
    author = "Dulat, Sayipjamal and Hou, Tie-Jiun and Gao, Jun and Guzzi, Marco and Huston, Joey and Nadolsky, Pavel and Pumplin, Jon and Schmidt, Carl and Stump, Daniel and Yuan, C. P.",
    title = "{New parton distribution functions from a global analysis of quantum chromodynamics}",
    eprint = "1506.07443",
    archivePrefix = "arXiv",
    primaryClass = "hep-ph",
    doi = "10.1103/PhysRevD.93.033006",
    journal = "Phys. Rev. D",
    volume = "93",
    number = "3",
    pages = "033006",
    year = "2016"
}

@article{Harland-Lang:2014zoa,
    author = "Harland-Lang, L. A. and Martin, A. D. and Motylinski, P. and Thorne, R. S.",
    title = "{Parton distributions in the LHC era: MMHT 2014 PDFs}",
    eprint = "1412.3989",
    archivePrefix = "arXiv",
    primaryClass = "hep-ph",
    reportNumber = "LCTS-2014-47, IPPP-14-97, DCPT-14-194",
    doi = "10.1140/epjc/s10052-015-3397-6",
    journal = "Eur. Phys. J. C",
    volume = "75",
    number = "5",
    pages = "204",
    year = "2015"
}

@article{Alekhin:2017kpj,
    author = {Alekhin, S. and Bl{\"u}mlein, J. and Moch, S. and Placakyte, R.},
    title = "{Parton distribution functions, $\alpha_s$, and heavy-quark masses for LHC Run II}",
    eprint = "1701.05838",
    archivePrefix = "arXiv",
    primaryClass = "hep-ph",
    reportNumber = "DESY-16-179, DO-TH-16-13",
    doi = "10.1103/PhysRevD.96.014011",
    journal = "Phys. Rev. D",
    volume = "96",
    number = "1",
    pages = "014011",
    year = "2017"
}

@article{Accardi:2016qay,
    author = "Accardi, A. and Brady, L. T. and Melnitchouk, W. and Owens, J. F. and Sato, N.",
    title = "{Constraints on large-$x$ parton distributions from new weak boson production and deep-inelastic scattering data}",
    eprint = "1602.03154",
    archivePrefix = "arXiv",
    primaryClass = "hep-ph",
    reportNumber = "JLAB-THY-16-2215",
    doi = "10.1103/PhysRevD.93.114017",
    journal = "Phys. Rev. D",
    volume = "93",
    number = "11",
    pages = "114017",
    year = "2016"
}

@article{H1:2015ubc,
    author = "Abramowicz, H. and others",
    collaboration = "H1, ZEUS",
    title = "{Combination of measurements of inclusive deep inelastic ${e^{\pm }p}$ scattering cross sections and QCD analysis of HERA data}",
    eprint = "1506.06042",
    archivePrefix = "arXiv",
    primaryClass = "hep-ex",
    reportNumber = "DESY-15-039",
    doi = "10.1140/epjc/s10052-015-3710-4",
    journal = "Eur. Phys. J. C",
    volume = "75",
    number = "12",
    pages = "580",
    year = "2015"
}

@article{Butterworth:2015oua,
    author = "Butterworth, Jon and others",
    title = "{PDF4LHC recommendations for LHC Run II}",
    eprint = "1510.03865",
    archivePrefix = "arXiv",
    primaryClass = "hep-ph",
    reportNumber = "OUTP-15-17P, SMU-HEP-15-12, TIF-UNIMI-2015-14, LCTS-2015-27, CERN-PH-TH-2015-249",
    doi = "10.1088/0954-3899/43/2/023001",
    journal = "J. Phys. G",
    volume = "43",
    pages = "023001",
    year = "2016"
}

@article{Burger:2021knd,
    author = {B{\"u}rger, S. and Wurm, T. and L{\"o}ffler, M. and G{\"o}ckeler, M. and Bali, G. and Collins, S. and Sch{\"a}fer, A. and Sternbeck, A.},
    collaboration = "RQCD",
    title = "{Lattice results for the longitudinal spin structure and color forces on quarks in a nucleon}",
    eprint = "2111.08306",
    archivePrefix = "arXiv",
    primaryClass = "hep-lat",
    doi = "10.1103/PhysRevD.105.054504",
    journal = "Phys. Rev. D",
    volume = "105",
    number = "5",
    pages = "054504",
    year = "2022"
}

@article{Gao:2026wlz,
    author = "Gao, Xiang and Hanlon, Andrew D. and Mukherjee, Swagato and Petreczky, Peter and Shu, Hai-Tao and Yao, Fei and Zhang, Rui and Zhao, Yong",
    title = "{Proton isovector helicity PDF at NNLO and the twist-3 moment $\tilde{d}_2$ from lattice QCD at physical quark masses}",
    eprint = "2604.00143",
    archivePrefix = "arXiv",
    primaryClass = "hep-lat",
    reportNumber = "MIT-CTP/6027",
    month = "3",
    year = "2026",
    journal = ""
}

@article{Pang:2024kza,
    author = "Pang, Zhuoyi and Zhang, Jian-Hui and Zhao, Dian-Jun",
    title = "{Moments from momentum derivatives in lattice QCD}",
    eprint = "2412.19862",
    archivePrefix = "arXiv",
    primaryClass = "hep-lat",
    doi = "10.1088/1674-1137/aded04",
    journal = "Chin. Phys.",
    volume = "49",
    number = "10",
    pages = "101001",
    year = "2025"
}

@article{Cocuzza:2023oam,
    author = "Cocuzza, C. and Metz, A. and Pitonyak, D. and Prokudin, A. and Sato, N. and Seidl, R.",
    collaboration = "JAM",
    title = "{Transversity Distributions and Tensor Charges of the Nucleon: Extraction from Dihadron Production and Their Universal Nature}",
    eprint = "2306.12998",
    archivePrefix = "arXiv",
    primaryClass = "hep-ph",
    reportNumber = "JLAB-THY-23-3859",
    doi = "10.1103/PhysRevLett.132.091901",
    journal = "Phys. Rev. Lett.",
    volume = "132",
    number = "9",
    pages = "091901",
    year = "2024"
}

@article{Cocuzza:2023vqs,
    author = "Cocuzza, C. and Metz, A. and Pitonyak, D. and Prokudin, A. and Sato, N. and Seidl, R.",
    collaboration = "Jefferson Lab Angular Momentum (JAM)",
    title = "{First simultaneous global QCD analysis of dihadron fragmentation functions and transversity parton distribution functions}",
    eprint = "2308.14857",
    archivePrefix = "arXiv",
    primaryClass = "hep-ph",
    reportNumber = "JLAB-THY-23-3901",
    doi = "10.1103/PhysRevD.109.034024",
    journal = "Phys. Rev. D",
    volume = "109",
    number = "3",
    pages = "034024",
    year = "2024"
}

@misc{Cocuzza:2023colab,
  author       = {Cocuzza, C. and Metz, A. and Pitonyak, D. and Prokudin, A. and Sato, N. and Seidl, R.},
  title        = {{Google Colab Notebook}},
  year         = {2023},
  howpublished = {\url{https://colab.research.google.com/github/prokudin/JAMDiFF_library/blob/main/JAMDiFF_Library.ipynb}},
  note         = {Accessed: 2023}
}

@misc{Pochinsky:QLua,
  author       = {A. Pochinsky},
  title        = {{QLua}},
  howpublished = {\url{https://github.com/usqcd-software/qlua}},
  year         = {},
  note         = {Accessed: 2026-04-08}
}

@article{Babich:2010qb,
    author = "Babich, R. and Brannick, J. and Brower, R. C. and Clark, M. A. and Manteuffel, T. A. and McCormick, S. F. and Osborn, J. C. and Rebbi, C.",
    title = "{Adaptive multigrid algorithm for the lattice Wilson-Dirac operator}",
    eprint = "1005.3043",
    archivePrefix = "arXiv",
    primaryClass = "hep-lat",
    doi = "10.1103/PhysRevLett.105.201602",
    journal = "Phys. Rev. Lett.",
    volume = "105",
    pages = "201602",
    year = "2010"
}

@misc{Osborn:QOPQDP,
  author       = {J. Osborn and others},
  title        = {{QOPQDP}},
  howpublished = {\url{https://github.com/usqcd-software/qopqdp}},
  year         = {},
  note         = {Accessed: 2026-04-08}
}

@article{Juelich:2018JURECA,
  author       = {{J\"ulich Supercomputing Centre}},
  title        = {{JURECA: Modular supercomputer at J\"ulich Supercomputing Centre}},
  journal      = {Journal of Large-Scale Research Facilities},
  volume       = {4},
  pages        = {A132},
  year         = {2018},
  doi          = {10.17815/jlsrf-4-121-1},
  url          = {http://dx.doi.org/10.17815/jlsrf-4-121-1}
}

@article{Juelich:2015JUQUEEN,
  author       = {{J\"ulich Supercomputing Centre}},
  title        = {{JUQUEEN: IBM Blue Gene/Q Supercomputer System at the J\"ulich Supercomputing Centre}},
  journal      = {Journal of Large-Scale Research Facilities},
  volume       = {1},
  pages        = {A1},
  year         = {2015},
  doi          = {10.17815/jlsrf-1-18},
  url          = {http://dx.doi.org/10.17815/jlsrf-1-18}
}

@article{Juelich:2019JUWELS,
  author       = {{J\"ulich Supercomputing Centre}},
  title        = {{JUWELS: Modular Tier-0/1 Supercomputer at the J\"ulich Supercomputing Centre}},
  journal      = {Journal of Large-Scale Research Facilities},
  volume       = {5},
  pages        = {A171},
  year         = {2019},
  doi          = {10.17815/jlsrf-5-171},
  url          = {http://dx.doi.org/10.17815/jlsrf-5-171}
}

\end{document}